\documentclass[%
  reprint,
  nofootinbib,
 superscriptaddress, 
 prd, 
 twocolumn, 
  aps,
floatfix
 ]{revtex4-2}

\usepackage{graphicx}
\usepackage{color}
\usepackage{natbib}
\usepackage{epsfig}
\usepackage{float}
\usepackage{xcolor}
\usepackage{subfigure}
\usepackage{amsmath,amssymb,amsfonts}
\usepackage[normalem]{ulem}
\usepackage[colorlinks,citecolor=blue,urlcolor=blue,linkcolor=blue]{hyperref}

\def\prl{Phys. Rev. Lett.}

\def\apj{Astrophys. J.}

\def\apjl{Astrophys. J. Lett.}

\def\mnras{Mon. Not. R. Astron. Soc.}

\def\part_n{\partial_\perp}

\def\ga{\,\,\raise0.14em\hbox{$>$}\kern-0.76em\lower0.28em\hbox
{$\sim$}\,\,}
\def\la{\,\,\raise0.14em\hbox{$<$}\kern-0.76em\lower0.28em\hbox
{$\sim$}\,\,}

\begin{document}

\preprint{APS/123-QED}

\title[Impact of pions on BNS mergers]{Impact of pions on binary neutron star mergers}

\author{Vimal Vijayan}
\email{v.vijayan@gsi.de}
\affiliation{GSI Helmholtzzentrum f\"ur Schwerionenforschung, Planckstra{\ss}e 1, 64291 Darmstadt, Germany}

\affiliation{Department of Physics and Astronomy, Ruprecht-Karls-Universität Heidelberg, Im Neuenheimer feld 226, 69120 Heidelberg, Germany}

\author{Ninoy Rahman}
\email{n.rahman@gsi.de}
\affiliation{GSI Helmholtzzentrum f\"ur Schwerionenforschung, Planckstra{\ss}e 1, 64291 Darmstadt, Germany}

\author{Andreas Bauswein}
\email{a.bauswein@gsi.de}
\affiliation{GSI Helmholtzzentrum f\"ur Schwerionenforschung, Planckstra{\ss}e 1, 64291 Darmstadt, Germany}
\affiliation{Helmholtz Forschungsakademie Hessen f{\"u}r FAIR, GSI Helmholtzzentrum für Schwerionenforschung, Planckstr. 1, 64291 Darmstadt, Germany}

\author{Gabriel Martínez-Pinedo}
\email{g.martinez@gsi.de}
\affiliation{GSI Helmholtzzentrum f\"ur Schwerionenforschung,
  Planckstra{\ss}e 1, 64291 Darmstadt, Germany} 
\affiliation{Institut f{\"u}r Kernphysik (Theoriezentrum),
    Fachbereich Physik, Technische Universit{\"a}t Darmstadt,
    Schlossgartenstra{\ss}e 2, 64289 Darmstadt, Germany} 
\affiliation{Helmholtz Forschungsakademie Hessen f{\"u}r FAIR, GSI
  Helmholtzzentrum für Schwerionenforschung, Planckstr. 1, 64291
  Darmstadt, Germany} 

\author{Ignacio L. Arbina}
\affiliation{Institut f{\"u}r Kernphysik (Theoriezentrum),
    Fachbereich Physik, Technische Universit{\"a}t Darmstadt,
    Schlossgartenstra{\ss}e 2, 64289 Darmstadt, Germany} 
\affiliation{GSI Helmholtzzentrum f\"ur Schwerionenforschung,
  Planckstra{\ss}e 1, 64291 Darmstadt, Germany} 

\begin{abstract}
  We study the impact of pions in simulations of neutron star mergers
  and explore the impact on gravitational-wave observables. We
  model charged and neutral pions as a non-interacting Boson gas with
  a chosen, constant effective mass. We add the contributions of
  pions, which can occur as a condensate or as a thermal population,
  to existing temperature and composition dependent equations of
  state. Compared to the models without pions, the presence of a pion
  condensate decreases the characteristic properties of cold,
  non-rotating neutron stars such as the maximum mass, the radius and
  the tidal deformability. We conduct relativistic hydrodynamical
  simulations of neutron star mergers for these modified equations of
  state models and compare to the original models, which ignore
  pions. Generally, the inclusion of pions leads to a softening of the
  equation of state, which is more pronounced for smaller effective
  pion masses. We find a shift of the dominant postmerger
  gravitational-wave frequency by up to 150~Hz to higher frequencies
  and a reduction of the threshold binary mass for prompt black-hole
  formation by up to 0.07~$M_\odot$. These quantitative changes
  compared to the equation of state model without pions are stronger
  for smaller effective pion masses and for underlying baryonic models
  which result in softer equations of state. We evaluate empirical relations between the threshold mass or the dominant postmerger
  gravitational-wave frequency and stellar parameters of nonrotating
  neutron stars. These relations are constructed to extract these
  stellar properties from merger observations and are built based on
  large sets of equation of state models which do not include
  pions. Comparing to our calculations with pions, we find that these
  empirical relations remain valid to good accuracy, which justifies
  their use although they neglect a possible impact of pions. Pions
  simultaneously modify merger characteristics and the properties of
  cold, nonrotating neutron stars. We also address the mass ejection
  by neutron star mergers and observe a moderate enhancement of the
  ejecta mass by a few ten per cent. While such variations may be
  within statistical uncertainties of the simulations, the increase is
  slightly more pronounced than one would have expected from empirical
  relations predicting the ejecta mass based on stellar parameters of
  nonrotating neutron stars. 
\end{abstract}

\maketitle

\section{Introduction}
The equation of state (EOS) of high-density matter determines the
stellar structure of neutron stars (NSs) and the dynamics and outcome
of binary neutron star (BNS) mergers. Therefore, the EOS is
particularly important to understand the gravitational-wave (GW)
emission, nucleosynthesis and electromagnetic radiation of merger
events~\citep{Fernandez2016,Baiotti2017,Shibata2019,Metzger2019,Bauswein_2019,Radice2020,Nakar2020,Chatziioannou2020,Cowan2021,Dietrich2021,Rosswog2022,Janka2022}.

The EOS of neutron star (NS) matter is not fully known and various
theoretical models have been put forward along with efforts to obtain
insights from observations of, for instance, BNS mergers. The
theoretical models are based on different methods to solve the nuclear
many-body problem and make different assumptions about the
constituents of high-density matter,
e.g.~\citep{Lattimer2012,Oertel2017,Baym2018,Raduta2021,Piekarewicz2022,Raduta2022,Keller.Hebeler.Schwenk:2022}. Apart
from neutrons, protons and electrons, some models additionally
consider for instance hyperons, deconfined quarks, muons, quarks,
pions, kaons, and at finite temperatures positrons, photons, neutrinos
and anti-neutrinos. The finite-temperature regime of the EOS is
relevant for phenomena such as core-collapse supernovae and binary
mergers, which can reach temperatures of several 10~MeV (see reviews
above).

Already for decades it has been conjectured that negatively charged
pions could form a Bose-Einstein condensate in the cores of
NSs~\citep{Sawyer1972,Migdal1973,Baym1974,Weise1975,Baeckman1975,Migdal1978,Ericson.Weise:1988,Migdal1990,Akmal1997}. If
and at which density such a condensate occurs, depends on the detailed
interactions between pions and nucleons and is currently not
known. One expects that pions are associated with an effective pion
mass which differs from its vacuum value of about 140~MeV. At finite
temperature one may in addition expect that a population of thermal
pions is present, where positively and negatively charged pions as
well as neutral pions may play a
role~\citep{Mayle1993,Ishizuka2008,Nakazato2008,Nakazato2010,Oertel2012,Peres2013,Fore_2020}.

The influence of pions (either as condensate or thermal population) in
BNS mergers is largely unexplored. But see
refs~\citep{Nakazato2008,Nakazato2010,Peres2013}, which investigate
pions in simulations of core-collapse supernovae. In fact only a few
currently available temperature-dependent EOS tables for astrophysical
applications include
pions~\citep{Ishizuka2008,Oertel2012,Schneider2019,Fore_2020}. These
calculations indicate a softening of the EOS compared to models
without pions, which results from a change of the proton fraction if
pions are included.

In this study we provide a first assessment of the potential impact of
pions in BNS merger simulations. We adopt a relatively simple model
describing pions as a non-interacting Boson gas with a chosen
effective mass, and by this consider both condensate and thermal
populations of pions. We add the contributions by pions to existing
EOS tables and compute the properties of isolated NSs by considering
matter in neutrinoless beta-equilibrium at zero temperature. We then
focus on the impact in dynamical simulations of BNS mergers and
analyze the influence of pions on the dynamics, on black-hole
formation, on the GW signal, and on the mass ejection in these events.

A major motivation of this exploration is to evaluate to which extent
neglecting pions in previous studies affects empirical relations for
the dominant postmerger GW frequency and the threshold mass for prompt
black-hole formation,
e.g.~\citep{Bauswein2012,Bauswein_2012,Bauswein2013,Hotokezaka2013,Bauswein2014,Bernuzzi2015,Takami2015,Rezzolla2016,Lehner2016,Koeppel2019,Breschi2019,Tsang2019,Agathos2020,Blacker_2020,Vretinaris_2020,Bauswein_2021,Tootle2021,Kashyap2022,Koelsch2022}. These
relations have been derived from calculations without
pions\footnote{Some of these studies do include a few models with
  pions such as the APR EOS~\citep{Akmal1998,Schneider2019}. However,
  the majority of employed EOS models ignores pions and thus the
  models with pions have only a negligible impact on the resulting
  fits. For simplicity, in the following we will refer to such
  empirical relations as being obtained from models without pions.}
and, partly, they are already employed to interpret observations and
to infer EOS information or they may be used in future
measurements. We find that most empirical relations also hold for models which include pions although the relations have been built based on calculations that neglect pions. This is because these relations connect properties of isolated, cold, nonrotating NSs with observable properties of the merger. Since the addition of pions modifies both the properties of isolated stars and of the merger, the empirical relations remain to good accuracy unaltered. We find this to hold for different chosen effective pion masses, which supposedly cover a reasonable range and to some extent reflects the more complicated physics of pions in the dense medium of
NSs~\citep{1996PhRvC..53.3057R,2017PhRvD..95c6020J,2021PhRvD.104e4005T,2020PhRvC.101c5809F,Fore2023}, which we do not include here. Based on these findings we conclude that it is safe to employ such relations in
existing and upcoming studies.

This paper is organized as follows. In Sect.~\ref{EOS} we describe the
inclusion of pions using in existing EOS models which originally do
not consider pions. Section~\ref{NS setup} briefly discusses the
impact on the stellar structure of isolated, nonrotating stars and
provides technical information on the hydrodynamical
simulations. Section~\ref{Results} addresses the dynamics of BNS
mergers with pions and the detailed analysis of the simulation results
including the GW signal, the threshold binary mass for prompt
black-hole formation, and the ejecta. In Sect.~\ref{sec:conclusions}
we summarize and discuss the conclusions of our findings. The appendix
briefly assess stellar equilibrium models which include also muons and
neutrinos in addition to pions. 

\section{Pionic equation of state} \label{EOS}

In this section, we discuss our procedure of adding pions in an existing non-pionic equation of state (EOS). The tabulated Steiner, Fischer, and Hempel (SFHo) EOS \citep{Hempel2010,2012ApJ...748...70H,2013ApJ...774...17S} and the DD2 EOS \citep{2005PhRvC..71f4301T, Hempel2010,2010PhRvC..81a5803T} are used in this study as base EOSs. These EOSs include neutrons, protons, light nuclei (deuterium, tritium, $^3$He, etc.), alpha particles, heavy nuclei (charge number $Z\ge6$), electrons, and positron as their constituent and assume nuclear statistical equilibrium (NSE) over the whole ranges of density and temperature. The base non-pionic EOSs tabulate various thermodynamic quantities against the baryonic density $\rho$, temperature $T$, and number fraction of positively charged particles $Y_\mathrm{p}$. We note that in the absence of pions, the net electron fraction $Y_\mathrm{e} = Y_\mathrm{e^-} - Y_\mathrm{e^+}$ is equal to $Y_\mathrm{p}$, where $Y_\mathrm{e^-}$, $Y_\mathrm{e^+}$ are the number fractions of electrons and positrons, respectively.

The charged pions $\pi^-,\,\pi^+$ and the neutral pion $\pi^0$ have
vacuum rest masses of $m_{\pi^{\pm}}=$139.57039 MeV and
$m_{\pi^0}=$134.9768 MeV \citep{2020PTEP.2020h3C01P}. We treat pions
as a free Bose gas and assume their masses do not change with baryonic
density however, see for instance~\citep{2014PhRvD..90k5027N,2017PhRvD..95c6020J,2021PhRvD.104e4005T,Fore2023} for detailed discussions about the pion mass variation in dense medium. Pions are in thermal and chemical equilibrium with
nucleons through the strong interaction. Therefore, the chemical
potentials of the pions follow
$\mu_{\pi^\pm}=\mp(\mu_\mathrm{n} - \mu_\mathrm{p})$ and
$\mu_{\pi^0} = 0$, where
$\mu_\mathrm{n},\,\mu_\mathrm{p},\,\mu_{\pi^-},\,\mu_{\pi^+}$, and
$\mu_{\pi^0}$ are the chemical potentials of neutrons, protons,
negatively charged pions, positively charged pions, and neutral pions,
respectively. Since we assume pions are a non-interacting bose gas,
the chemical potential of negatively charged pions should be equal to
or smaller than their vacuum rest mass. Additionally, charge
neutrality requires
$Y_\mathrm{p} = Y_\mathrm{e} + Y_{\pi} = Y_\mathrm{e} + Y_{\pi^-} -
Y_{\pi^+}$, where $Y_{\pi}$, $Y_{\pi^-}$, and $Y_{\pi^+}$ are the net
number fraction of charged pions, the number fraction of negatively
charged pions, and the number fraction of positively charged pions.

\begin{figure*}[t]
	\includegraphics[width=0.99\textwidth]{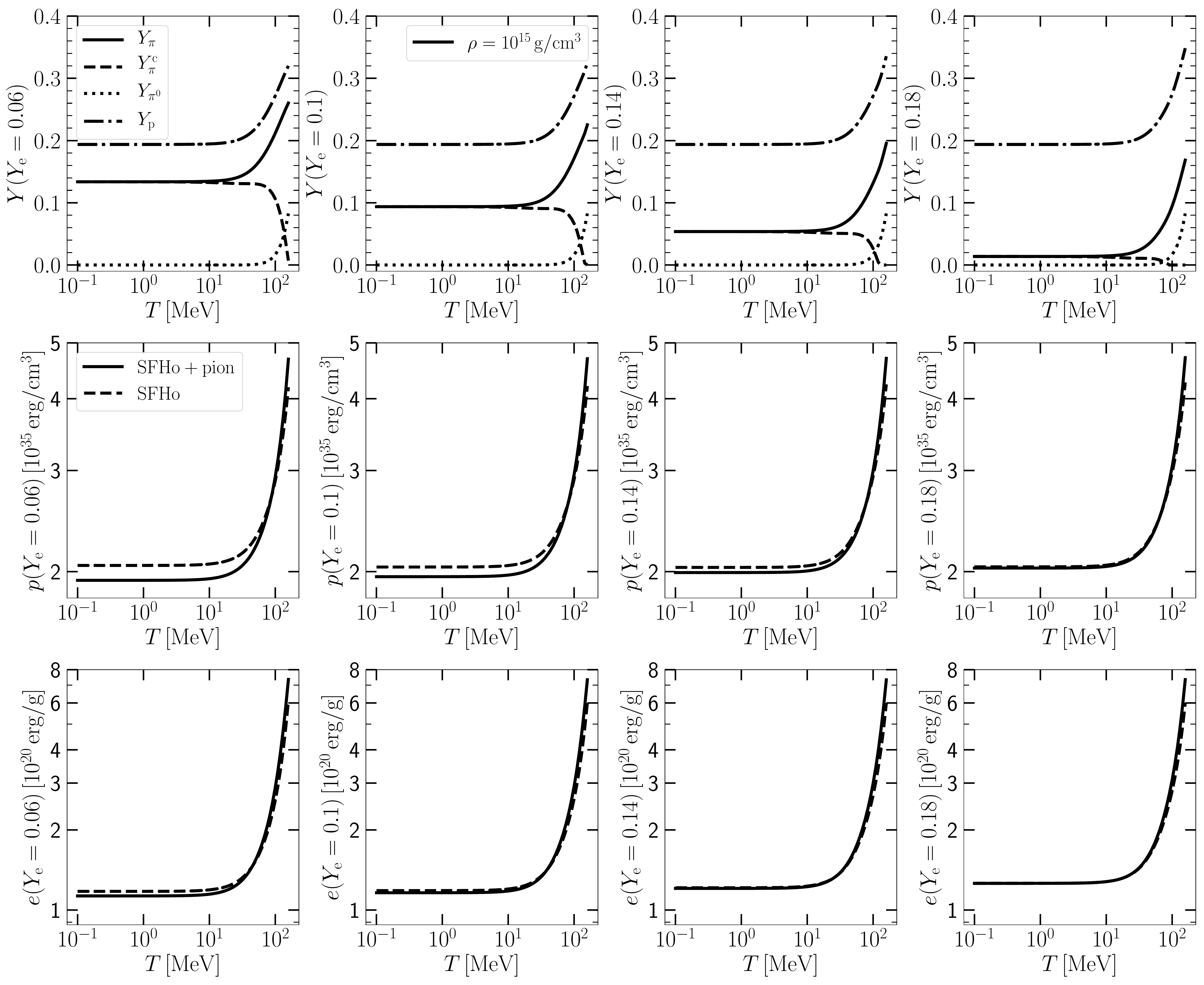}%
	\vspace*{-4mm}
	\caption{Comparison between the SFHo EOS and the modified SFHo EOS with pion masses equal to their vacuum values at a baryon density of $10^{15}\,\mathrm{g/cm^3}$. The net number fraction of charged pions (solid lines) and number fraction of condensed negative pions (dashed lines), neutral pions (dotted lines), and positive charges (dotted-dashed lines) are shown in the top row. The total pressure and total specific internal energy are shown for the SFHo EOS (dashed lines) and the modified SFHo EOS with pions (solid lines) in the second and third rows, respectively. Different columns represent different electron fractions.}	
	\label{fig:pion_EOS_T_rho_1e15_mpi_139}
\end{figure*}

To add charged pions to a base EOS, we apply the following
procedure. For a given baryonic density $\rho$, temperature $T$, and
net electron fraction $Y_\mathrm{e}$, we iteratively determine
$Y_\mathrm{p}$ until charge neutrality $Y_\mathrm{p} - Y_\mathrm{\pi}
= Y_\mathrm{e}$ is fulfilled. The base EOS provides $\hat \mu =
\mu_\mathrm{n} - \mu_\mathrm{p}$ and in case $|\hat \mu| <
m_{\pi^\pm}$, the net number fraction of pions $Y_{\pi}$ is evaluated employing the Bose-Einstein statistics for a given $T$ and $\mu_{\pi^\pm} = \pm \hat \mu$ in each iteration.

In neutron-rich conditions when $Y_\mathrm{p} < 0.5$, the base EOS can yield $\hat \mu > m_{\pi^-}$. However, the maximum allowed value of negatively charged pions' chemical potential $\mu_{\pi^-}^\mathrm{max}$ is their rest mass as pions are a Bose gas. Under such thermodynamic conditions when $\hat \mu > m_{\pi^-}$, the negatively charged pions can form the Bose-Einstein condensate which, unlike thermal pions, have zero kinetic energy. In such conditions, we iteratively solve the equation $\hat \mu = \mu_{\pi^-}^\mathrm{max} = m_{\pi^-}$ to find a new value of $Y_\mathrm{p}$ under fixed baryonic density, temperature, and electron fraction. Afterwards, we employ the charge neutrality condition to obtain the number fraction of negatively charged condensed pions $Y^\mathrm{c}_{\pi} = Y_\mathrm{p} - Y_\mathrm{e} - Y^\mathrm{thermal}_{\pi}$ where the number fraction of thermal pions $Y^\mathrm{thermal}_{\pi}$ is evaluated from the Bose-Einstein statistics with $\mu_{\pi^-} = m_{\pi^-}$. 

This condensed part of $\pi^-$ does not contribute to the total pressure and the total thermal energy. In proton-rich conditions, a condensate of positively charged pions can form; however, such scenarios are not expected for zero-temperature $\beta$-equilibrium neutron stars (NSs) and hypermassive neutron stars (HMNSs). The number fraction of neutral pions can be evaluated easily since it only depends on the temperature of the Bose gas. Finally, we determine the contribution of pions to the pressure, the internal energy, and the total entropy per baryon using the Bose-Einstein statistics.

We note that the base SFHo and DD2 EOSs employed in this study include the electronic contributions corresponding to $Y_\mathrm{e} = Y_\mathrm{p}$ in the total pressure, the total energy, and the total entropy of the gas, and the base EOSs tabulate these thermodynamic quantities against the $\rho,\,T,\,\mathrm{and}\,Y_\mathrm{p}$. Therefore, while building our modified EOSs with pions, which tabulate thermodynamic quantities as a function of the $\rho,\,T,\,\mathrm{and}\,Y_\mathrm{e}$ with $Y_\mathrm{e} \ne Y_\mathrm{p}$, we first subtract the electronic contributions corresponding to $Y_\mathrm{p}=Y_\mathrm{e}+Y_\pi$ from the total pressure, the total energy, and the total entropy, and then we add the electronic contributions corresponding to $Y_\mathrm{e}$ to the total pressure, the total energy, and the total entropy of the gas.

The base EOS considers the Coulomb interactions between charged
particles and includes their contributions to the pressure, the
internal energy, and the proton chemical potential. The Coulomb
contributions in these mentioned thermodynamic quantities can be
different for pions and electrons as the mean separation length
between positive charges and pions can be smaller than the mean
separation length between positive charges and electrons. Despite
that, we do not modify the Coulomb contributions to thermodynamic
quantities in this paper, which is a caveat of our
scheme. However, at densities relevant for pion production, i.e., densities above the nuclear saturation density, the nuclear contributions dominate over the Coulomb contributions. Therefore, our approximation regarding the Coulomb interactions may have a minor impact on the structure of NSs and the GWs emitted during merger, which are the two main topics of this paper.

In a dense medium, the effective pion mass may vary with the baryon density, temperature, and composition. due to the pion-pion interaction and the pion-nucleon interaction (see, e.g., \citep{2014PhRvD..90k5027N,2017PhRvD..95c6020J,2021PhRvD.104e4005T,Fore2023}). Reference~\citep{Fore2023} has shown a significant increase in the mass of the negatively charged pions at densities above saturation density and zero temperature. However, the behavior at typical temperatures of a few 10~MeV, which is the relevant regime for postmerger remnants, requires further studies. For simplicity, we use fixed pion masses adopting the vacuum mass, 170~MeV, and 200~MeV.
To probe the consequences of the effective pion mass variation on various thermodynamic quantities, we construct and study EOSs with different constant pion masses. We assume that neutral and charged pions have the same masses and already remark that positively charged pions are strongly suppressed under the relevant conditions such that we do not expect major effects by a possible mass splitting between negatively and positively charged pions. Our treatment may not capture all intricacies of how the dense medium may change the pion EOS (see, e.g., \citep{1996PhRvC..53.3057R,2017PhRvC..96a5207L,2020PhRvC.101c5809F,Fore2023} for discussions about the medium modification of pionic EOSs), however, it provides a simple way to study the probable impact of pion mass variations on binary neutron star (BNS) mergers.

In this paper, we use the following notation: ``SFHo'' for the base SFHo EOS, ``$\mathrm{SFHo}+\pi$, $m_\pi=\mathrm{Vac. \, mass}$'', ``$\mathrm{SFHo}+\pi$, $m_\pi=\mathrm{170 \, MeV}$'', ``$\mathrm{SFHo}+\pi$, $m_\pi=\mathrm{200 \, MeV}$'' for SFHo based EOSs that include pions with effective pion masses equal to the vacuum mass, 170~MeV and 200~MeV, respectively. A similar notation is used for the models based on DD2. 

In Fig.~\ref{fig:pion_EOS_T_rho_1e15_mpi_139}, we compare the SFHo EOS
and the modified SFHo EOS with pion masses equal to their vacuum
values at a baryon density of $10^{15}\,\mathrm{g/cm^3}$. The top
panels show the net number fraction of the charged pions $Y_{\pi}$
(solid lines), the number fraction of the condensed negatively charged
pions $Y_{\pi}^\mathrm{c}$ (dashed lines), the neutral pions
$Y_\mathrm{\pi^0}$ (dotted lines), and the positive charges
$Y_\mathrm{p}$ (dotted-dashed lines). The middle and bottom panels
show the total pressure and the specific internal energy,
respectively, for the SFHo EOS (dashed lines) and the modified SFHo
EOS with pion masses equal to their vacuum values (solid lines). The
panels show the aforementioned quantities as functions of the gas
temperature, and different columns represent different electron
fractions, namely $Y_\mathrm{e}=$ 0.06 (first column), 0.1 (second
column), 0.14 (third column), and 0.18 (fourth column). Such values of
the electron fraction and the baryon density of
$10^{15}\,\mathrm{g/cm^3}$ are representative of conditions in an HMNS
(see, e.g.,
\citep{2016PhRvD..93l4046S,Lehner2016,Foucart2016,Radice2018,Ardevol_Pulpillo_2019}).

\begin{figure}[htb]
	\includegraphics[width=\linewidth]{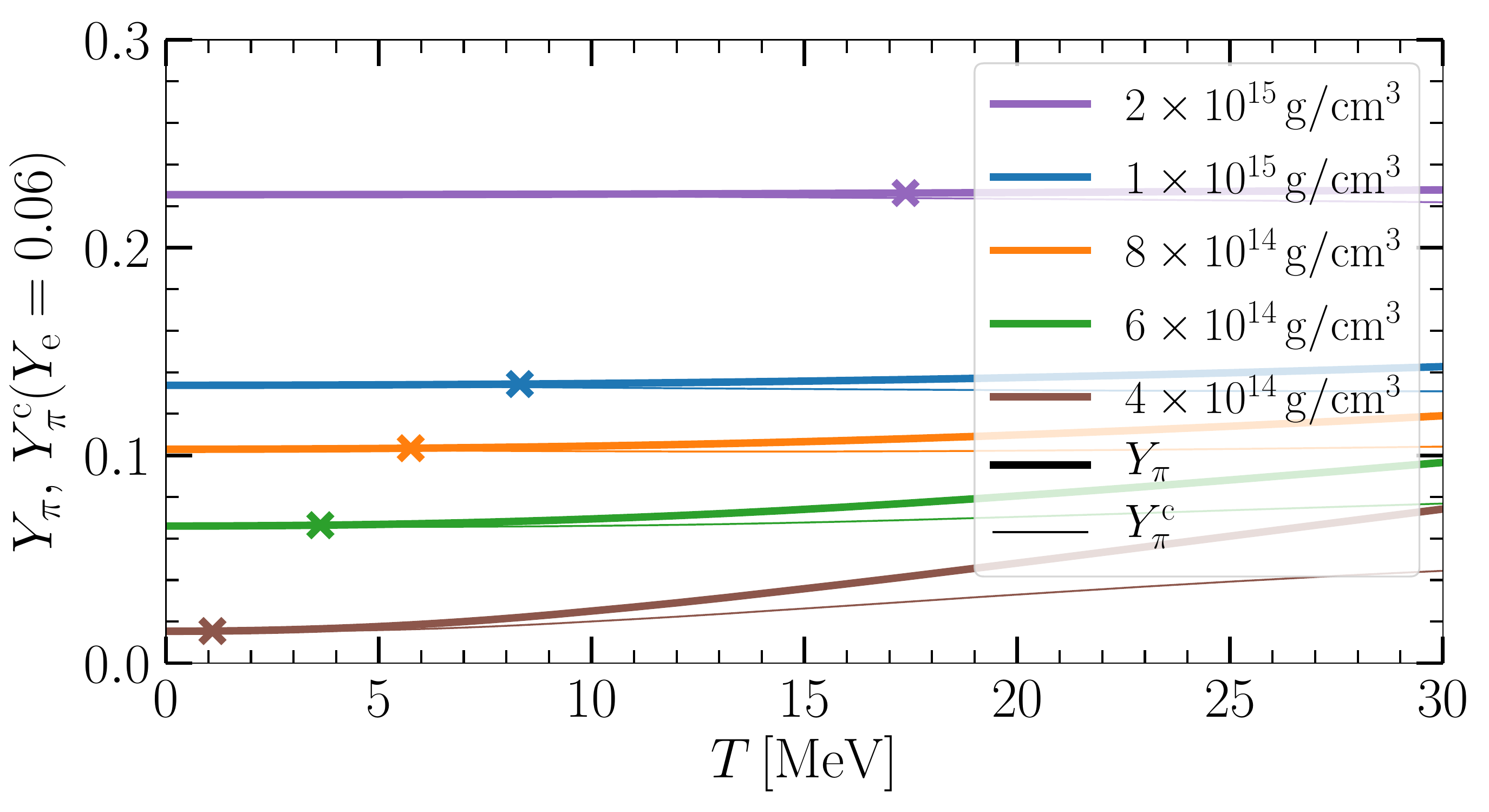}
	\includegraphics[width=\linewidth]{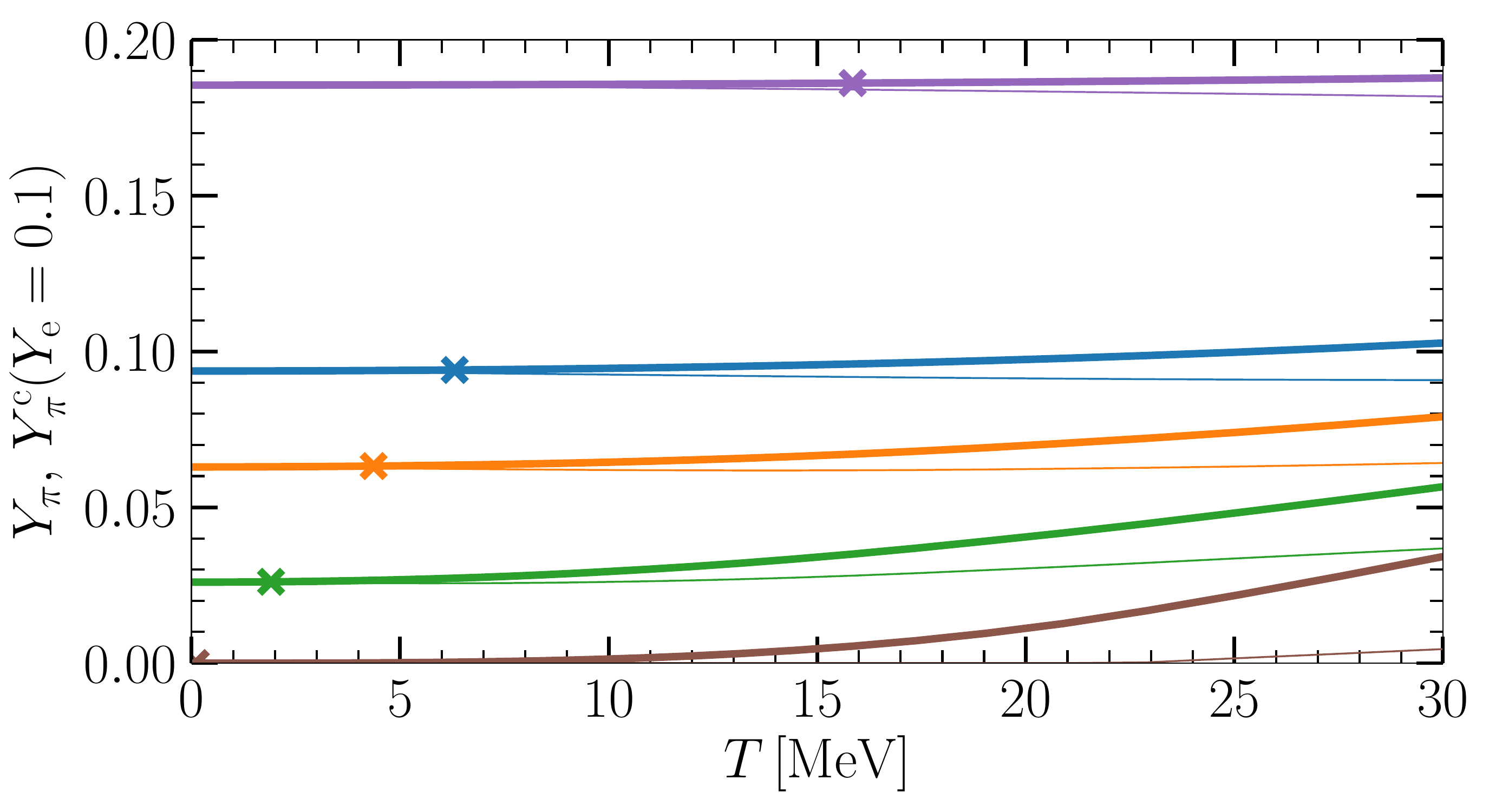}
	\caption{Net number fraction of charged pions and number
          fraction of negative pions in the condensate versus
          temperature at different densities for the modified SFHo EOS
          with pion masses equal to their vacuum values. The top panel
          and the bottom panel show these number fractions at electron
          fractions of 0.06 and 0.1, respectively. The crosses mark
          the threshold temperatures. We define these threshold
          temperatures as the points where the number fraction of
          negative pions in the condensate differs by one percent from
          the net number fraction of charged
          pions.\label{fig:pion_EOS_Tthres_mpi_139}} 
\end{figure}

\begin{figure*}[htb]
	\includegraphics[width=\linewidth]{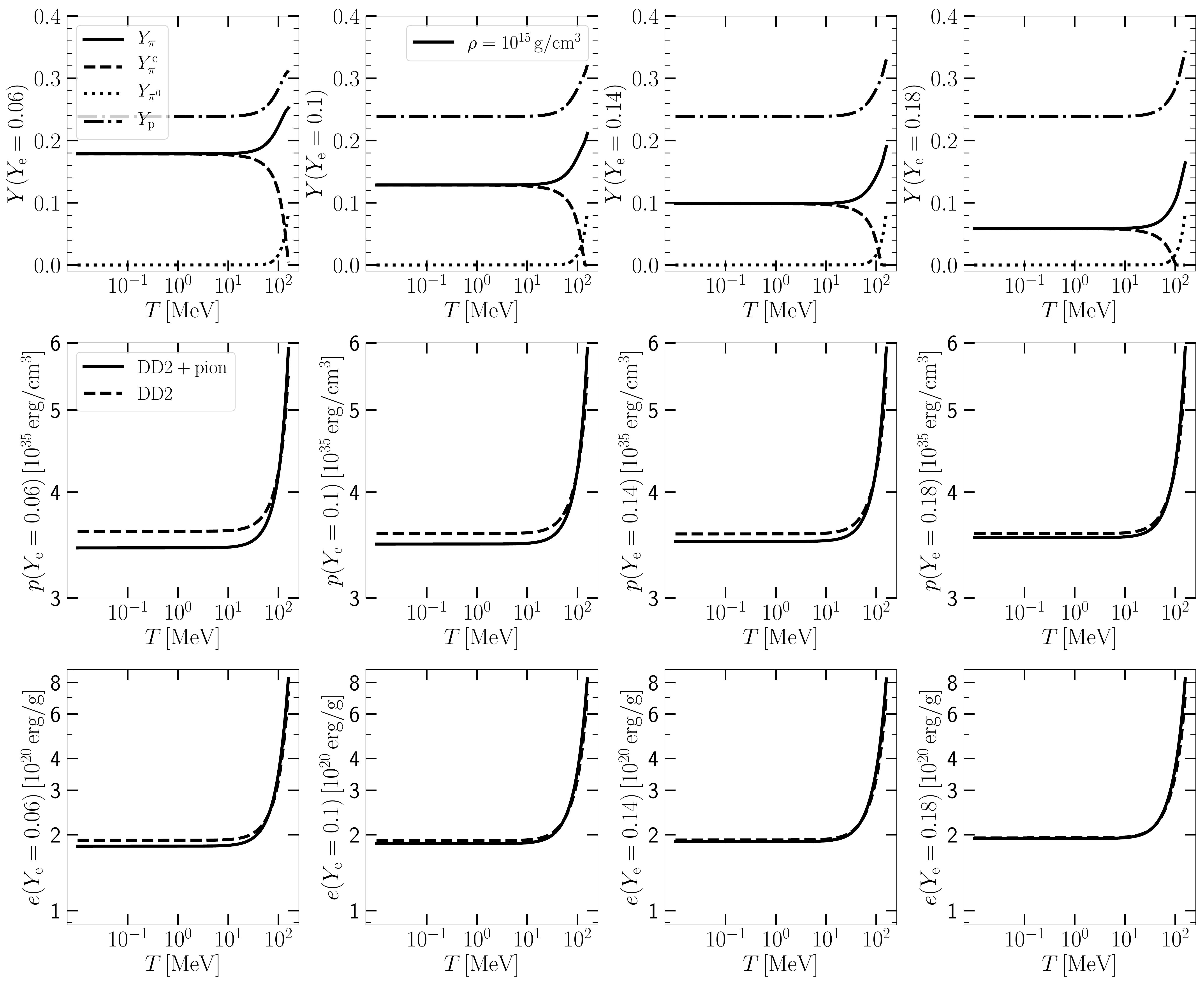}
	\caption{Comparison between the DD2 EOS and the modified DD2
          EOS with pion masses equal to their vacuum
          values. Quantities shown here are the same as in
          Fig.~\ref{fig:pion_EOS_T_rho_1e15_mpi_139} \label{fig:pion_EOS_T_rho_1e15_mpi_139_dd2}}   
\end{figure*}

At these thermodynamic conditions, the net number fraction of charged
pions and the number fraction of condensed negatively charged pions
remain constant below a threshold temperature
$T_\mathrm{thres}$. Below $T_\mathrm{thres}$, pions are predominately
in the Bose-Einstein condensate form ($Y_{\pi} \approx
Y_{\pi}^\mathrm{c}$). As the temperature rises beyond
$T_\mathrm{thres}$, more and more pions go out of the condensate form;
moreover, the net number fraction of charged pions grows with the
temperature as thermal pions, i.e. pions with non-zero kinetic
energies, are produced copiously. We also notice that the threshold
temperature rises with growing baryon density and diminishes with
increasing electron fraction as shown in
Fig.~\ref{fig:pion_EOS_Tthres_mpi_139}, where $T_\mathrm{thres}$ are
marked by crosses for various densities and electron fractions. We
approximate $T_\mathrm{thres}$ as the temperature where the number
fraction of condensed negatively charged pions differs by one percent
from the net number fraction of charged pions. We also observe that
the neutral pions are only produced copiously at high temperatures and
their number fraction does not depend significantly on the electron
fraction since their chemical potential is zero. In contrast, the net
number fraction of charged pions decreases with the growing electron
fraction. The number fraction of positively charged particles follows
the charge neutrality condition $Y_\mathrm{p} = Y_\mathrm{e} +
Y_{\pi}$. In neutron-rich conditions shown in
Fig. \ref{fig:pion_EOS_T_rho_1e15_mpi_139}, the $\pi^-$ are more
abundant than $\pi^+$ ($Y_{\pi} >0$) which leads to $Y_\mathrm{p} >
Y_\mathrm{e}$; therefore, the inclusion of pions production moves
baryonic contribution matter towards a more symmetric condition
($Y_\mathrm{p} \rightarrow 0.5$) and thus affects the pressure.  

The total pressure shown in the middle row of Fig. \ref{fig:pion_EOS_T_rho_1e15_mpi_139} contains contributions from baryons, electrons, positrons, photons for both EOSs and, in addition, includes contributions from thermal charged pions and the neutral pions for the modified SFHo EOS. The bottom panels of Fig. \ref{fig:pion_EOS_T_rho_1e15_mpi_139} display the specific internal energy, i.e. the relativistic specific internal energies from which the vacuum rest-mass energies of baryons are subtracted. It contains contributions from baryons, electrons, positrons, photons, and, if present, from pions. The condensed negatively charged pions do not contribute to the pressure as they possess zero kinetic energy and only contribute to the specific internal energies.

At temperatures below several 10 MeV the pressure as well as the specific internal energy of the EOS which includes pions, is smaller than those of the base EOS. This is because a fraction of the pions exists in condensed form and does not contribute to the thermal pressure and thermal energy. Furthermore, the difference between the two EOSs becomes smaller with increasing electron fraction because of the decreasing net number fraction of charged pions. We also notice that the decrease in the pressure due to pion production is more pronounced compared to that of the specific internal energy. 

\begin{figure}[htb]
	\includegraphics[width=\linewidth]{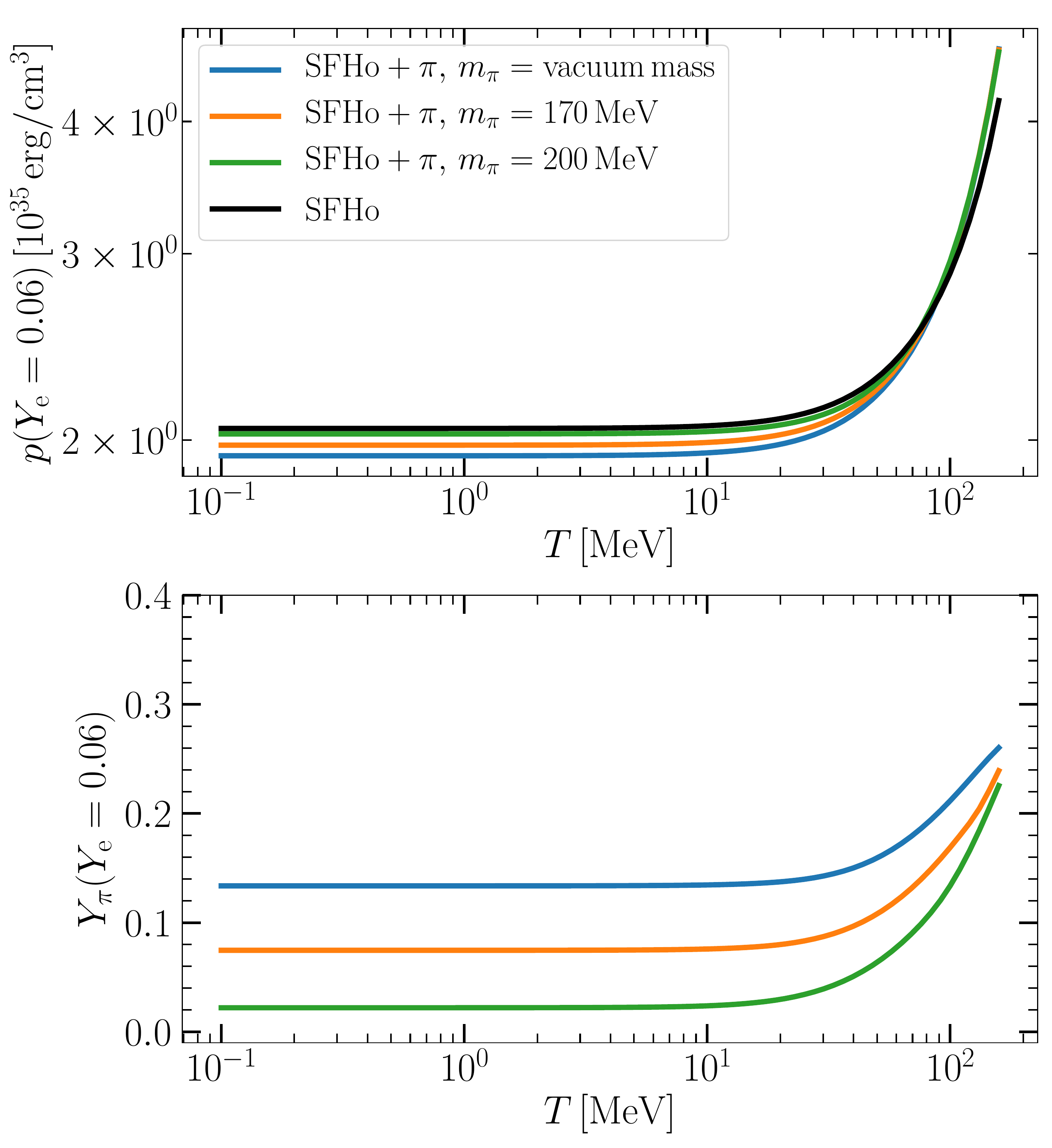}
	\caption{Total pressure of matter (top panel) and net number fraction of charged pions (bottom panel) versus temperature at a baryon density of $10^{15}\,\mathrm{g/cm^3}$ and at an electron fraction of 0.06 for the SFHo EOS (black lines) and the modified SFHo EOSs with different pion masses, namely, vacuum mass (blue lines), 170 MeV (orange lines), and 200 MeV (green lines). The total pressure of matter for the modified EOSs with pions converges towards the SFHo EOS values as the mass of the pion increases. Moreover, the net number fraction of charged pions decreases with the growing pion mass.}	
	\label{fig:pion_EOS_T_rho_1e15_SFHo_all_mass_compare}
\end{figure}

As temperature rises and pions go out of the condensed form, these
reductions in the pressure and the specific internal energy vanish,
and at temperatures above several 10 MeV, the pressure and the
specific internal energy of the modified SFHo EOS become larger than
that of the base SFHo EOS due to the contributions from the
  thermal charged pions and the neutral pions, which are produced
abundantly at high temperature. 

\begin{figure}[htb]
	\includegraphics[width=\linewidth]{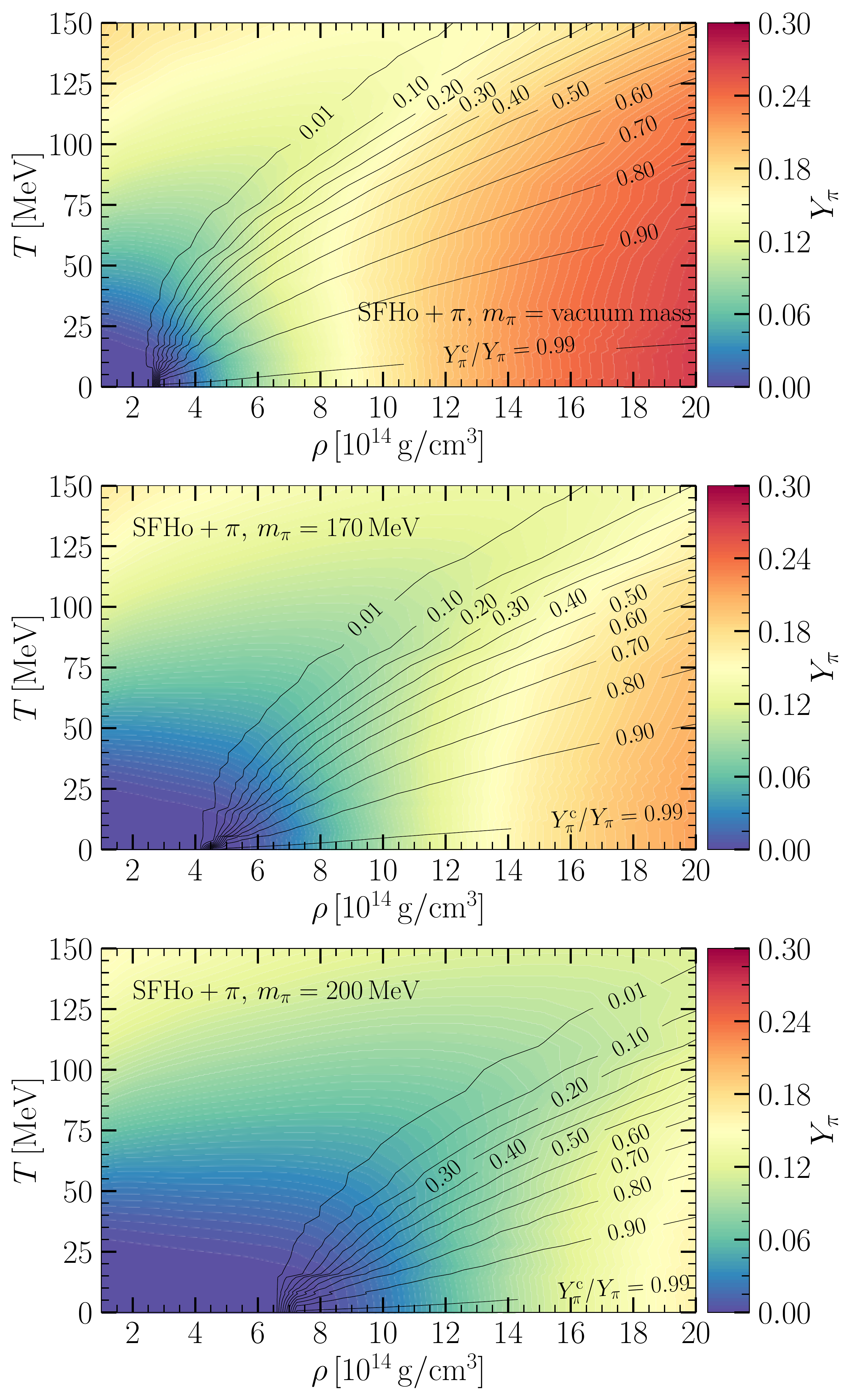}
	\caption{Equilibrium net number fractions of charged pions, i.e., the net number fraction of charged pions when the charged pions, nucleons, and electrons are in chemical equilibrium, are shown colour-coded on the baryon density-temperature plane ($\rho$-$T$ plane) for the modified SFHo EOSs with pion masses equal to their vacuum values (top panel), 170 MeV (middle panel), and 200 MeV (bottom panel). The ratios of the equilibrium number fraction of negatively charged condensed pions and the equilibrium net number fraction of charged pions are marked by black lines.}
	\label{fig:pion_eos_T_rho_ypi_eq}
\end{figure}

In Fig.~\ref{fig:pion_EOS_T_rho_1e15_mpi_139_dd2}, we show the
comparison between the DD2 EOS and the modified DD2 EOS with pion
masses equal to their vacuum values. The quantities shown in
Fig.~\ref{fig:pion_EOS_T_rho_1e15_mpi_139_dd2} are the same as in
Fig.~\ref{fig:pion_EOS_T_rho_1e15_mpi_139}. The qualitative trends
discussed above for the modified SFHo EOS are also observed for the
modified DD2 EOS. However, there are some quantitative differences
between the modified DD2 EOS and the modified SFHo EOS, which are
expected since the base EOSs are based on different nuclear models
\citep{2013ApJ...774...17S,2005PhRvC..71f4301T, 2010PhRvC..81a5803T}.  

\begin{figure}[htb]
	\includegraphics[width=\linewidth]{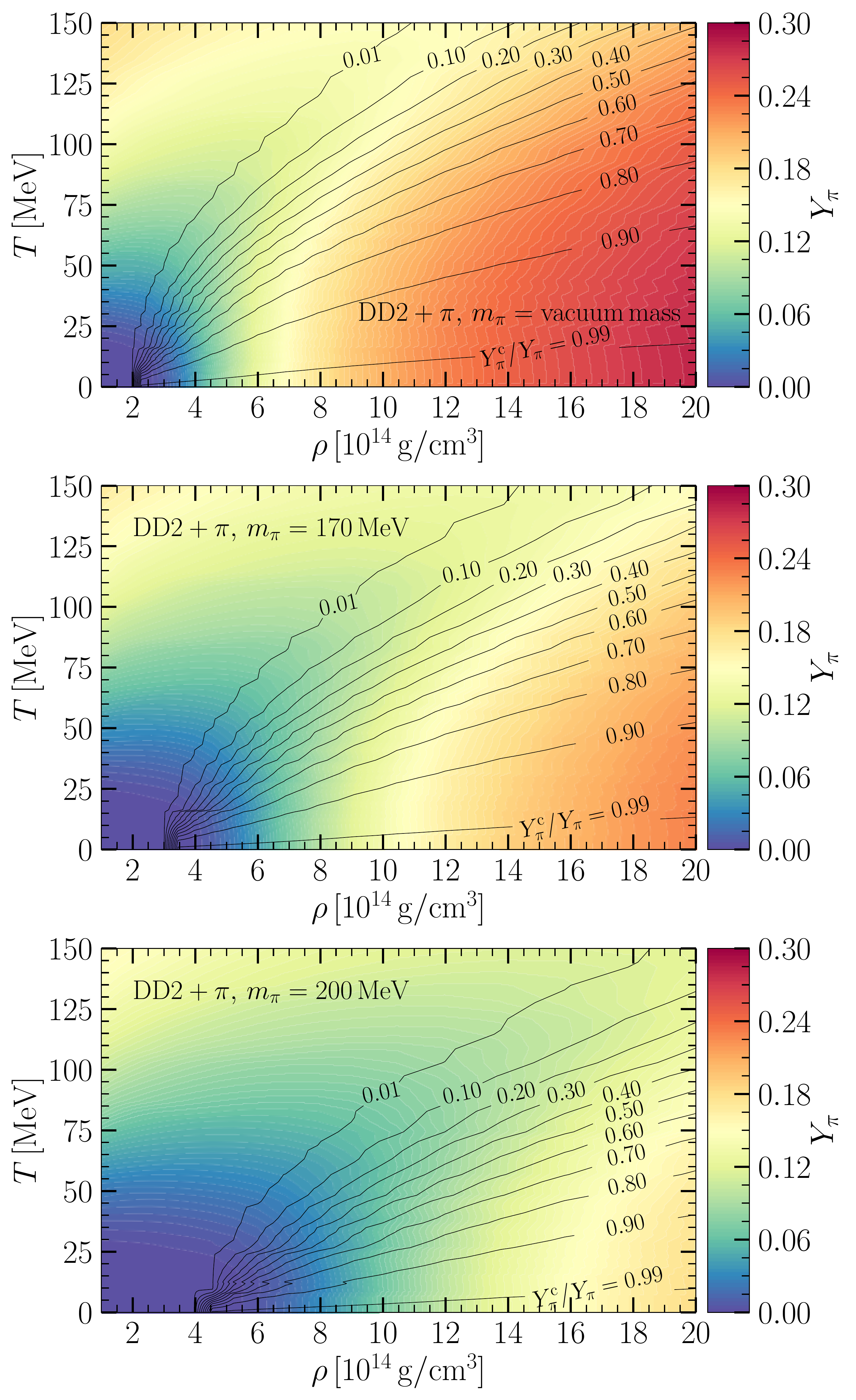}
	\caption{Same as Fig.~\ref{fig:pion_eos_T_rho_ypi_eq} but for the EOSs based on the DD2 model.}
	\label{fig:pion_eos_T_rho_ypi_eq_dd2}
\end{figure}

In Fig.~\ref{fig:pion_EOS_T_rho_1e15_SFHo_all_mass_compare}, we depict the total pressures (top panel) and the net number fractions of charged pions (bottom panel), respectively, versus temperature at a baryon density of $10^{15}\,\mathrm{g/cm^3}$ and an electron fraction of 0.06 for the SFHo EOS (black lines) and the modified SFHo EOS with various pion masses, namely the vacuum masses (blue lines), 170 MeV (orange lines), and 200 MeV (green lines). For large pion masses, the total pressure of the modified SFHo EOSs approaches that of the base EOS. This reduction in the total pressure difference between the modified SFHo EOSs and the base SFHo EOS is expected since the charged pion production diminishes with increasing pion mass. The reduced pion production is reflected by the shift of the net number fraction of charged pions to smaller values as the pion masses increase(lower panel). Recently, \citep{Fore2023} discusses the density dependence of pion mass at $T$=0. They showed the negatively charged pion mass can be greater than 200\,MeV at densities relevant for the pion production, i.e., at densities greater than nuclear saturation density.
The net pion fraction will be minuscule for such a high value of
negatively charged pion mass and the impact of pion production on the
total pressure, the total energy, etc. will be negligible. The
dependence of pion effective mass on the density and temperature and
how pion's energy-momentum dispersion relation changes in a dense
medium require further studies.

In Fig.~\ref{fig:pion_eos_T_rho_ypi_eq}, we show the net number fraction of charged pions colour-coded on the baryon density-temperature plane ($\rho$-$T$ plane) for modified SFHo EOSs with pion masses equal to their vacuum values (top panel), 170\,MeV (middle panel), and 200\,MeV (bottom panel). Additionally, we mark the ratios of the number fraction of condensed charged pions and the net number fraction of charged pions $Y_\pi^\mathrm{c}/Y_\pi$ by black lines. $Y_\pi$ and $Y_\pi^\mathrm{c}$ are evaluated under the assumption that the charged pions are in chemical equilibrium with the nucleons and electrons $\mu_{\pi^-}=\mu_\mathrm{n}-\mu_\mathrm{p}=\mu_\mathrm{e}$, where $\mu_\mathrm{e}$ is the chemical potential of electrons. The $Y_\pi^\mathrm{c}/Y_\pi=0.99$ line corresponds to the threshold temperature $T_\mathrm{thres}$ (see also Fig.~\ref{fig:pion_EOS_Tthres_mpi_139}). We notice from Fig.~\ref{fig:pion_eos_T_rho_ypi_eq} that for a given temperature, the net number fraction of charged pions typically rises with the baryon density; moreover, the ratio between $Y_\pi^\mathrm{c}$ and $Y_\pi$ increases with the baryon density for all modified EOSs. A larger fraction of negatively charged pions enters the condensed state as the temperature decreases, which is also evident from Figs.~\ref{fig:pion_EOS_T_rho_1e15_mpi_139} and \ref{fig:pion_EOS_Tthres_mpi_139}. By comparing the panels for different modified SFHo EOSs, i.e. different assumed pion masses, in Fig.~\ref{fig:pion_eos_T_rho_ypi_eq}, we see that the value of $Y_\pi$ drops with increasing charged pion mass for a given baryon density and temperature. As the charged pion masses grow, a larger density and a lower temperature are needed to achieve the same value of $Y_\pi^\mathrm{c}$/$Y_\pi$ ratio because a larger value of $\hat \mu$ is required to form the pion condensate. Similar qualitative trends in $Y_\pi$ and $Y_\pi^\mathrm{c}/Y_\pi$ with varying density, temperature, and pion masses can be observed for the modified DD2 EOSs in Fig.~\ref{fig:pion_eos_T_rho_ypi_eq_dd2}.

\section{Stellar structure and merger modelling}\label{NS setup}

To understand the impact of pions in BNS mergers, we conduct a number of simulations using the EOSs discussed in section \ref{EOS} with effective pion masses equal to their vacuum masses, 170\,MeV, and 200\,MeV based on the SFHo and DD2 model, respectively. As a reference, we also consider simulations using the base EOSs without pions.

\subsection{Stellar structure of isolated NS star}

\begin{figure}[!htb]
\includegraphics[width=\linewidth]{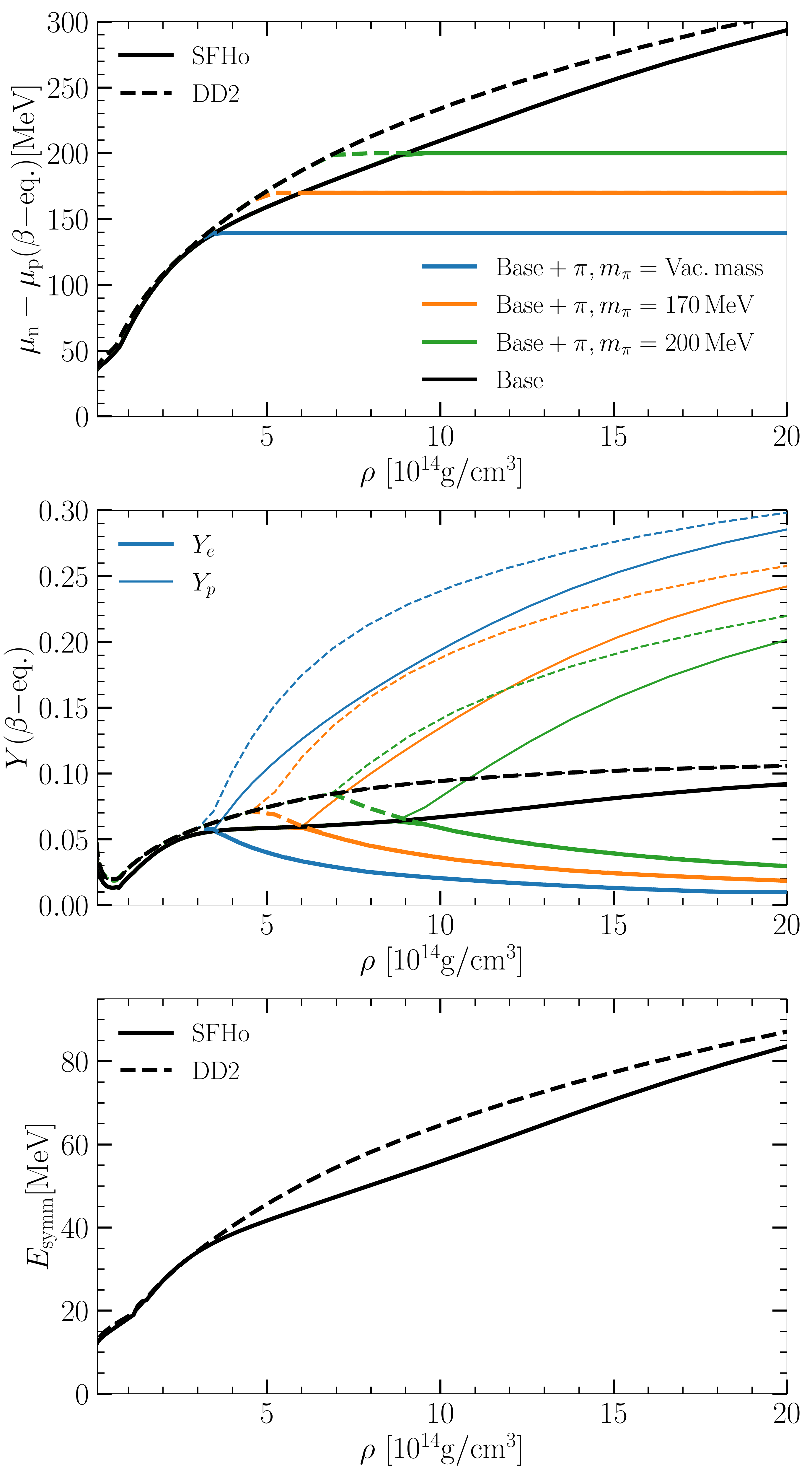}
\caption{Top panel shows the difference between neutron and proton
  chemical potential as function of density at zero temperature for beta-equilibrium. Middle panel provides the electron fraction and proton fraction in beta-equilibrium. Bottom panel displays symmetry energy of nuclear matter as the energy difference between neutron matter and symmetric matter. In all panels black curves refer to
  the base EOS without pions. Colored curves show the aforementioned
  quantities for EOSs with pions assuming different effective pion
  masses. Solid lines are for SFHo based models, while dashed curves
  visualize DD2 based models. In the middle panel thin lines display
  the proton fraction and thick lines indicate the electron
  fraction. For the base models both coincide. \label{fig:Esymm}} 
\end{figure}

\begin{figure*}[htb]
    \centering
     	\subfigure[SFHo]{\includegraphics[width=0.49\linewidth]{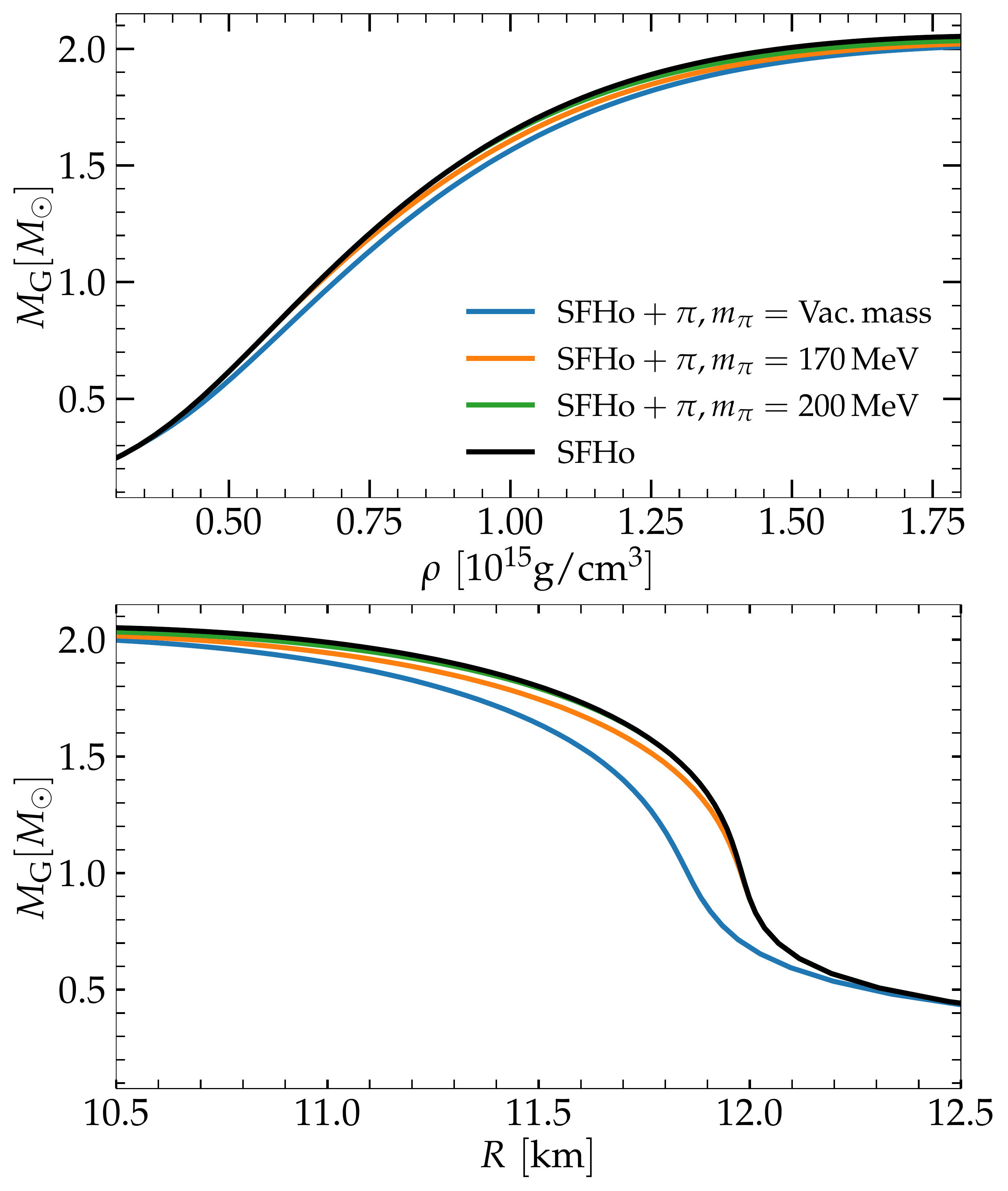}\label{fig:TOV_M_R_SFHo}}%
        \subfigure[DD2]{\includegraphics[width=0.49\linewidth]{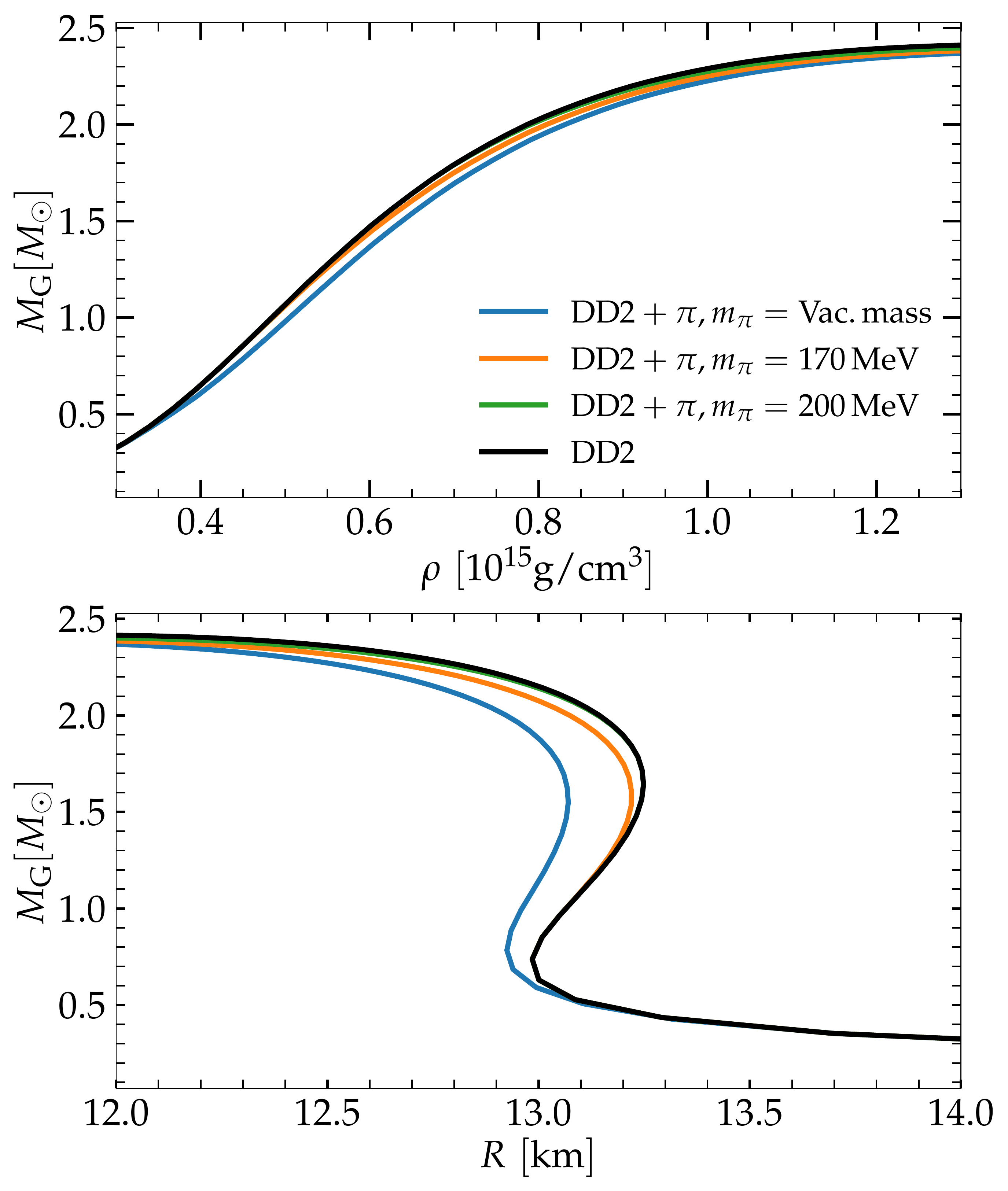}\label{fig:TOV_M_R_DD2}}%
    \caption{Left: Gravitational mass versus central baryon density (top panels) and gravitational mass versus the radius of NSs (bottom panels) for the base SFHo EOS and the ones including pions, SFHo, $\mathrm{SFHo}+\pi$, $m_\pi=\mathrm{Vac. \, mass}$, $\mathrm{SFHo}+\pi$, $m_\pi=\mathrm{170 \, MeV}$ and $\mathrm{SFHo}+\pi$, $m_\pi=\mathrm{200 \, MeV}$. Right: Same plots as in the left panel, but for DD2, $\mathrm{DD2}+\pi$,$m_\pi=\mathrm{Vac. \, mass}$, $\mathrm{DD2}+\pi$, $m_\pi=\mathrm{170 \, MeV}$ and $\mathrm{DD2}+\pi$, $m_\pi=\mathrm{200 \, MeV}$.}  
    \label{fig:TOV_M_R}
\end{figure*}

Before we describe merger simulations with the new EOS models, we
briefly discuss the impact of pions on the stellar structure by
considering the zero temperature\footnote{For the sake of simplicity we write ``zero temperature'' although the lowest temperature listed in the EOS tables is 0.1~MeV, and we compute TOV profiles for this temperature.} neutrinoless beta-equilibrium slices of the EOS models and by solving the Tolman–Oppenheimer–Volkoff (TOV) equations
\citep{1939PhRv...55..364T,1939PhRv...55..374O} for these barotropic relations.

Considering pions in NS matter has a strong impact on the
conditions for chemical equilibrium, and by this it affects the
stellar structure of NSs in equilibrium. It also determines the
initial conditions for binary mergers. The conditions for beta-equilibrium at zero temperature are clarified in
Fig.~\ref{fig:Esymm}. The upper panel shows the difference between the
chemical potentials of neutrons and protons as function of
density. Once the $\mu_\mathrm{n}-\mu_\mathrm{p}$ reaches the
effective pion mass, the difference remains constant. Hence, the
electron chemical potential is constant, which implies that the
electron fraction $Y_\mathrm{e}$ decreases with density because
$\mu_\mathrm{e}=\mathrm{const.}\propto n_\mathrm{e}^{1/3}\propto (\rho
Y_\mathrm{e})^{1/3}$ for relativistic electrons. This can be seen in
the middle panel, where the thick lines exhibit $Y_\mathrm{e}$. Since
the electron chemical potential is given by the effective pion mass,
the electron fraction is independent of the nuclear EOS in the regime
where $\mu_\mathrm{n}-\mu_\mathrm{p}$ remains constant. For larger
effective pion masses, this behavior sets in at a higher
density. Also, at this threshold density the proton fraction starts to
deviate from the electron fraction (middle panel) because of the
occurrence of condensed pions. In comparison to the base EoS,
$Y_\mathrm{p}$ is increased and charge neutrality is established
between protons, electrons and pions. 

The proton fraction does show a dependence on the nuclear EOS at all densities, i.e.~also in the regime with pion condensation. Hence, the fraction of pions does also depend on the EOS fulfilling charge neutrality.

This behavior can be understood by considering the symmetry energy of
the EOS defined as the energy difference between neutron matter and
symmetric matter. The symmetry energy is known to determine the proton
fraction in beta-equilibrium for matter composed of neutrons, protons
and electrons by connecting the difference of chemical potentials
between neutrons and protons,
$\hat{\mu}=\mu_\mathrm{n}-\mu_\mathrm{p}$, and $Y_\mathrm{p}$ (see
e.g.~\citep{Haensel.Potekhin.ea:2007,Lattimer:2023}; noting that slightly different definitions of the symmetry energy are used in the literature).  In the regime of pion condensation with $\hat{\mu}=\mathrm{const.}$, the proton fraction is thus entirely determined by the symmetry energy. The symmetry energy is given in the bottom panel of Fig.~\ref{fig:Esymm}. As the symmetry energy increases for constant $\hat{\mu}$, the proton fraction increases, which is clearly visible in the middle panel. The middle panel also shows that the proton fraction is larger for the DD2 based models, which is a consequence of their larger symmetry energy (bottom panel). 

In Fig.~\ref{fig:TOV_M_R_SFHo}, we show the NS gravitational masses
versus central densities (top panel) and NS radii (bottom panel) at
zero temperature for the SFHo EOS and the modified SFHo EOSs with pions, respectively. Figure.~\ref{fig:TOV_M_R_DD2} displays the same
relations for the DD2 EOS and modified DD2 EOSs, respectively. The
production of condensed pions in a NS lowers the pressure
support against gravity and leads to a reduced mass for the same
central density as we can notice from the top panel of
Figs.~\ref{fig:TOV_M_R_SFHo} and \ref{fig:TOV_M_R_DD2}. The maximum
mass is also reduced by the inclusion of pions for both EOS
models. The effect amounts to at most one to two percent for the
models adopting the vacuum masses. For higher effective pion masses
the impact is smaller. See Tabs. \ref{tab:progenitor_property_SFHo}
and \ref{tab:progenitor_property_dd2}. Similar trends are found for NS
radii, which are reduced by about one percent for models with the pion
vacuum masses.

In general, the softening from pions leads to a higher compactness. For a given EOS with fixed effective pion mass, the softening at higher densities is inversely related to the effective mass of the pion used in that EOS. For large effective pion masses, the stellar structure is very similar to the one resulting from the EOSs without pions. 

\subsection{Simulation details}

\begin{table*}[htb]
	\centering
    \caption{Properties of the simulated models with the SFHo EOS and
      the modified SFHo EOSs with pions. First column shows names of
      the simulated models. Second column shows the gravitational
      masses of BNSs, $m_\pi$ is the effective pion mass employed in
      the modified EOS, $R_\mathrm{ns}$ and $\Lambda$ are the radius
      and the dimensionless tidal deformability of a single NS
      corresponding to one of the star in the equal-mass
      BNS system. $M_\mathrm{chirp}$ represents the chirp mass of the
      BNS system. The column ``Prompt collapse'' indicates whether a
      BH is formed immediately after the merger or not,
      $f_\mathrm{peak}$ denotes the peak frequency of the GW
      spectrum. The $f_\mathrm{peak}$ is shown only for the models
      without prompt collapse. \label{tab:progenitor_property_SFHo}}
  \begin{ruledtabular}
	\begin{tabular*}{0.95\textwidth}{lccccccc}
		Model & BNS masses & $m_\pi$ & $R_\mathrm{ns}$ & $\Lambda$ & $M_\mathrm{chirp}$ & Prompt collapse & $f_\mathrm{peak}$ \\  
        & $[M_\odot]$ & [$\mathrm{MeV}$] & $[\mathrm{km}]$ & & $[M_\odot]$ & & $[\mathrm{kHz}]$  \\ 
        \hline
        SFHo\_Pions\_VacMass\_1.35\_1.35   & 1.35--1.35 & vacuum mass     & 11.727 & 377.902 & 1.175 & no  & 3.462 \\
        SFHo\_Pions\_170MeV\_1.35\_1.35    & 1.35--1.35 & 170 MeV         & 11.872 & 413.490 & 1.175 & no  & 3.351 \\
        SFHo\_Pions\_200MeV\_1.35\_1.35    & 1.35--1.35 & 200 MeV         & 11.896 & 420.776 & 1.175 & no  & 3.318  \\
        SFHo\_Base\_1.35\_1.35          & 1.35--1.35 & no pion         & 11.896 & 420.813 & 1.175 & no  & 3.310  \\
        SFHo\_Pions\_VacMass\_1.40\_1.40   & 1.40--1.40 & vacuum mass     & 11.700 & 296.937 & 1.219 & no  & 3.599 \\
        SFHo\_Pions\_170MeV\_1.40\_1.40    & 1.40--1.40 & 170 MeV         & 11.845 & 324.561 & 1.219 & no  & 3.530 \\
        SFHo\_Pions\_200MeV\_1.40\_1.40    & 1.40--1.40 & 200 MeV         & 11.874 & 332.950 & 1.219 & no  & 3.454 \\
        SFHo\_Base\_1.40\_1.40          & 1.40--1.40 & no pion         & 11.874 & 332.970 & 1.219 & no  & 3.419  \\
        SFHo\_Pions\_VacMass\_1.42\_1.42   & 1.42--1.42 & vacuum mass     & 11.687 & 270.635 & 1.236 & yes & ---  \\
        SFHo\_Pions\_170MeV\_1.42\_1.42    & 1.42--1.42 & 170 MeV         & 11.833 & 296.199 & 1.236 & no  & 3.621  \\
        SFHo\_Pions\_200MeV\_1.42\_1.42    & 1.42--1.42 & 200 MeV         & 11.864 & 303.134 & 1.236 & no  & 3.552  \\
        SFHo\_Base\_1.42\_1.42          & 1.42--1.42 & no pion         & 11.864 & 303.089 & 1.236 & no  & 3.499  \\
        SFHo\_Pions\_VacMass\_1.425\_1.425 & 1.425--1.425 & vacuum mass   & 11.684 & 264.060 & 1.241 & yes & ---  \\
        SFHo\_Pions\_170MeV\_1.425\_1.425  & 1.425--1.425 & 170 MeV       & 11.830 & 289.202 & 1.241 & yes & ---  \\
        SFHo\_Pions\_200MeV\_1.425\_1.425  & 1.425--1.425 & 200 MeV       & 11.862 & 295.680 & 1.241 & no  & 3.622  \\
        SFHo\_Base\_1.425\_1.425        & 1.425--1.425 & no pion       & 11.862 & 295.619 & 1.241 & no  & 3.573  \\
        SFHo\_Pions\_VacMass\_1.43\_1.43   & 1.43--1.43 & vacuum mass     & 11.681 & 257.484 & 1.245 & yes & ---  \\
        SFHo\_Pions\_170MeV\_1.43\_1.43    & 1.43--1.43 & 170 MeV         & 11.827 & 282.205 & 1.245 & yes & ---  \\
        SFHo\_Pions\_200MeV\_1.43\_1.43    & 1.43--1.43 & 200 MeV          & 11.859 & 288.226 & 1.245 & yes & ---  \\
        SFHo\_Base\_1.43\_1.43          & 1.43--1.43 & no pion         & 11.859 & 288.156 & 1.245 & no  & 3.614  \\
	\end{tabular*}
      \end{ruledtabular}
\end{table*}

We simulate the hydrodynamical evolution of NS mergers using a general relativistic smoothed particle hydrodynamics (SPH) code that employs the conformal flatness condition  to solve the general relativistic Einstein field equations~\citep{1980grg1.conf...23I, Wilson_1996, Oechslin_2002, Oechslin_2007,Bauswein+2010}. The initial data are constructed by imposing neutrinoless $\beta$-equilibrium conditions in the respective EOSs where neutrons, protons, electrons and pions (if present) are in chemical and thermal equilibrium. The stars are assumed not to have any intrinsic rotation. Our simulations begin with two NSs in a quasi-circular equilibrium orbit with an orbital separation such that the merger takes place after a few revolutions. We advect the initial electron fraction with the fluid i.e. $\frac{dY_\mathrm{e}}{dt}= 0$, which represents a rather crude treatment of weak interaction. 

In this study we only consider equal mass BNS systems. We choose 1.35-1.35$M_\odot$ binaries and 
systems with a total mass close to the threshold binary mass for
prompt black hole formation, which we estimate for the EOSs with pions
from empirical relations \citep{Bauswein_2021}. We provide a complete
list of simulations in table~\ref{tab:progenitor_property_SFHo} for SFHo based models and table~\ref{tab:progenitor_property_dd2} for DD2 based models\footnote{In these tables we do not list all simulations performed for this study but provide only those for which we run the same total binary mass for all four versions of the EOS model. For determining the threshold binary mass for the individual EOS models we conducted additional simulations for specific total binary masses.}. These include information on the total mass of the binaries{, the mass of pions, the radius of cold, isolated NSs with half of the total mass of the BNS, the tidal deformability of the binary system, and the chirp mass. 
The tidal deformability is defined as $\Lambda = \frac{2}{3} k_2 \left(\frac{R}{M}\right)^5$, where $R$ and $M$ are the radius and the gravitational mass of the individual non-rotating NSs and $k_2$ is the tidal Love number~\citep{Hinderer_2008}. The chirp mass is given by $M_\mathrm{chirp} = \frac{\left(M_1M_2\right)^{3/5}}{\left(M_1+M_2\right)^{1/5}}$, where $M_1$ and $M_2$ are BNS masses}. We also indicate whether the system undergoes a prompt gravitational collapse and provide the dominant GW frequency of the postmerger phase. The dominant GW frequency is not shown for the models which experience prompt collapse. 

\begin{table*}
	\centering
    \caption{Properties of the simulated models with the DD2 EOS and
      the modified DD2 EOSs with pions. Same as Tab.\ref{tab:progenitor_property_SFHo} but for the models based on DD2.}
    \begin{ruledtabular}
	\begin{tabular*}{0.95\textwidth}{lccccccc}
		Model & BNS masses & $m_\pi$ & $R_\mathrm{ns}$ & $\Lambda$ & $M_\mathrm{chirp}$ & Prompt collapse & $f_\mathrm{peak}$ \\
        & $[M_\odot]$ & [$\mathrm{MeV}$] & $[\mathrm{km}]$ & & $[M_\odot]$ & & $[\mathrm{kHz}]$ \\
        \hline
        DD2\_Pions\_VacMass\_1.35\_1.35     & 1.35--1.35 & vacuum mass      & 13.048 & 796.159 & 1.175 & no  & 2.681  \\
        DD2\_Pions\_170MeV\_1.35\_1.35      & 1.35--1.35 & 170 MeV          & 13.189 & 854.982 & 1.175 & no  & 2.642  \\
        DD2\_Pions\_200MeV\_1.35\_1.35      & 1.35--1.35 & 200 MeV          & 13.199 & 871.076 & 1.175 & no  & 2.597  \\
        DD2\_Base\_1.35\_1.35            & 1.35--1.35 & no pion          & 13.199 & 871.135 & 1.175 & no  & 2.608  \\
        DD2\_Pions\_VacMass\_1.62\_1.62     & 1.62--1.62 & vacuum mass      & 13.068 & 262.744 & 1.410 & no  & 3.099  \\
        DD2\_Pions\_170MeV\_1.62\_1.62      & 1.62--1.62 & 170 MeV          & 13.219 & 287.197 & 1.410 & no  & 3.030  \\
        DD2\_Pions\_200MeV\_1.62\_1.62      & 1.62--1.62 & 200 MeV          & 13.247 & 294.408 & 1.410 & no  & 3.008  \\
        DD2\_Base\_1.62\_1.62            & 1.62--1.62 & no pion          & 13.247 & 294.465 & 1.410 & no  & 2.997  \\
        DD2\_Pions\_VacMass\_1.64\_1.64     & 1.64--1.64 & vacuum mass      & 13.066 & 243.524 & 1.428 & yes & ---  \\
        DD2\_Pions\_170MeV\_1.64\_1.64      & 1.64--1.64 & 170 MeV          & 13.217 & 266.835 & 1.428 & no  & 3.137  \\
        DD2\_Pions\_200MeV\_1.64\_1.64      & 1.64--1.64 & 200 MeV          & 13.248 & 270.335 & 1.428 & no  & 3.024  \\
        DD2\_Base\_1.64\_1.64            & 1.64--1.64 & no pion          & 13.248 & 270.407 & 1.428 & no  & 3.036  \\
        DD2\_Pions\_VacMass\_1.65\_1.65     & 1.65--1.65 & vacuum mass      & 13.065 & 234.778 & 1.436 & yes & ---  \\
        DD2\_Pions\_170MeV\_1.65\_1.65      & 1.65--1.65 & 170 MeV          & 13.216 & 256.653 & 1.436 & yes & ---  \\
        DD2\_Pions\_200MeV\_1.65\_1.65      & 1.65--1.65 & 200 MeV          & 13.248 & 260.521 & 1.436 & no  & 3.127  \\
        DD2\_Base\_1.65\_1.65            & 1.65--1.65 & no pion          & 13.248 & 260.531 & 1.436 & no  & 3.047  \\
        DD2\_Pions\_VacMass\_1.66\_1.66     & 1.66--1.66 & vacuum mass      & 13.064 & 226.031 & 1.445 & yes & ---  \\
        DD2\_Pions\_170MeV\_1.66\_1.66      & 1.66--1.66 & 170 MeV          & 13.216 & 246.472 & 1.445 & yes & ---  \\
        DD2\_Pions\_200MeV\_1.66\_1.66      & 1.66--1.66 & 200 MeV          & 13.248 & 251.637 & 1.445 & yes & ---  \\
        DD2\_Base\_1.66\_1.66            & 1.66--1.66 & no pion          & 13.248 & 251.658 & 1.445 & no & 3.139  \\
	\end{tabular*}
      \end{ruledtabular}
	\label{tab:progenitor_property_dd2}
\end{table*}

\section{Simulation results} \label{Results}

In this section we describe the general dynamics and the evolution of the pion contributions in our binary merger simulations. We address the postmerger GW signal and the threshold mass for prompt black-hole formation and discuss to which extent these features are affected by the presence of pions. 

\subsection{Dynamics and pion production in NS mergers}\label{Evolution} 

\begin{figure*}
    \centering
    \subfigure[SFHo]{\includegraphics[width=0.49\linewidth]{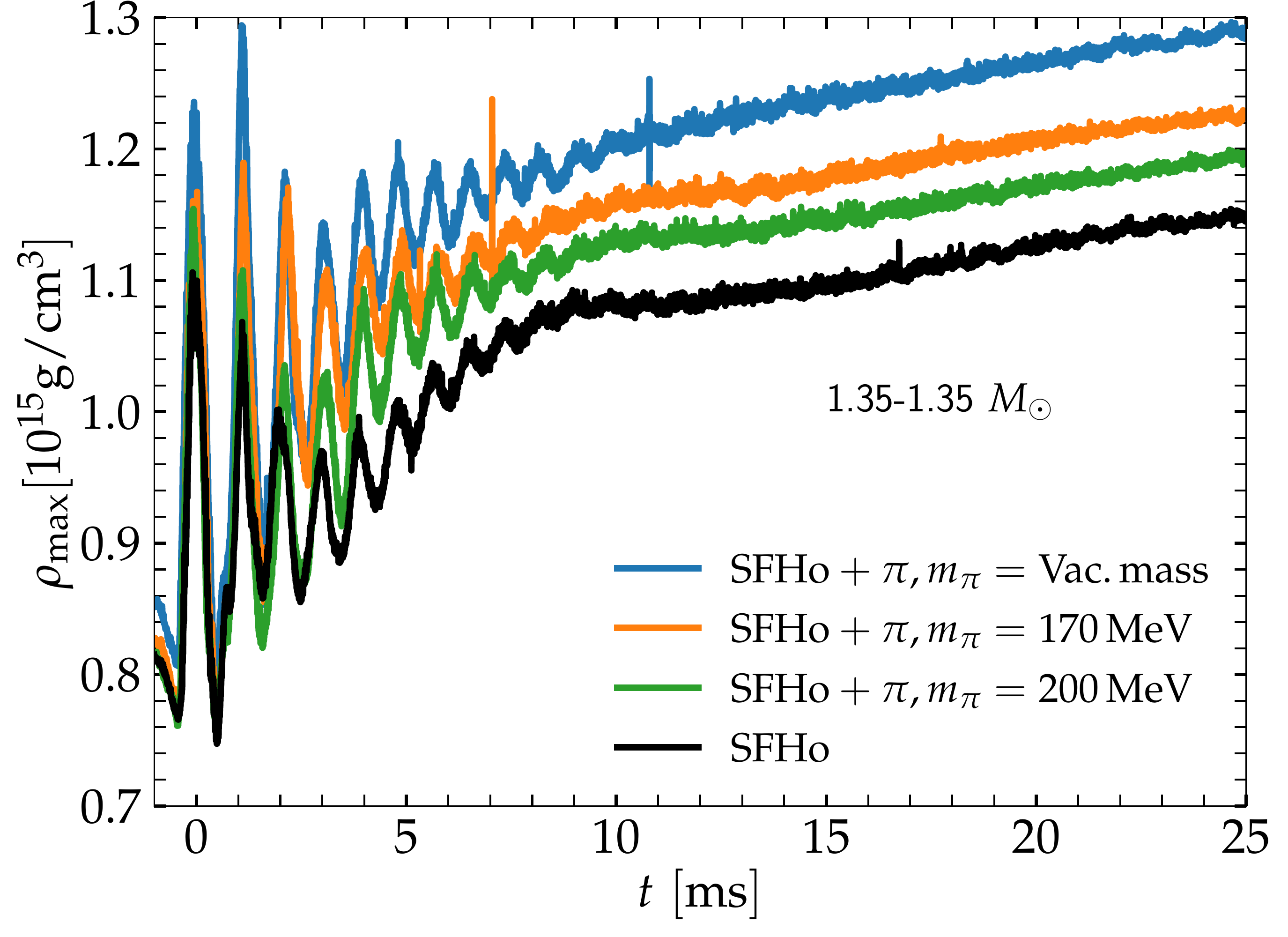}\label{fig:Rhomax_SFHo}}
    \subfigure[DD2]{\includegraphics[width=0.49\linewidth]{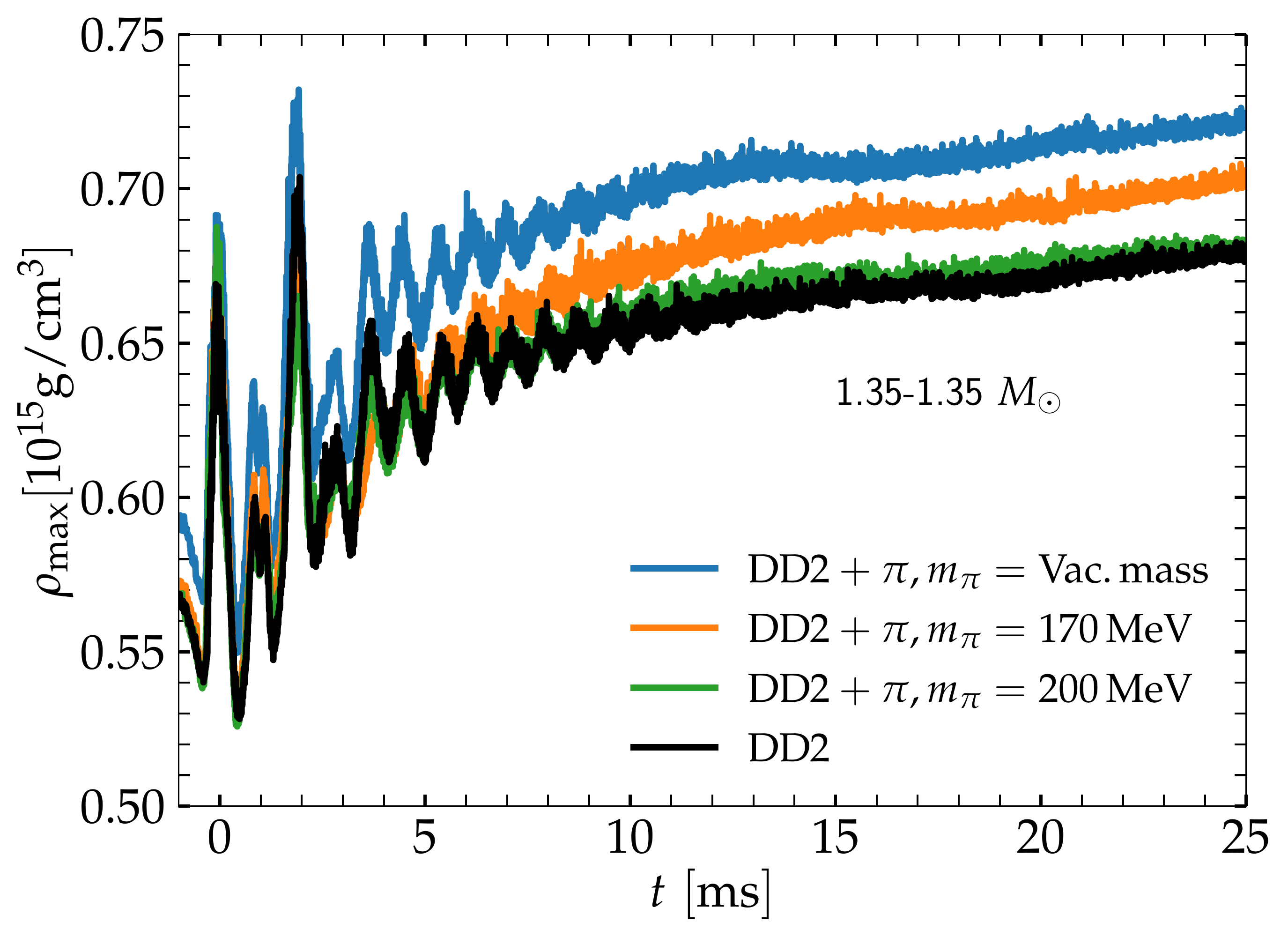}\label{fig:Rhomax_DD2}}
    \caption{Time evolution of the maximum baryonic rest-mass density for 1.35-1.35$M_\odot$ BNS systems using SFHo EOS (left panel) and DD2 EOS (right panel). The time zero corresponds to the instant of the maximum compression during the first bounce after merging (minimum in the lapse function).We evaluate the maximum baryonic rest-mass density on the SPH particle level, which is why some noise is visible.} 
    \label{fig:Rhomax}
\end{figure*}

\begin{figure*}
    \centering
    \subfigure[SFHo]{\includegraphics[width=0.49\linewidth]{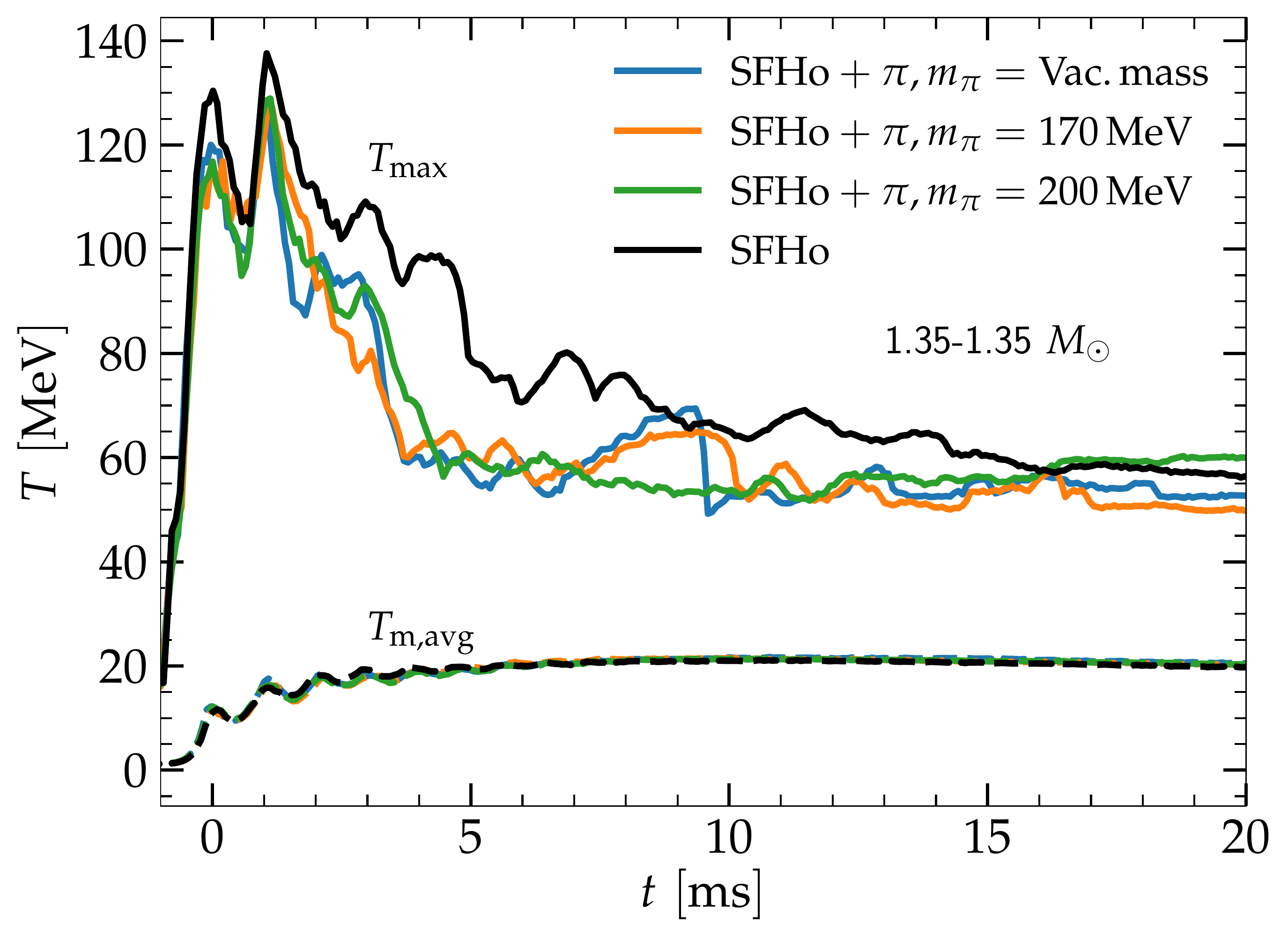}\label{fig:Tmax_SFHo}}
    \subfigure[DD2]{\includegraphics[width=0.49\linewidth]{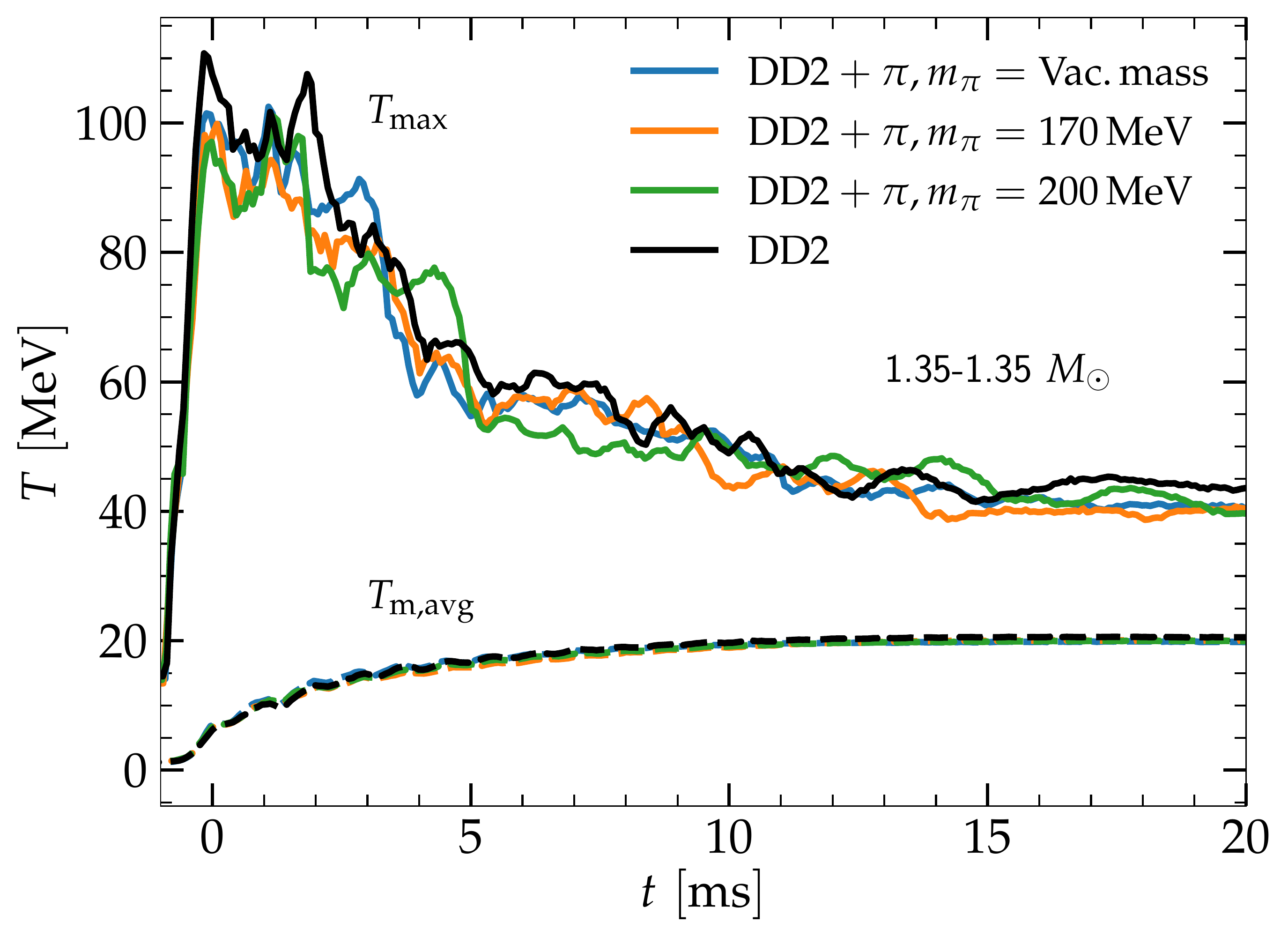}\label{fig:Tmax_DD2}}
    \caption{Time evolution of the maximum and mass-averaged temperature for 1.35-1.35$M_\odot$ BNS systems using SFHo EOS (left panel) and DD2 EOS (right panel).}
    \label{fig:Tmax}
\end{figure*}

\begin{figure*}
    \centering
     	\subfigure[Base DD2]{\includegraphics[width=0.495\linewidth]{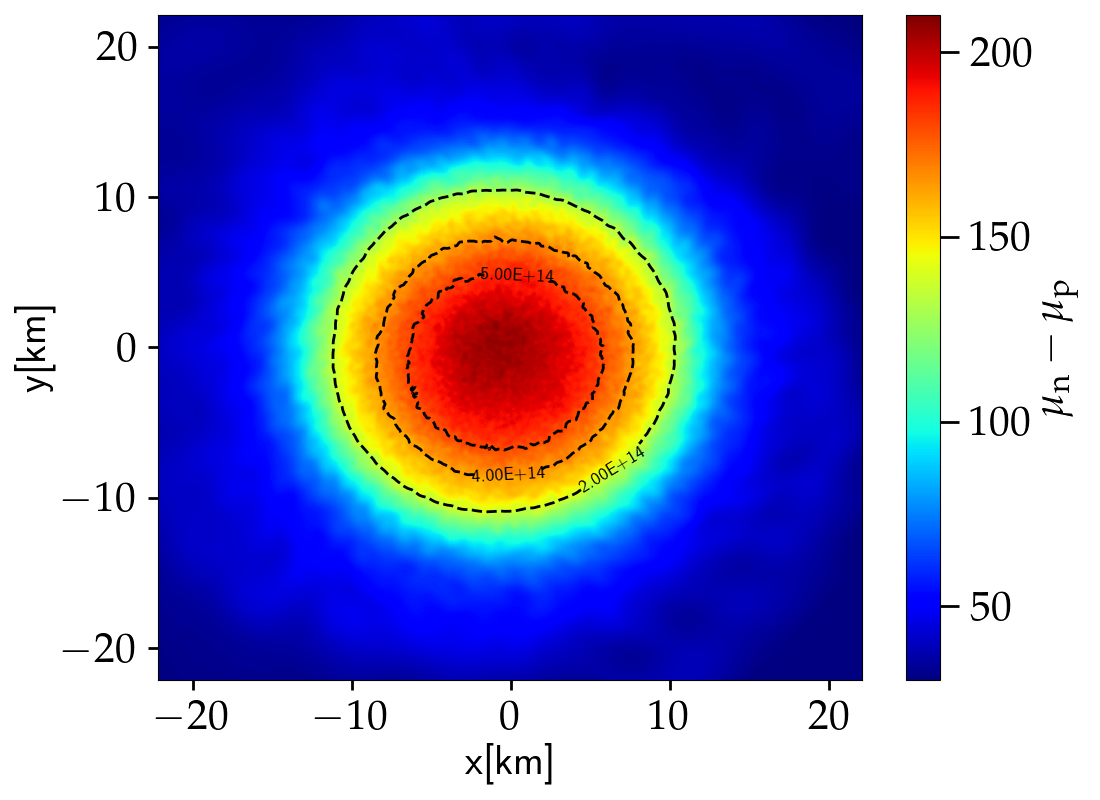}\label{fig:mun-mup-nopion}}
     	\subfigure[$\mathrm{DD2}+\pi, m_\pi=\mathrm{Vac. \, mass}$]{\includegraphics[width=0.495\linewidth]{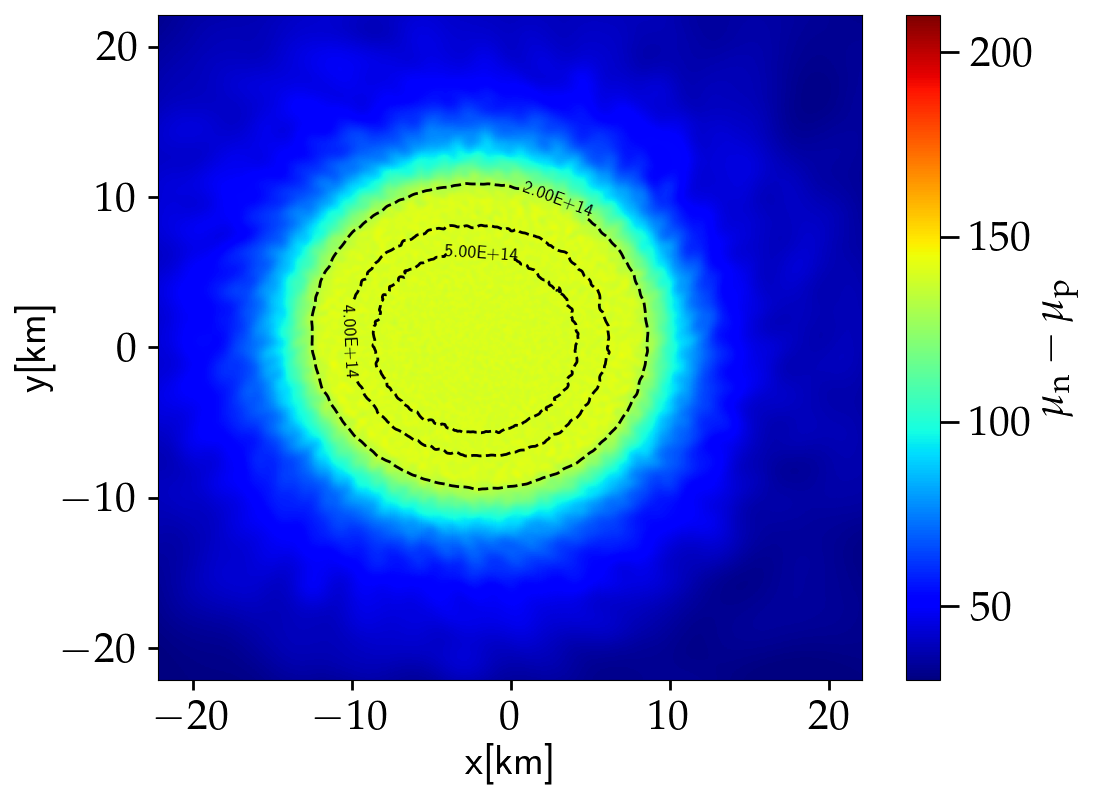}\label{fig:mun-mup-pion}}
   \caption{The panels (a) and (b) show snapshots of neutron and proton chemical potential difference $\mu_\mathrm{n}-\mu_\mathrm{p}$ in the equatorial plane for a 1.35-1.35$M_\odot$ BNS model after the merger, using the base DD2 EOS without pions and the modified DD2 EOS that includes pions with an effective pion masses equal to their vacuum masses, $\mathrm{DD2}+\pi, m_\pi=\mathrm{Vac. \, mass}$. Density regions of $2\times 10^{14} \mathrm{g/cm^3}$, $4\times 10^{14}\mathrm{g/cm^3}$ and $5\times 10^{14}\mathrm{g/cm^3}$ are shown by dashed contour lines.}
   \label{fig:mun-mup}
\end{figure*}

Apart from models undergoing a prompt gravitational collapse to a black hole, all BNS merger simulations proceed qualitatively similar, and we do not recognize any specific differences between the calculations with the base EOSs and those with the modified EOSs with pions. Most of the simulated binaries result in the formation of a rotating NS remnant, and only a few systems with relatively large total binary masses lead to direct black hole formation after merging (to be discussed below).

In the simulations yielding a NS remnant, the merging stars form a rotating double-core structure and, while the densities in the postmerger phase increase, the remnant undergoes vivid oscillations. For the 1.35-1.35~$M_\odot$ models, the density oscillations can be seen in Fig.~\ref{fig:Rhomax} showing the evolution of the maximum baryonic rest-mass density. The quantitative behavior is inline with existing literature results, i.e., softer EOSs lead to higher densities. This concerns the comparison between the base EOSs, SFHo and DD2, and also the comparison between the respective base EOS and the versions with pions. Smaller effective pion masses lead to softer EOSs and more compact stellar configurations and thus higher $\rho_\mathrm{max}$ in the merger remnant. Quantitatively, the maximum densities in the merger remnant are consistent with empirical relations for $\rho_\mathrm{max}$ found for a large sample of candidate EOSs without pions (see discussion below in subsection \ref{Mass relations}).

\begin{figure*}
     \subfigure[$\mathrm{DD2}+\pi, m_\pi=\mathrm{Vac. \, mass}$]{\includegraphics[width=0.32\linewidth]{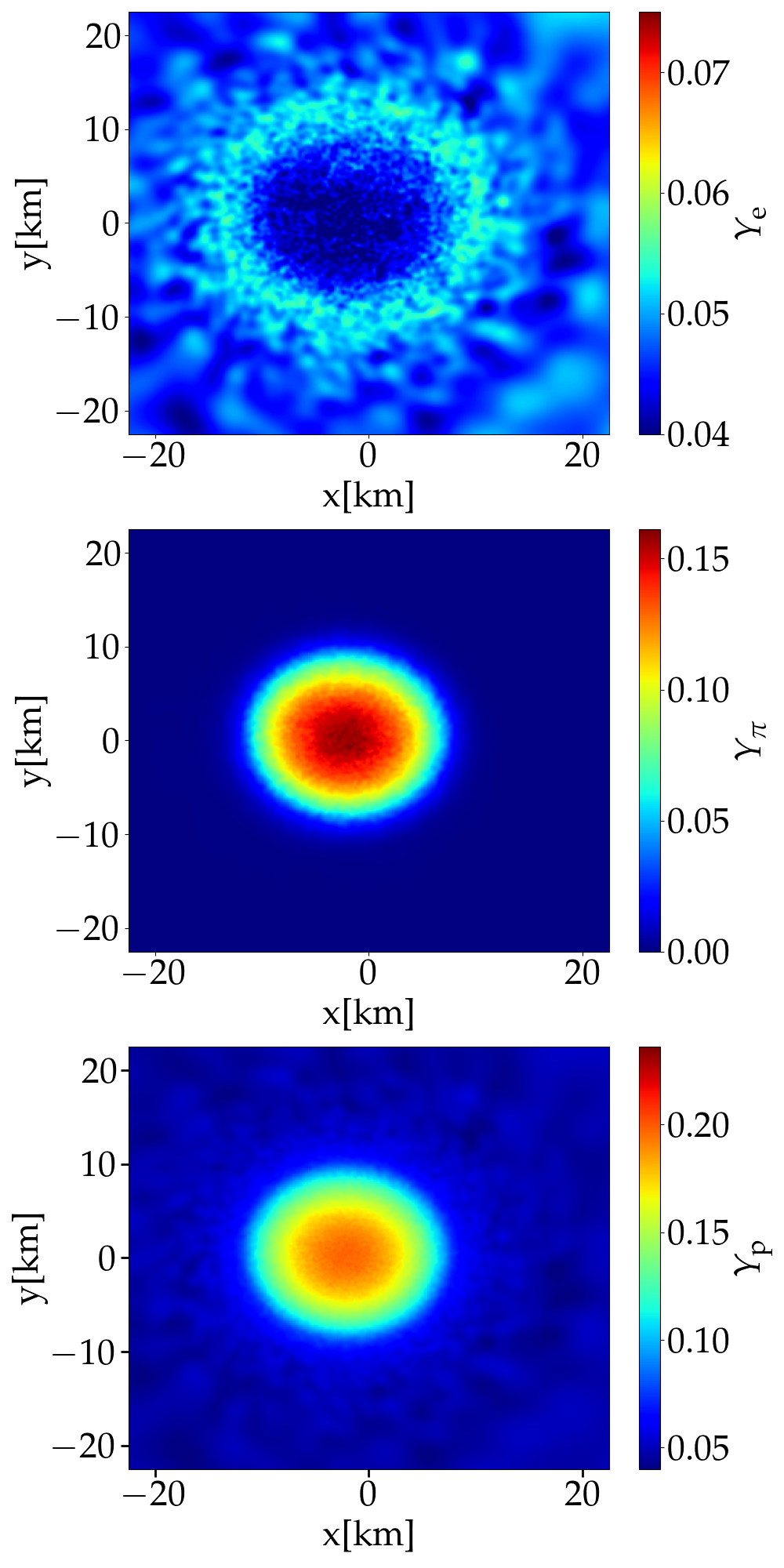}\label{fig:ye-ypi139}}
     \subfigure[$\mathrm{DD2}+\pi, m_\pi=\mathrm{170 \, MeV}$]{\includegraphics[width=0.32\linewidth]{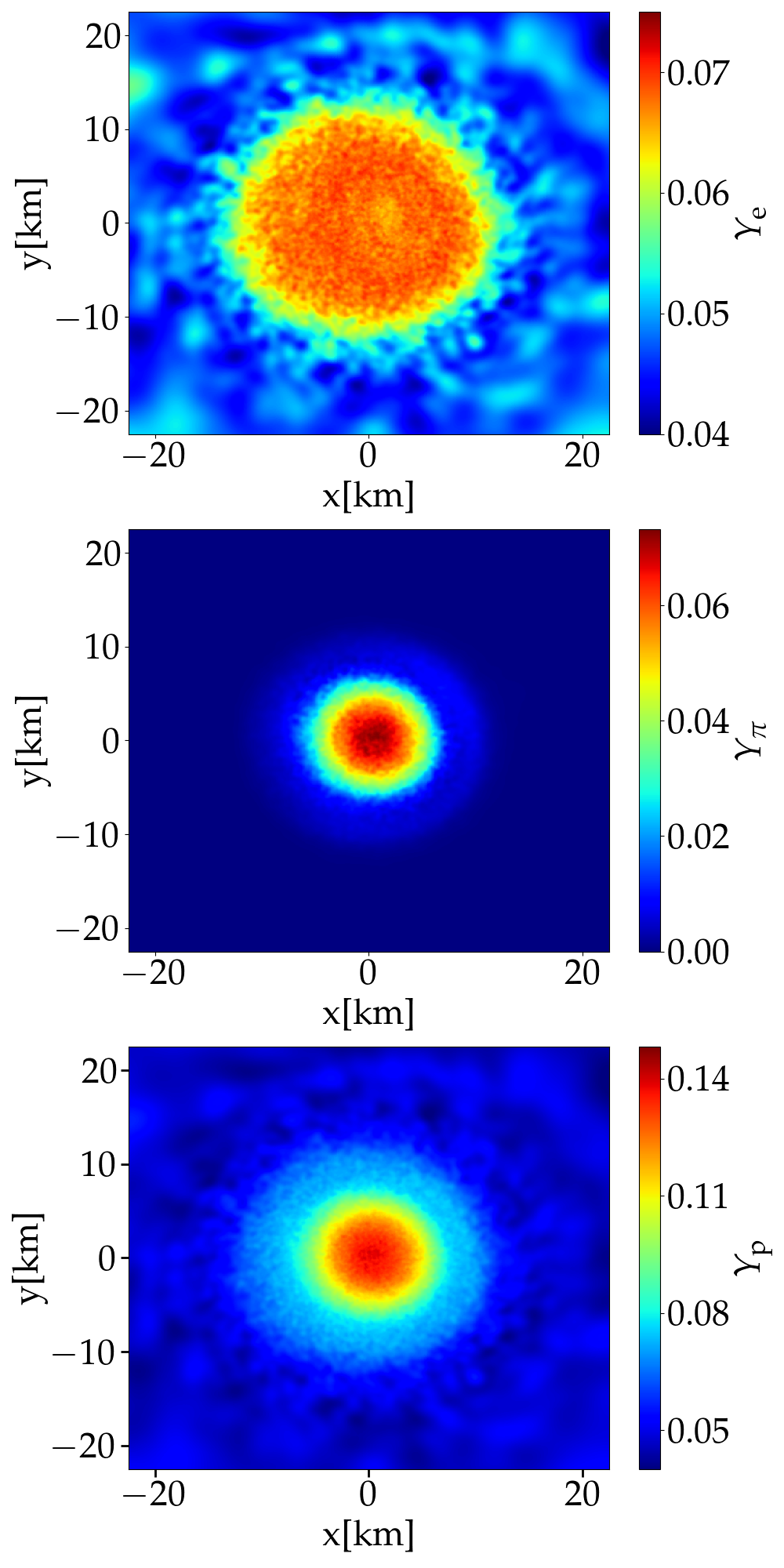}\label{fig:ye-ypi170}}
     \subfigure[$\mathrm{DD2}+\pi, m_\pi=\mathrm{200 \, MeV}$]{\includegraphics[width=0.32\textwidth]{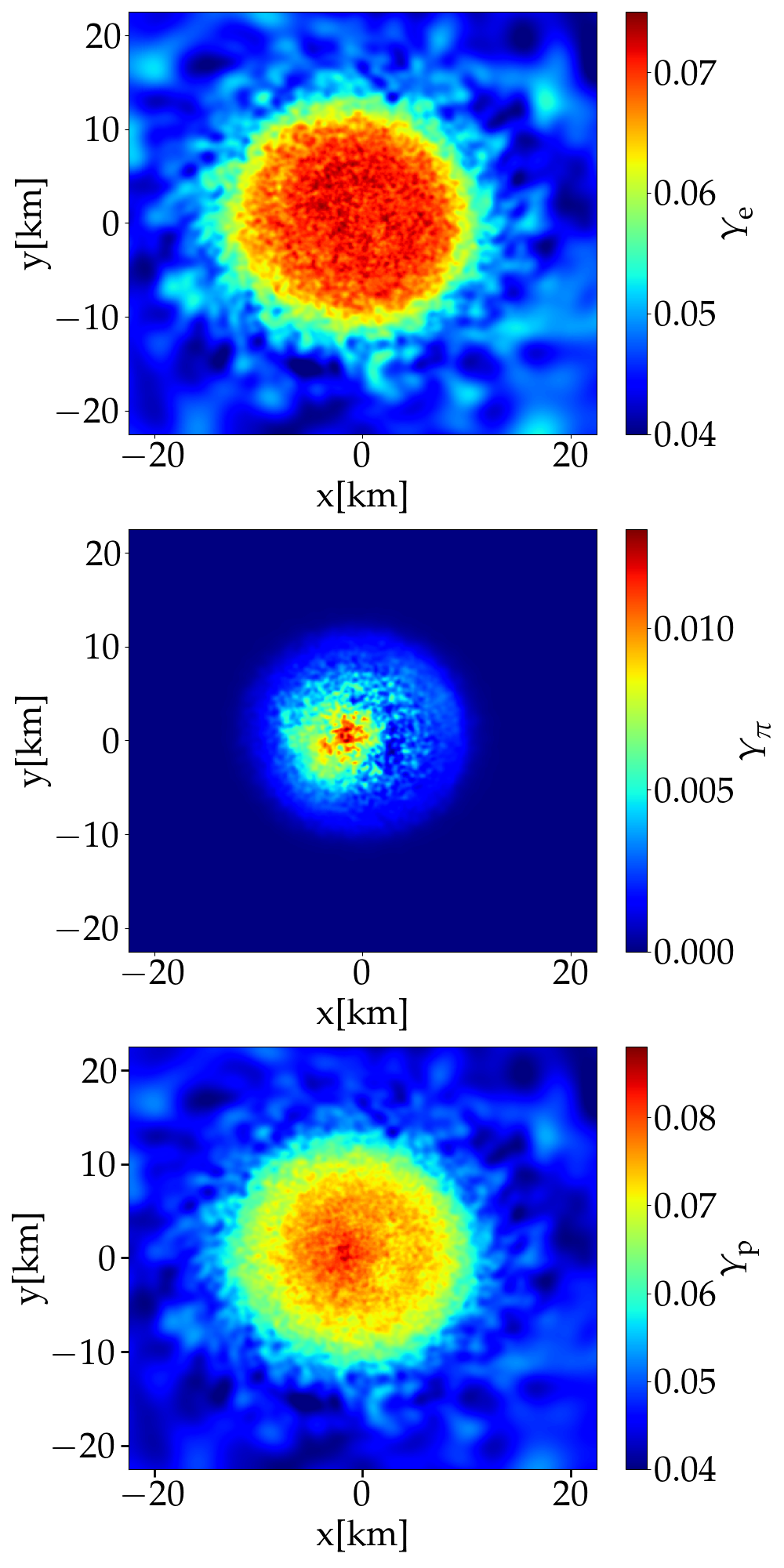}\label{fig:ye-ypi200}}
     \caption{$Y_e$ distribution (top row panels), $Y_\pi$ distribution (middle row panels) and $Y_\mathrm{p}$ distribution (bottom row panels) in the equatorial plane for 1.35-1.35$M_\odot$ BNS systems using $\mathrm{DD2}+\pi, m_\pi=\mathrm{Vac. \, mass}$ (left column panels), $\mathrm{DD2}+\pi, m_\pi=\mathrm{170 \, MeV}$ (middle column panels) and $\mathrm{DD2}+\pi, m_\pi=\mathrm{200 \, MeV}$ (right column panels).}
     \label{fig:ye-ypi}
 \end{figure*}

\begin{figure*}
     \subfigure[$\mathrm{T} ~\mathrm{[MeV]}$ 
     ]{\includegraphics[width=0.475\textwidth]{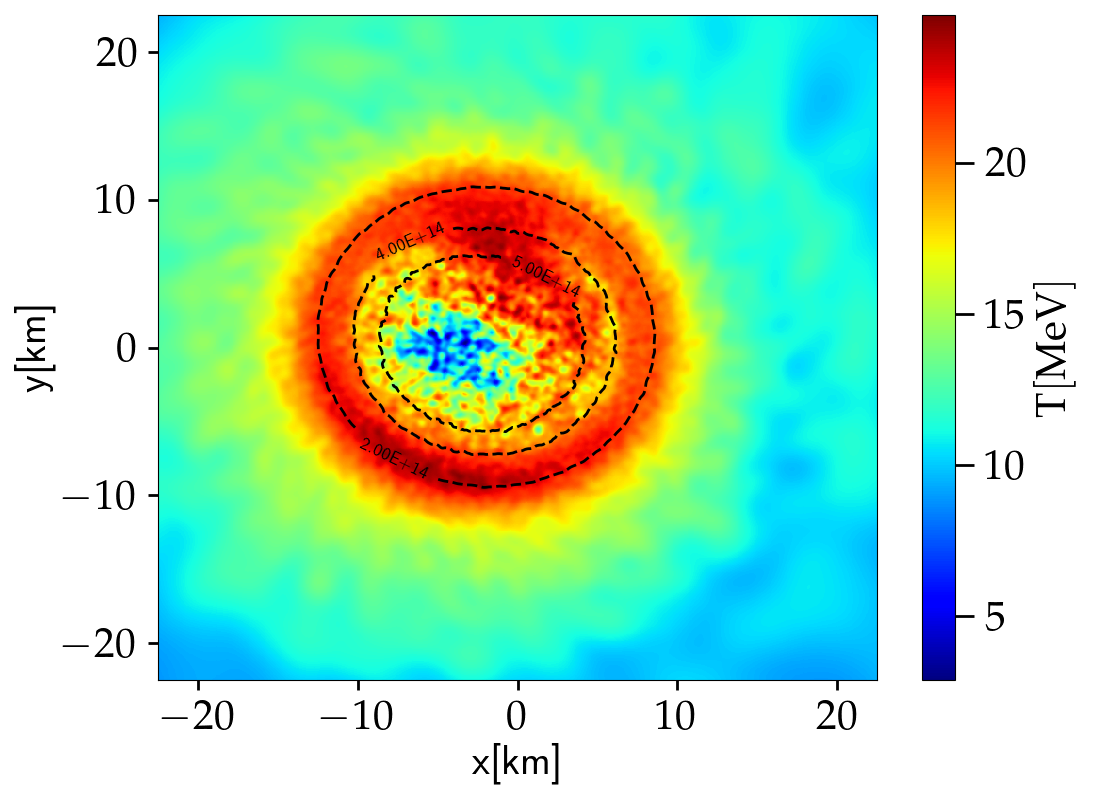}\label{fig:T_dd2_p139}}     
     \subfigure[$Y_\pi^\mathrm{c}$ 
     ]{\includegraphics[width=0.475\textwidth]{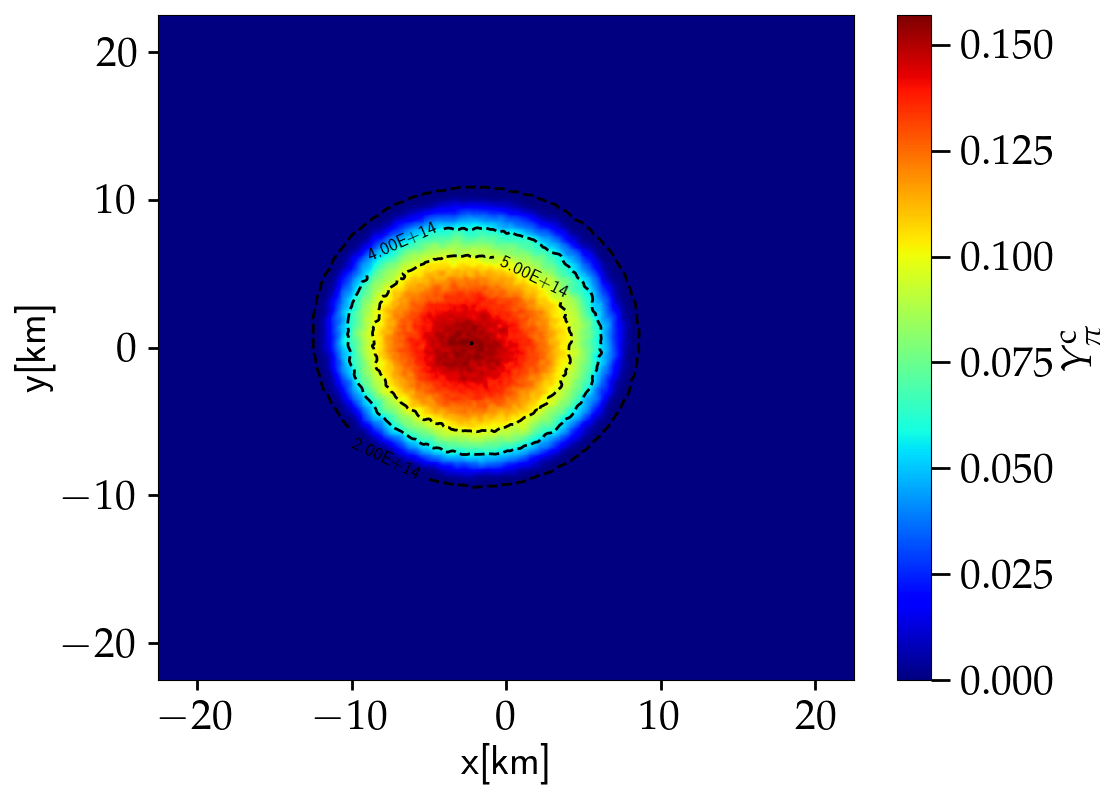}\label{fig:ypic139}}
     \\
     \subfigure[$Y_\pi^\mathrm{thermal}$ 
     ]{\includegraphics[width=0.475\textwidth]{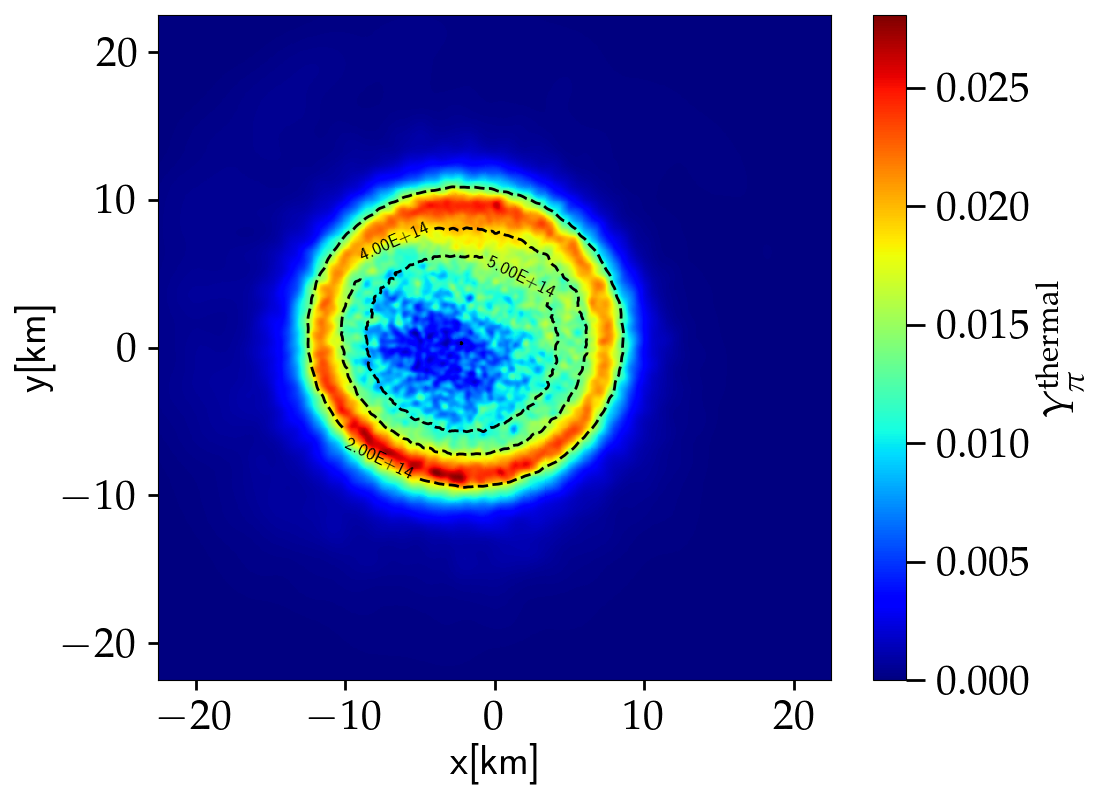}\label{fig:ypith139}}
     \subfigure[$Y_{\pi^0}$ 
     ]{\includegraphics[width=0.475\textwidth]{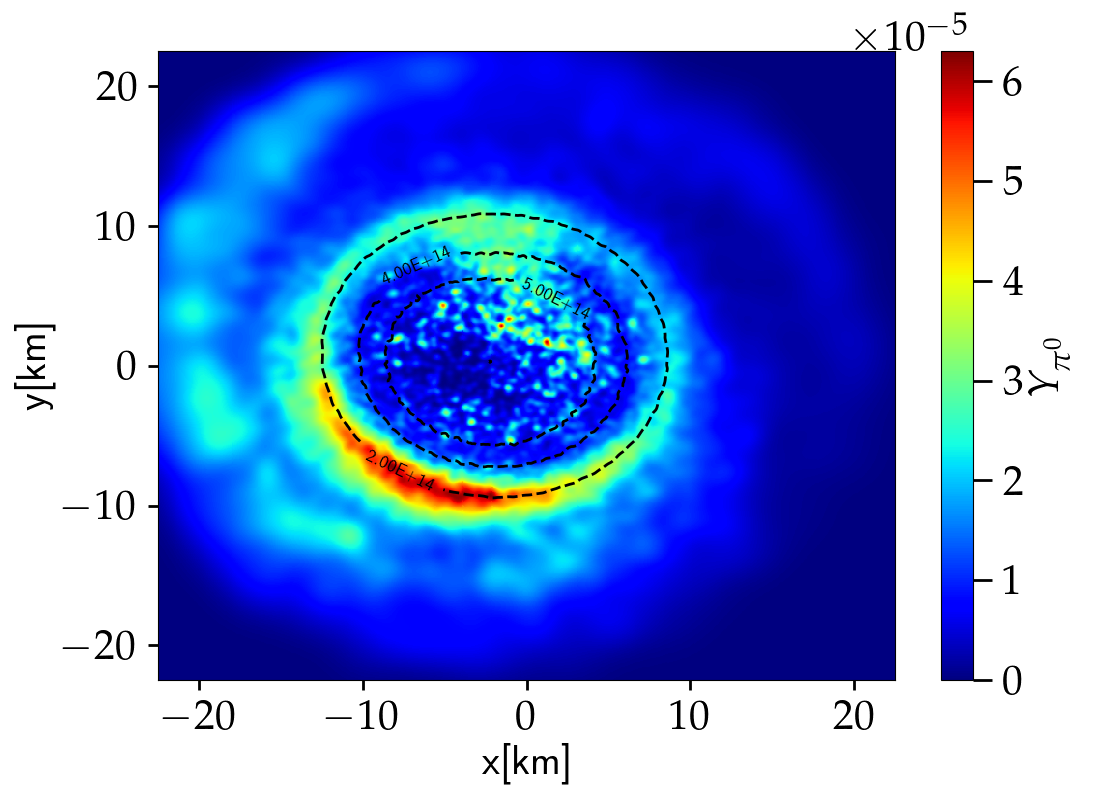}\label{fig:ypi0139}}
     \caption{Temperature, $Y_\pi^\mathrm{c}$, $Y_\pi^\mathrm{thermal}$ and $Y_{\pi^0}$ distribution in the equatorial plane for a 1.35-1.35$M_\odot$ BNS system using the $\mathrm{DD2}+\pi, m_\pi=\mathrm{Vac. \, mass}$ model. Densities of $2\times 10^{14}\mathrm{g/cm^3}$, $4\times 10^{14}\mathrm{g/cm^3}$ and $5\times 10^{14}\mathrm{g/cm^3}$ are shown by dashed contour lines. }
     \label{fig:Typicypithypi0}
 \end{figure*}

\begin{figure*}
     \subfigure[$\mathrm{SFHo}+\pi, m_\pi=\mathrm{Vac. \, mass}$]{\includegraphics[width=0.475\textwidth]{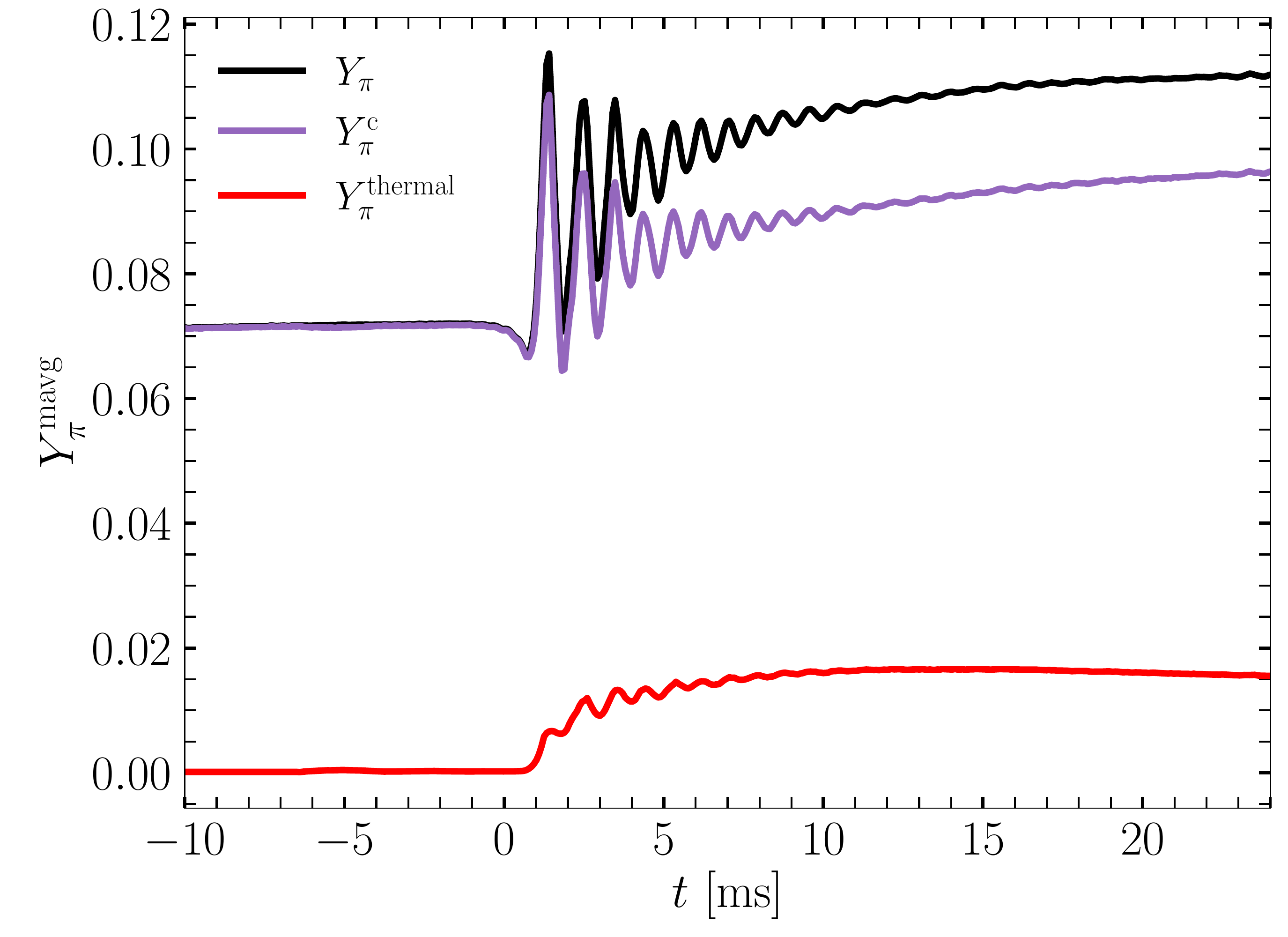}\label{fig:sfhoypi_ypic139}}
     \subfigure[$\mathrm{DD2}+\pi, m_\pi=\mathrm{Vac. \, mass}$]{\includegraphics[width=0.475\textwidth]{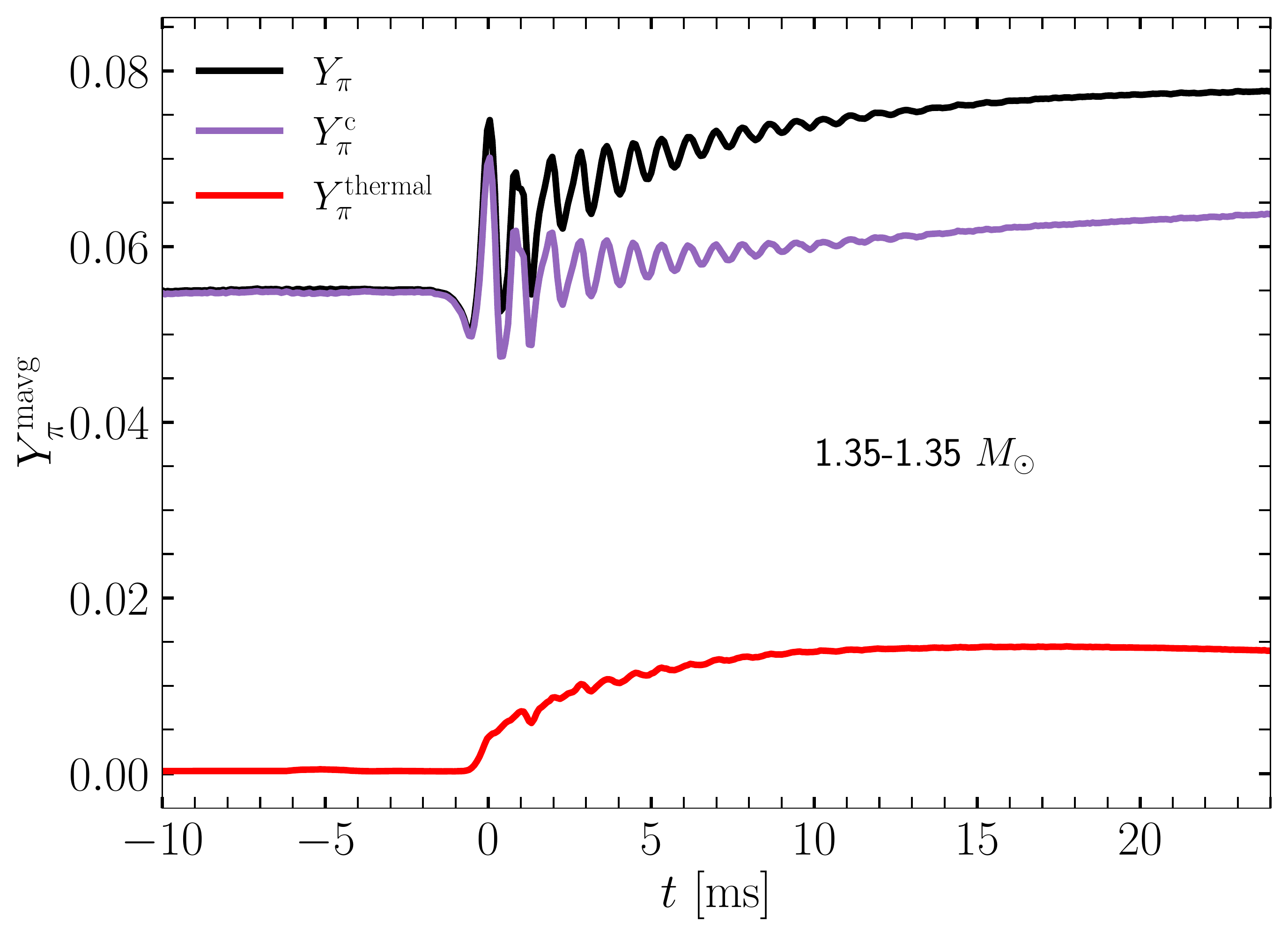}\label{fig:dd2ypi_ypic139}}
     \\
     \subfigure[$\mathrm{SFHo}+\pi, m_\pi=\mathrm{170 \, MeV}$]{\includegraphics[width=0.475\textwidth]{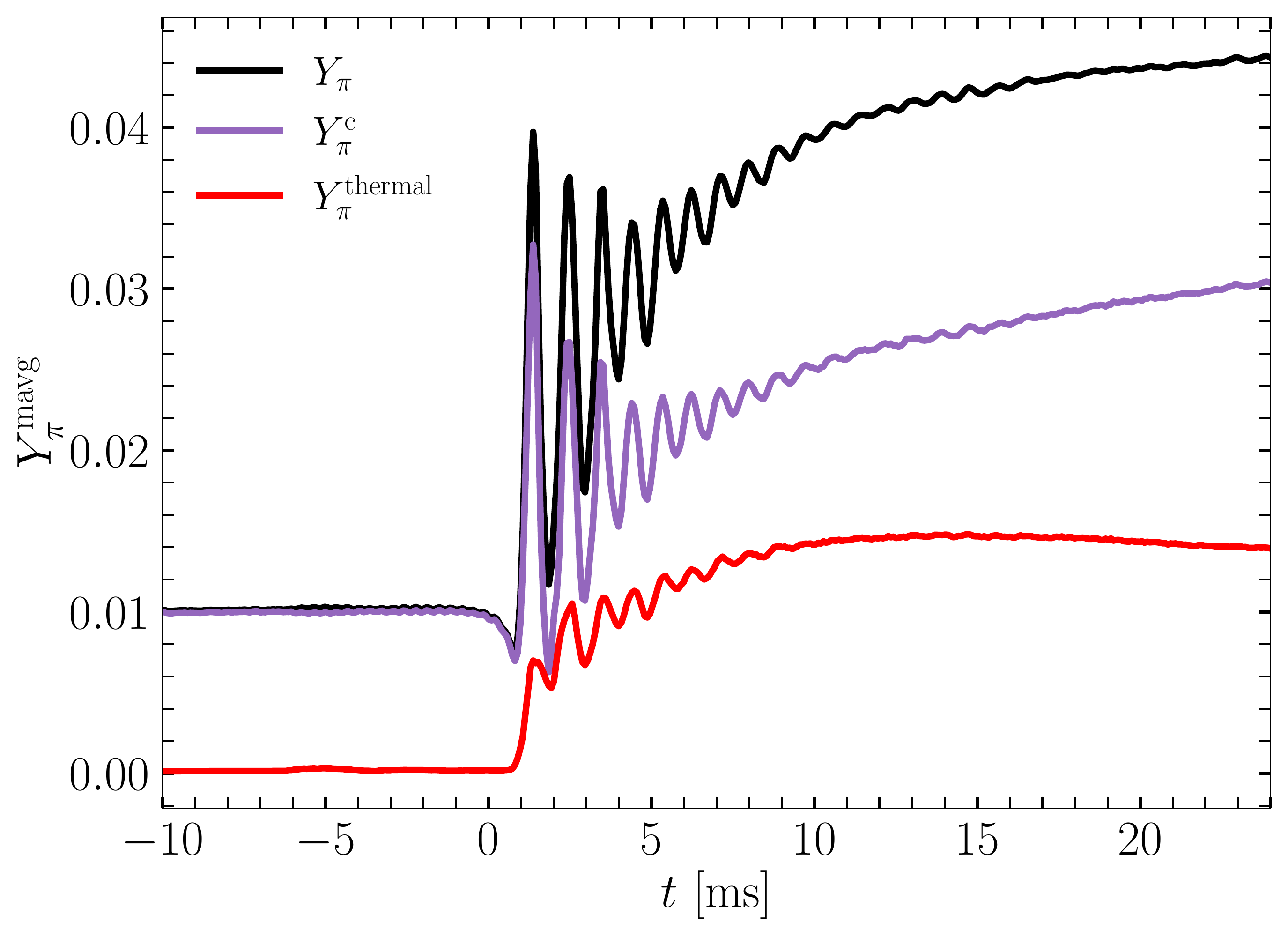}\label{fig:sfhoypi_ypic170}}
     \subfigure[$\mathrm{DD2}+\pi, m_\pi=\mathrm{170 \, MeV}$]{\includegraphics[width=0.475\textwidth]{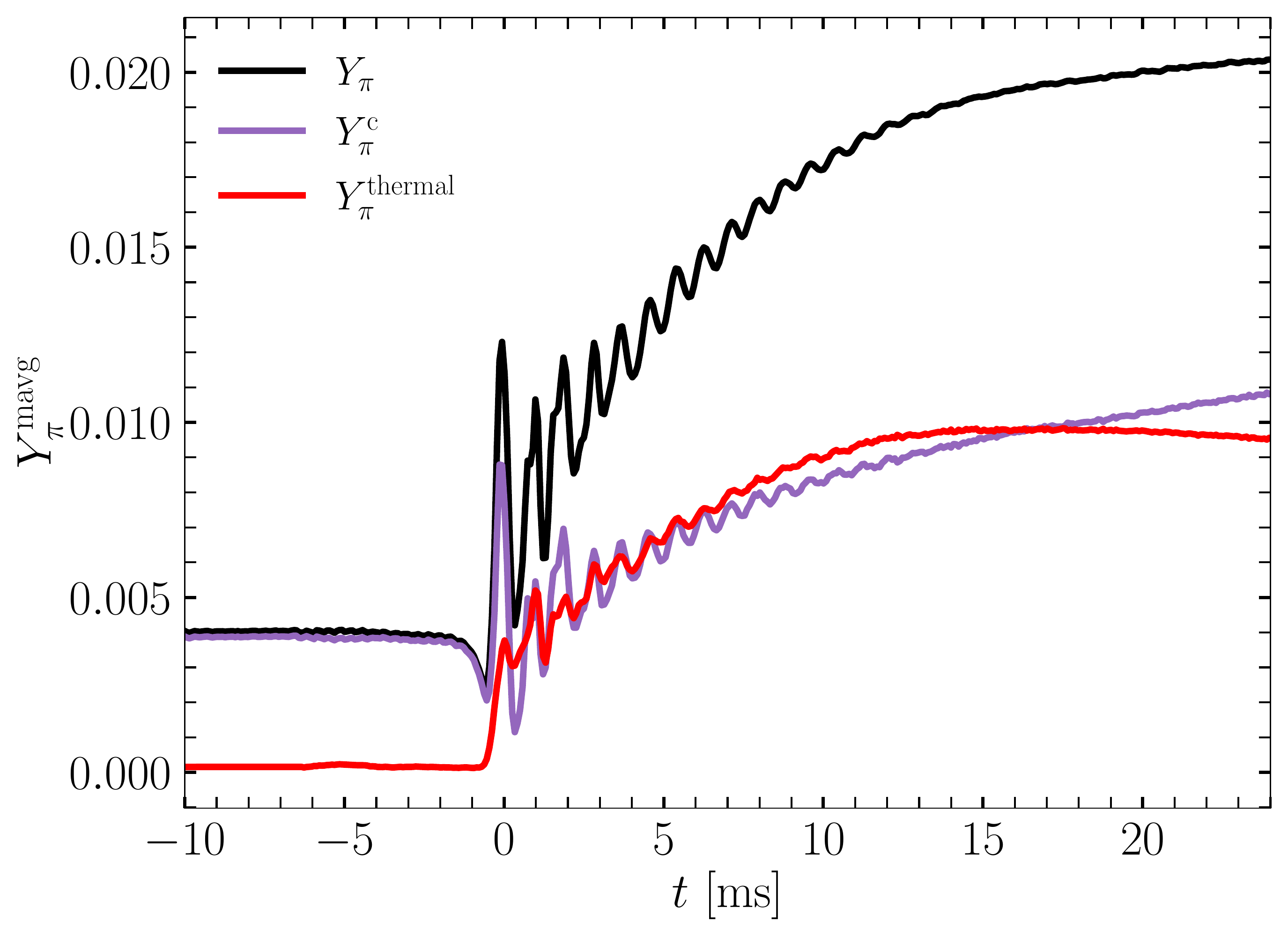}\label{fig:dd2ypi_ypic170}}     
     \\
     \subfigure[$\mathrm{SFHo}+\pi, m_\pi=\mathrm{200 \, MeV}$]{\includegraphics[width=0.475\textwidth]{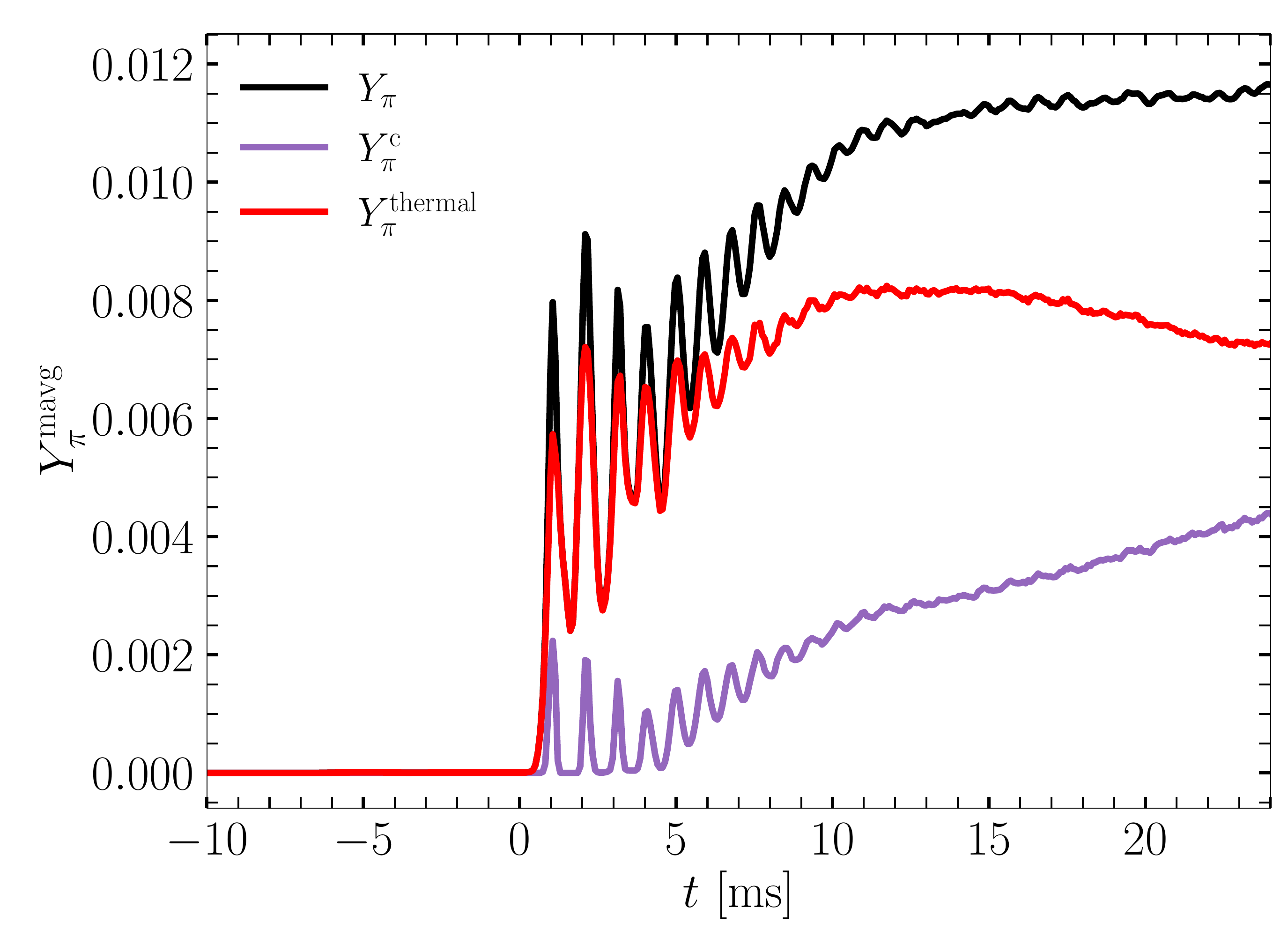}\label{fig:sfhoypi_ypic200}}
     \subfigure[$\mathrm{DD2}+\pi, m_\pi=\mathrm{200 \, MeV}$]{\includegraphics[width=0.475\textwidth]{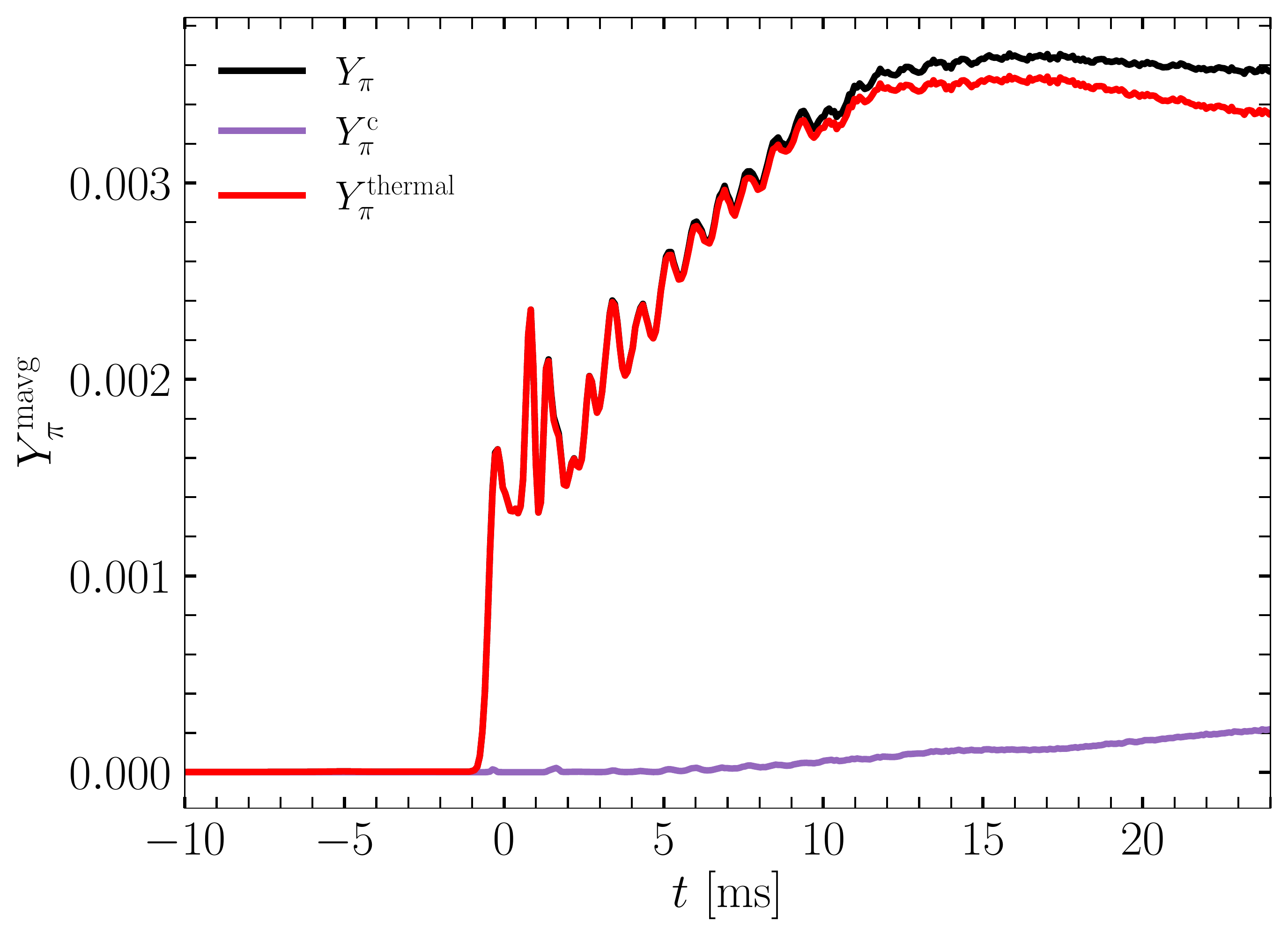}\label{fig:dd2ypi_ypic_200}}
     \caption{Time evolution of the mass-averaged pion fraction $Y^\mathrm{mavg}_{\pi}$ of total pion fraction $Y_{\pi}$, condensed pion fraction $Y_{\pi}^\mathrm{c}$ and thermal pion fraction $Y_{\pi}^\mathrm{thermal}$ for 1.35-1.35$M_\odot$ BNS systems. 
     }
     \label{fig:ypiypicypith}
 \end{figure*}

As discussed in Sect.~\ref{EOS}, the production of pions depends on the difference between the chemical potentials of neutrons and protons, and through this on the density and temperature. At typical densities of NSs and NS merger remnants, the contributions from thermal pions become dominant only beyond a threshold temperature of several 10~MeV (see Figs.~\ref{fig:pion_EOS_Tthres_mpi_139} and discussion in Sect.\ref{EOS}). We assess the temperature regime in the merger remnants of 1.35-1.35~$M_\odot$ binary simulations in Fig.~\ref{fig:Tmax} visualizing the evolution of the maximum temperature and the mass-averaged temperature\footnote{We define a mass-averaged quantity by summing the contribution from all SPH particles weighted by the mass of the particle relative to the total mass.}. The largest temperatures occur at the collisional interface between the merging NSs with this contact layer having a thickness of typically some 100~meters to km. We note that the exact values of $T_\mathrm{max}$ are affected by the numerical resolution with the general trend being an increase in the maximum temperature of a few tens of per cent \footnote{We have investigated the resolution dependence by running additional simulations with about 600,000, 1,200,000 and 2,400,000 SPH particles compared to our default resolution of 300,000 SPH particles. The maximum temperatures show a slight increase with higher resolution by at most a few 10~MeV, which is roughly comparable to the differences between models with and without pions.}. Furthermore, we find that relative changes on $T_{\mathrm{max}}$ with time are independent of the resolution. Although the maximum temperatures do reach about 100~MeV, such conditions are only found in a small volume in the remnant and not necessarily at the highest densities. Average temperatures in the merger remnant are significantly lower (of the order of about 20~MeV), which suggests that thermal pions likely play only a sub-dominant role in the merger remnant of the systems considered here. Hence, we anticipate that the inclusion of pions in BNS merger simulations overall has a softening effect.  

In Fig.~\ref{fig:Tmax_SFHo}, we observe that the models with the softer SFHo EOS lead to slightly higher maximum temperatures, which is understandable given that the stars described by this EOS collide more violently and the compression in the remnant is stronger. There is also the tendency that models with pions result in slightly lower maximum temperatures. Figure~\ref{fig:pion_EOS_T_rho_1e15_SFHo_all_mass_compare} clearly shows that at such high temperatures the inclusion of pions leads to a stiffening of the EOSs, which is roughly independent of the effective pion mass. We caution that one should not overinterpret the differences in the maximum temperature between EOSs because we evaluate the temperature on the basis of individual SPH particles, which are affected by noise\footnote{We also run simulations with about 600,000 SPH particles for all SFHo models with and without pions. The evolution of $T_\mathrm{max}$ in the first few milliseconds after merging follows approximately the one in the simulations with our default resolution. Only for the model with $m_\pi$ equal to the vacuum mass, we find that $T_\mathrm{max}$ is higher (by about 20~MeV) compared to the run with 300,000 particles. Comparing runs with 600,000 particles, the model with pion mass equal to the vacuum mass actually features the highest temperatures in the first milliseconds after merging somewhat in contrast to the finding for the default-resolution simulations, where the base models yields the highest $T_\mathrm{max}$.}.

Evaluating directly the ratio between the number of thermal pions and pions in the condensate, we do indeed find that thermal pions are subdominant even in the hot merger remnant and that the main effects of pions originate from the condensate (Figs.~\ref{fig:Typicypithypi0} and~\ref{fig:ypiypicypith} except for the case with  $m_\pi=200$~MeV). 

In this respect, it is also instructive to directly compare the difference $\hat{\mu}$ between the chemical potentials of neutrons and protons for the base model and the calculation including pions. Figure~\ref{fig:mun-mup} shows $\hat{\mu}$ in the equatorial plane for the DD2 base model and the model with pions adopting the vacuum pion mass. In the latter case,  $\hat{\mu}$ is limited by the pion mass ($\sim$139.6~MeV for this case), whereas  $\hat{\mu}$ reaches more than 200~MeV in the remnant's center if pions are ignored. If pions with an effective mass equal to their vacuum mass are included, a pion condensate is found throughout the whole high-density region of the remnant. One may already conclude from the range of $\hat{\mu}$ in the left panel of Fig.~\ref{fig:mun-mup} that the condensation of pions in the remnant depends on the adopted effective pion mass. $\hat{\mu}$ increases towards the center for the base model and reaches 200~MeV only in the very inner region. Hence, for $m_\pi=200$~MeV, pion condensation only takes place in the center of the remnant at the highest densities. This clearly demonstrates that the effective pion mass has a significant impact on the production of pions in BNS mergers.

These observations and their impact are further illustrated in Fig.~\ref{fig:ye-ypi}, which displays snapshots of the electron fraction (upper panel), the pion fraction (middle panel) and the proton fraction (bottom panel) in the equatorial plane for a 1.35-1.35$M_\odot$ BNS merger remnant about ($\sim 18.95$) milliseconds after merging, employing the pionic DD2 EOSs. The columns refer to different $m_\pi$ increasing from left to right.

It is obvious that the inclusion of pions and their effective mass have a very strong impact on the electron fraction in the remnant, which is determined by $\hat{\mu}$ and thus by the presence of pions and their properties in the progenitor (see Fig.~\ref{fig:Esymm}). For relatively small pion masses, the electron fraction in the remnant remains rather small while the proton fraction increases with the produced number of pions (Fig.~\ref{fig:ye-ypi139}). For $m_\pi=\mathrm{200 \, MeV}$ (Fig.~\ref{fig:ye-ypi200}) the highest $Y_\mathrm{e}$ and the lowest $Y_\mathrm{p}$ is found. We recall that in these simulations we employ a rather simplistic treatment of weak interactions by ignoring neutrinos and only advecting the electron fraction. Hence, the behavior in Fig.~\ref{fig:ye-ypi} can be easily understood by the behavior in Fig.~\ref{fig:Esymm}. For smaller effective pion masses, $Y_\mathrm{e}$ is strongly reduced in the initial stars, which thus yields a lower electron fraction in the remnant. The behavior of the pion fraction and the proton fraction is a consequence of the advected electron fraction. Both follow similar trends as for equilibrium NSs, i.e.~a lower $Y_\mathrm{e}$ is accompanied by a larger proton fraction and a larger pion fraction (see Fig.~\ref{fig:Esymm}). Despite the simple treatment of weak interactions, the panels indicate a potentially significant impact by pions on the microphysics of weak interactions as they affect the conditions for the chemical potentials in the remnant. 

The middle row panels in Fig.~\ref{fig:ye-ypi} visualize the pion content in the remnant depending on the effective pion mass. As expected, the volume of the region inside the HMNS where pions are present, shrinks as the effective pion mass increases. The pion content is strongly suppressed for $m_\pi=\mathrm{200 \, MeV}$, and only occurs in the center where $\hat{\mu}$ reaches the pion mass. This panel illustrates that the $\mathrm{DD2}+\pi, m_\pi=\mathrm{200 \, MeV}$ model should behave very similar to the base EoS. 

To address the distribution of thermal and condensed pions in different regions of the merger remnant, we show the distribution of $Y_\pi^\mathrm{c}$ and $Y_\pi^\mathrm{thermal}$ in Fig.~\ref{fig:Typicypithypi0} for the $\mathrm{DD2}+\pi, m_\pi=\mathrm{Vac. \, mass}$ model along with the corresponding temperature distribution. The correlation between thermal pions and the local temperature is clear from the hot ring like structure in the merger remnant shown in Figs.~\ref{fig:T_dd2_p139} and \ref{fig:ypith139} whereas $Y_\pi^\mathrm{c}$ roughly follows the density in Fig.~\ref{fig:ypic139}. Although the number of neutral pions increases with temperature, the relative magnitude is very low (Fig.~\ref{fig:ypi0139}). We expect the neutral pion contribution to become noticeable only at high temperatures around $\sim$100\,MeV (particles with maximum temperature), see Fig.~\ref{fig:pion_EOS_T_rho_1e15_mpi_139} and~\ref{fig:pion_EOS_T_rho_1e15_mpi_139_dd2}.

As temperatures and densities increase during the dynamical evolution, more pions are produced as compared to the pion content of the initial stars. We quantify the pion production in Fig.~\ref{fig:ypiypicypith}, which provides the mass averaged fraction of the total number of pions, condensed pions and thermal pions as function of time for both SFHo and DD2 EOSs. 

One can clearly recognize that the effective pion mass sensitively affects the production of pions. For the pion mass equal to the vacuum mass, there are already a lot of pions present in the initial stars and the additional increase of the pion fraction during merging is moderate. For $m_\pi=\mathrm{170 \, MeV}$ we see a significant increase of $Y_\pi$ by about a factor four. If the effective pion mass is as high as 200~MeV, there are no pions present during the inspiral and only the density and temperature increase after merging leads to the production of pions, which are predominantly thermal.

SFHo is softer than DD2 and thus results in higher densities in the inspiralling stars and in the postmerger remnant. Hence, the pion production is increased for the SFHo based models as compared to the EOSs relying on DD2. Considering this behavior, we anticipate that the softening of the EOS by pions is more pronounced for the softer SFHo model as compared to DD2. 

\subsection{Gravitational wave signal}\label{GW spectra}

\begin{figure*}
    \centering
            \subfigure[SFHo]{\includegraphics[width=0.48\textwidth]{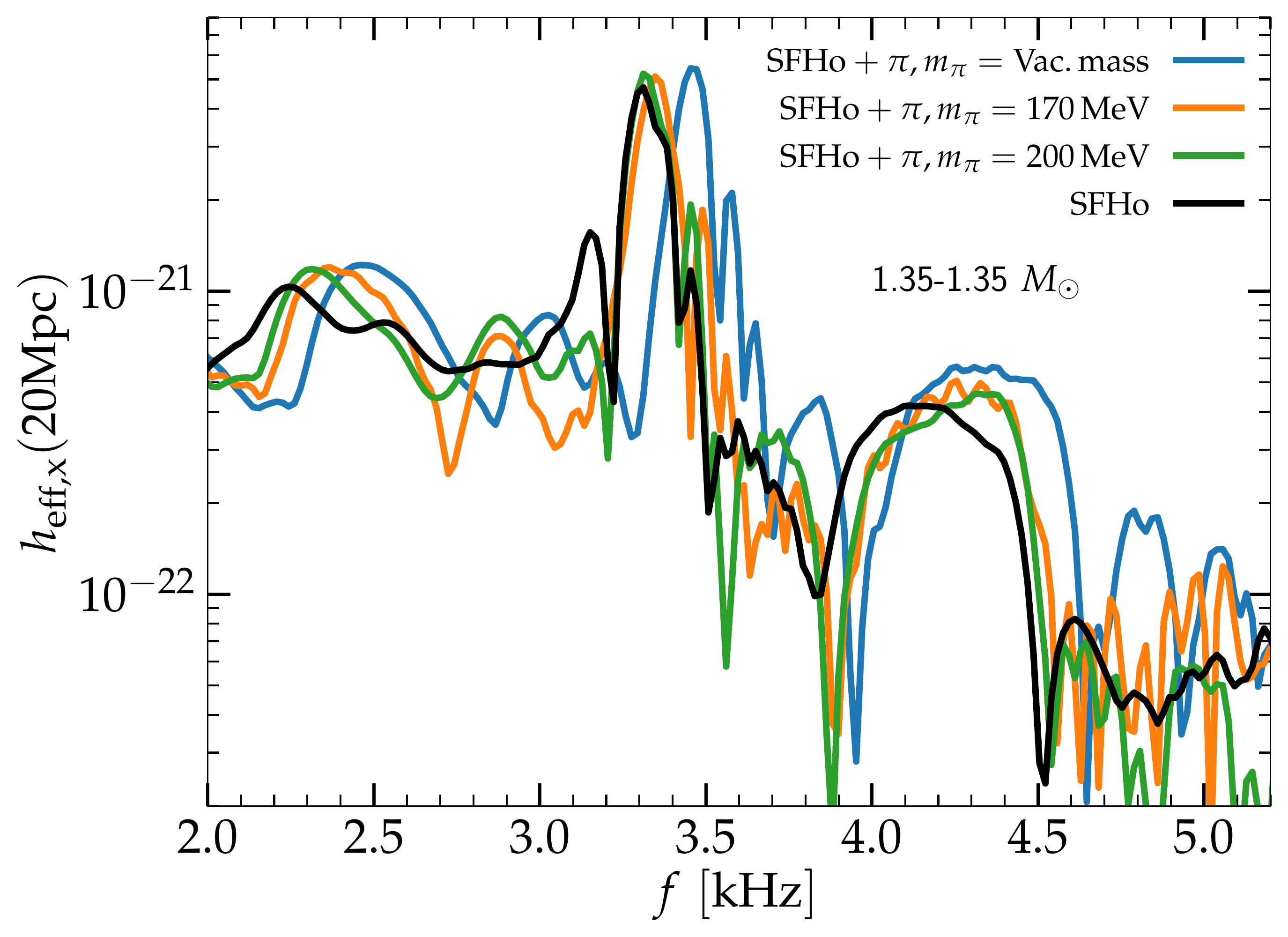}\label{fig:spectra_SFHO}}
	        \subfigure[DD2]{\includegraphics[width=0.48\textwidth]{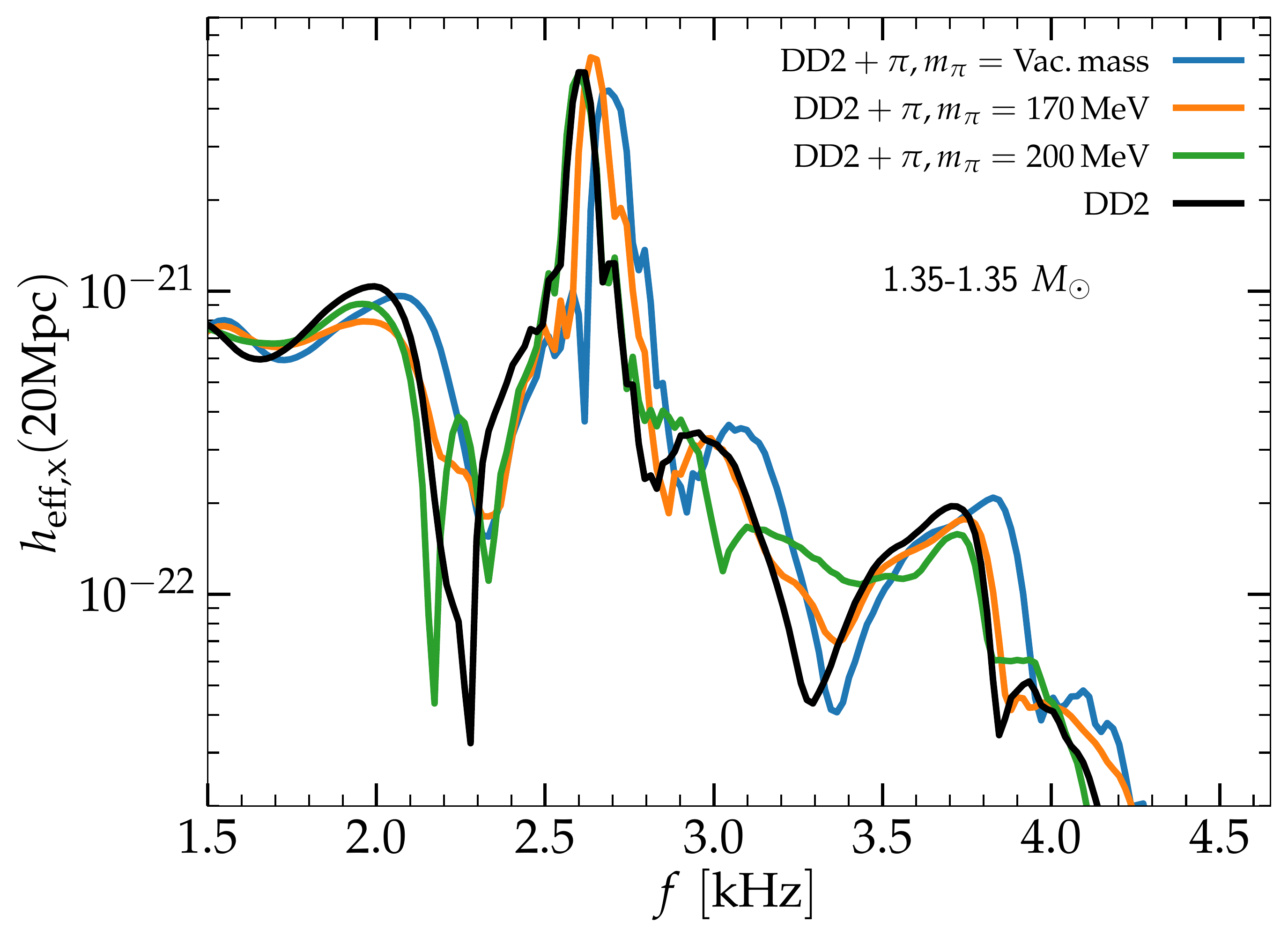}\label{fig:spectra_DD2}}
   
   	\caption{Comparison of the the cross-polarization amplitude of the GW spectra for an observer at 20 Mpc along the polar axis with the SFHo and DD2 EOSs for 1.35-1.35$M_\odot$ binaries. The color scheme is the same as in Figs.~\ref{fig:Rhomax} and~\ref{fig:Tmax}.}
   	\label{fig:spectra}
\end{figure*}

\subsubsection{Inspiral}
The GW signal reflects the dynamical evolution of BNS mergers and thus carries an imprint of the EOS. The GW inspiral signal is determined by the tidal deformability $\Lambda$, where we find a reduction of $\Lambda$ of up to about 10 per cent by the inclusion of pions comparing the base models without pions and the modified EOSs with an effective pion mass equal to the vacuum mass. As anticipated from the mass-radius relations (Fig.~\ref{fig:TOV_M_R}), the reduction of the tidal deformability depends on the effective pion mass, and larger effective pion masses have a smaller or even negligible impact on the tidal deformability. 

While these findings are not unexpected, they caution that ignoring pions in EOS calculations might potentially introduce a systematic bias if the actual effective pion mass in medium is close to its vacuum mass. This may apply to cases where microphysical parameters of the EOS as for instance the slope of the symmetry energy are inferred from the GW inspiral signal through microphysical EOS models which consider only the baryonic and leptonic contributions. Neglecting pions and thus an additional unmodelled softening of the EOS model would lead to a systematic bias of these quantities. Clearly, whether these are severe issues, cannot be judged in our present explorative study with the rather crude incorporation of pions and especially freely chosen effective pion masses. One may expect that the actual impact of pions may become stronger for high-mass NSs and BNSs.

\subsubsection{Postmerger}

The GW signal can also contain a high-frequency component originating from the postmerger phase, i.e. the dynamical evolution of the NS remnant, whereas the high-frequency emission from a directly forming black hole in a prompt gravitational collapse is very weak. If a rotating NS merger remnant forms, one may expect a pronounced impact of pions because the postmerger remnant features finite temperatures and higher densities than the inspiralling NSs. The GW emission of a NS postmerger remnant is dominated by a single, roughly constant frequency, $f_\mathrm{peak}$, which reflects the main oscillation mode of the central object and occurs as a strong peak in the GW spectrum.

Figure~\ref{fig:spectra} shows the amplitude of the postmerger GW spectra for an observer at 20~Mpc along the polar axis for 1.35-1.35$M_\odot$ BNS systems with the SFHo EOSs (left panel) and DD2 EOSs (right panel). It is known that the main frequency scales tightly with NS radii as a measure for the softness of the EOS~\citep{Bauswein_2012}. The DD2 models oscillate at significantly lower frequencies of about 2.6~kHz compared to more than 3~kHz for the SFHo based EOSs.

Considering the impact of pions, we find that their inclusion shifts the main peak to higher frequencies. This effect is more pronounced for smaller effective pion masses, which is not surprising considering for instance the mass-radius relations of these EOSs. The shift appears to be slightly more pronounced for the SFHo based models with a modification of about 200~Hz. For $m_\pi=\mathrm{200 \, MeV}$, the main peak is hardly affected by the inclusion of pions, which is consistent with the findings in Sect.~\ref{Evolution}, namely that there is only a small contribution by pions for this model. We also recognize an impact on the secondary features of the GW spectrum, i.e. subdominant frequency peaks that have for instance been associated with the formation of massive tidal bulges at the surface of the remnant or a non-linear coupling of the dominant mode and the quasi-radial oscillation of the central object (see~\citep{Bauswein_2019} for a detailed discussion). Interestingly, it seems that at least for the SFHo based models the influence of pions on those features is stronger.

\begin{figure}[htb]
	\includegraphics[width=\linewidth]{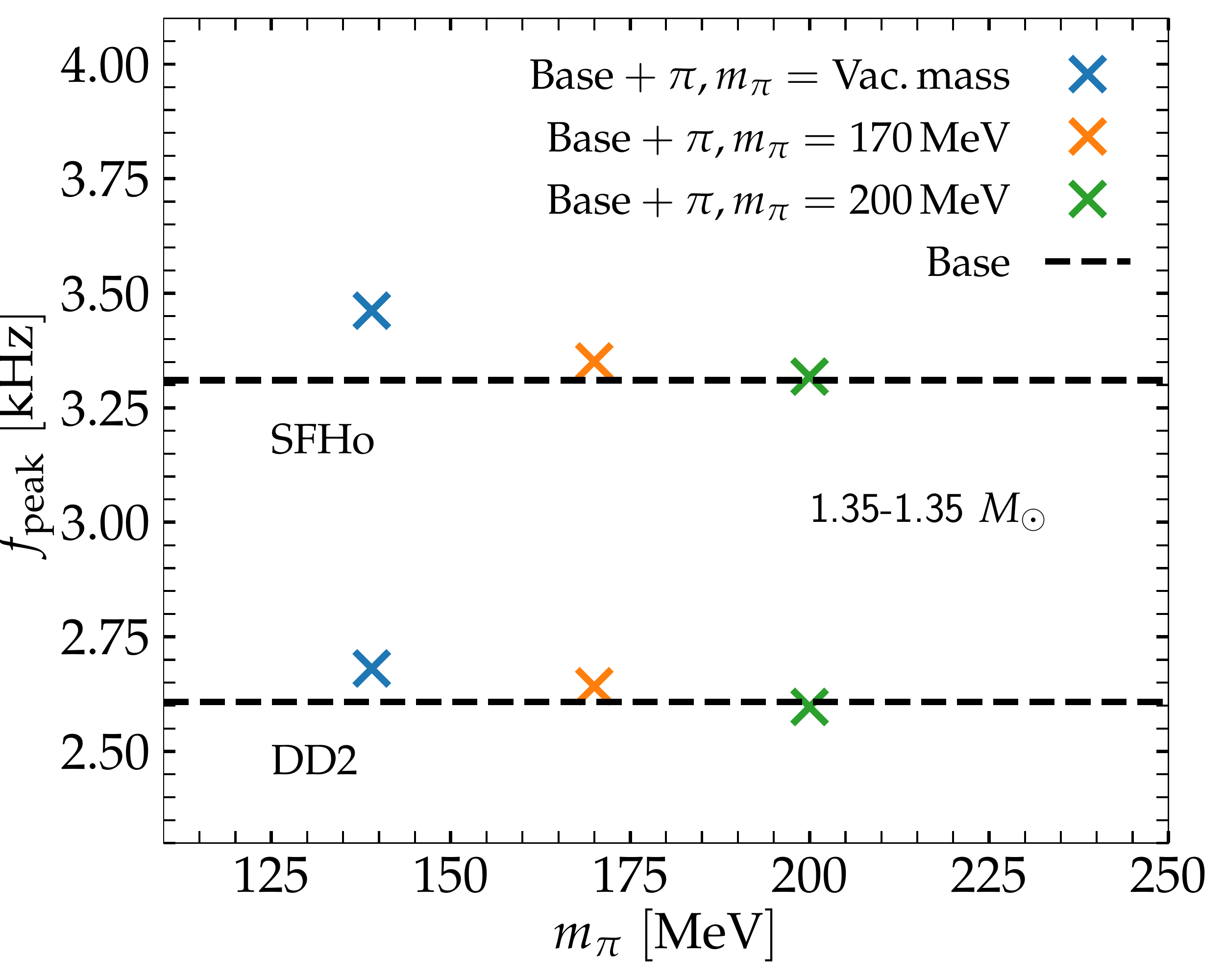}
	\caption{The dominant postmerger GW oscillation frequency with respect to the pion mass employed in the modified EOSs for 1.35-1.35$M_\odot$ simulations. $f_\mathrm{peak}$ for the respective base EOS is indicated by the black dashed line.}	
	\label{fig:fpeak_mpi}
\end{figure}

We display the frequency shift of the main peak and the quantitative impact of the chosen effective pion mass in Fig.~\ref{fig:fpeak_mpi}, which reveals a clear dependence on $m_\pi$. The shifts also indicate (based on only these two base EOSs) that there may be a stronger impact for softer EOSs. This may not be surprising since such models reach higher densities and temperatures, where pions may lead to a strong impact. We observe similar trends for the frequency shift also for other binary masses (see Tabs. ~\ref{tab:progenitor_property_SFHo} and ~\ref{tab:progenitor_property_dd2}).

\subsection{Empirical relations for postmerger GW frequencies} \label{Mass relations}
By surveying a large sample of EOS models without pions, a number of empirical relations have been identified, which relate the main postmerger frequency to stellar properties of isolated, cold, nonrotating NSs such as their radii or tidal deformability, e.g.~\citep{Bauswein2012,Bauswein_2012,Hotokezaka2013,Bernuzzi2015,Takami2015,Rezzolla2016,Lehner2016,Breschi2019,Tsang2019,Blacker_2020,Vretinaris_2020}. These relations are important because they are the basis for inferring these stellar parameters from a measurement of $f_\mathrm{peak}$, e.g.~\citep{Chatziioannou2017}.

Our results so far showed that the inclusion of pions affects both the stellar parameters of isolated NSs and the properties of the postmerger GW emission. As the empirical relations are built on models ignoring pions, it is thus important to investigate whether models with pions do follow these empirical relations or whether there are deviations. This is relevant because the use of the existing relations could introduce a bias when they are employed for EOS constraints.

We also recall that the impact of pions may be different in different regimes. The analysis in Sect.~\ref{EOS} shows that including pions can lead to a softening of the EOS compared to a base model without pions or to a stiffening for instance at high temperatures. Hence, the effect of pions is not straightforward to estimate.

We address this aspect by evaluating mass-dependent empirical relations, i.e. relations for a fixed total binary mass, since those yield the tightest relations and possible deviations can be identified more easily. Specifically, we check the results for 1.35-1.35~$M_\odot$ binaries.

\begin{figure}[htb]
	\includegraphics[width=0.48\textwidth]{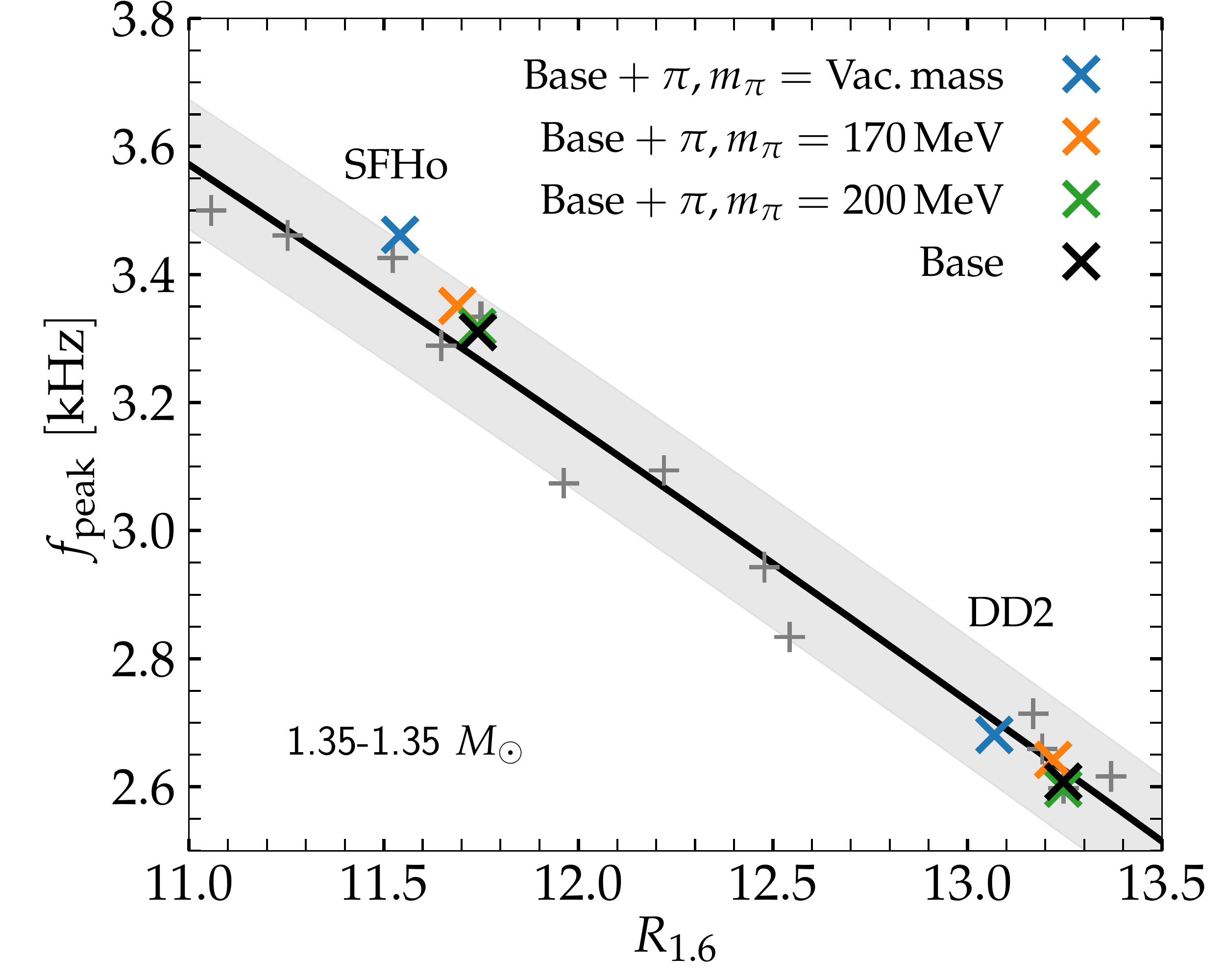}
	\caption{Dominant postmerger GW frequency, $f_\mathrm{peak}$ versus the radius of non-rotating 1.6$M_\odot$ NS for 1.35-1.35$M_\odot$ BNS models using the modified SFHo and DD2 EOSs (crosses) compared to models using different microphysical EOSs which do not include pions (plus signs).
	The solid curve is a least square fit to the data points of EOSs without pions, with the gray shaded area visualizing the largest deviation of the data from the least squares fit.}	
	\label{fig:fpeak_R16}
\end{figure}

Figure~\ref{fig:fpeak_R16} displays the relation between $f_\mathrm{peak}$ and the radius $R_{1.6}$ of nonrotating NSs with a mass of 1.6~$M_\odot$~\citep{Bauswein_2012}. The plot includes results from our new calculations and older data from \citep{Blacker_2020} for a larger set of EOSs none of which considers pions. The solid black line is a least square polynomial fit to the data from \citep{Blacker_2020}, i.e. it describes the behavior of EOS models without pions. The band illustrates the maximum scatter in this relation; we choose the width to equal the maximum residual between fit and data.

The figure shows that the inclusion of pions does not lead to strong deviations from the empirical $f_\mathrm{peak}$-$R_{1.6}$ relation. The presence of pions simultaneously changes the properties of the postmerger GW emission and the parameters of nonrotating NSs. The modifications by pions to a large extent cancel each other. Only for the SFHo model with $m_\pi$ equal to the mass in vacuum, $f_\mathrm{peak}$ is shifted to higher frequencies such that it is marginally compatible with the band defining the inherent scatter of this relation (see also \citep{Lioutas_2021} for an extensive discussion of frequency deviations in empirical relations for the postmerger GW signal). These findings indicate that at most there could be a slight bias for very soft EOS models in the sense that the actual relation lies at slightly higher frequencies if the actual effective pion mass is close to its vacuum value.

\begin{figure}[htb]
	\includegraphics[width=0.48\textwidth]{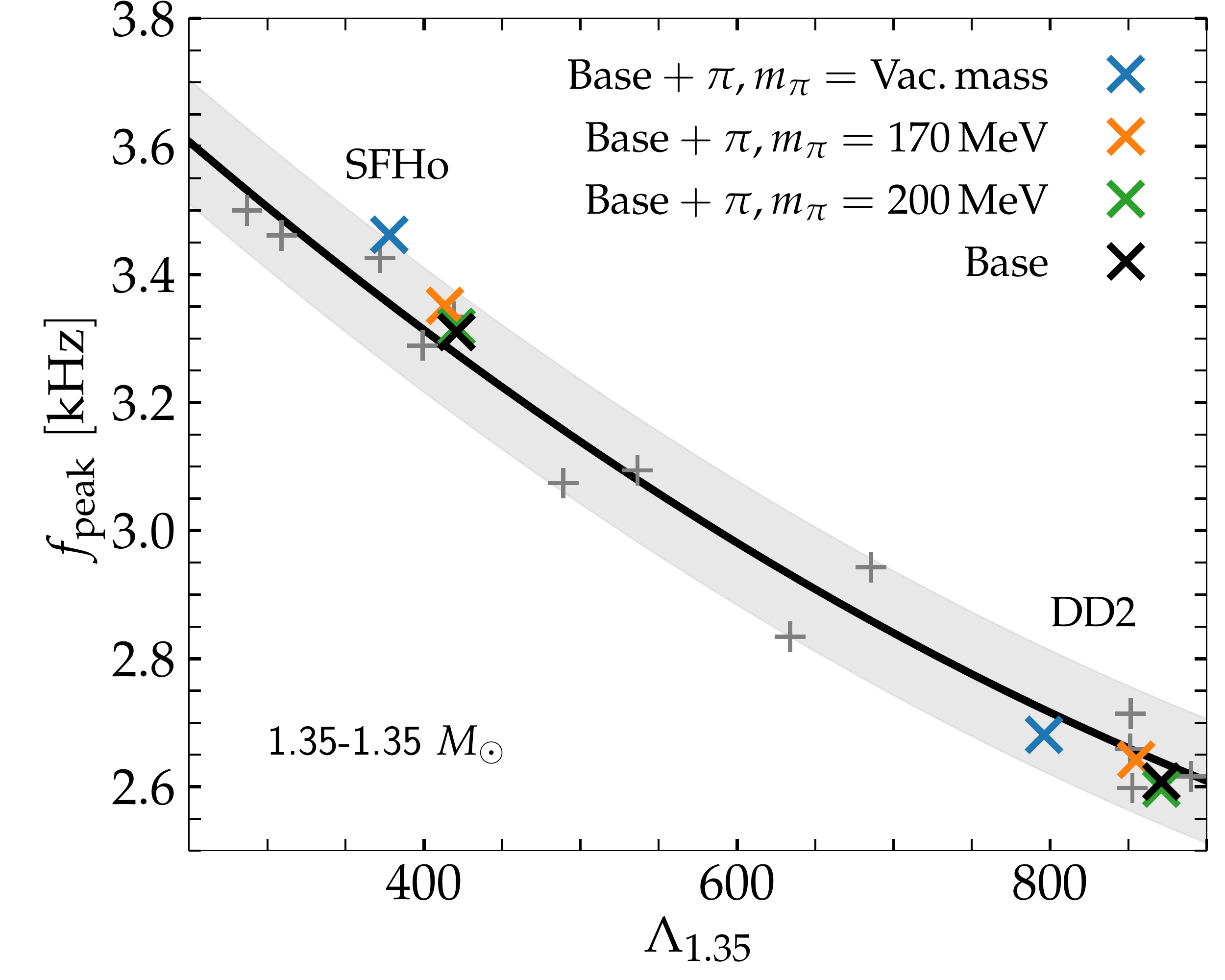}
	\caption{Dominant postmerger GW frequency, $f_\mathrm{peak}$, versus the tidal deformability $\Lambda$ of 1.35-1.35$M_\odot$ BNSs using modified SFHo and DD2 EOSs (crosses) compared to models using different microphysical EOSs which do not include pions (plus signs).	The solid curve is a least square fit to the data points of EOSs without pions, with the gray shaded area visualizing the largest deviation of the data from the least squares fit.}	
	\label{fig:fpeak_lambda}
\end{figure}

A similar picture arises for relations which link the dominant postmerger frequency and the tidal deformability~\citep{Takami2015,Bernuzzi2015,Blacker_2020}. This is demonstrated in Fig.~\ref{fig:fpeak_lambda}. Again, models with pions are in agreement with relations that are based on models without pions. $\mathrm{SFHo}+\pi, m_\pi=\mathrm{Vac.\, mass}$ model leads to a somewhat more pronounced frequency shift, and this case is only marginally compatible with the old relation.

\begin{figure}
	\includegraphics[width=0.48\textwidth]{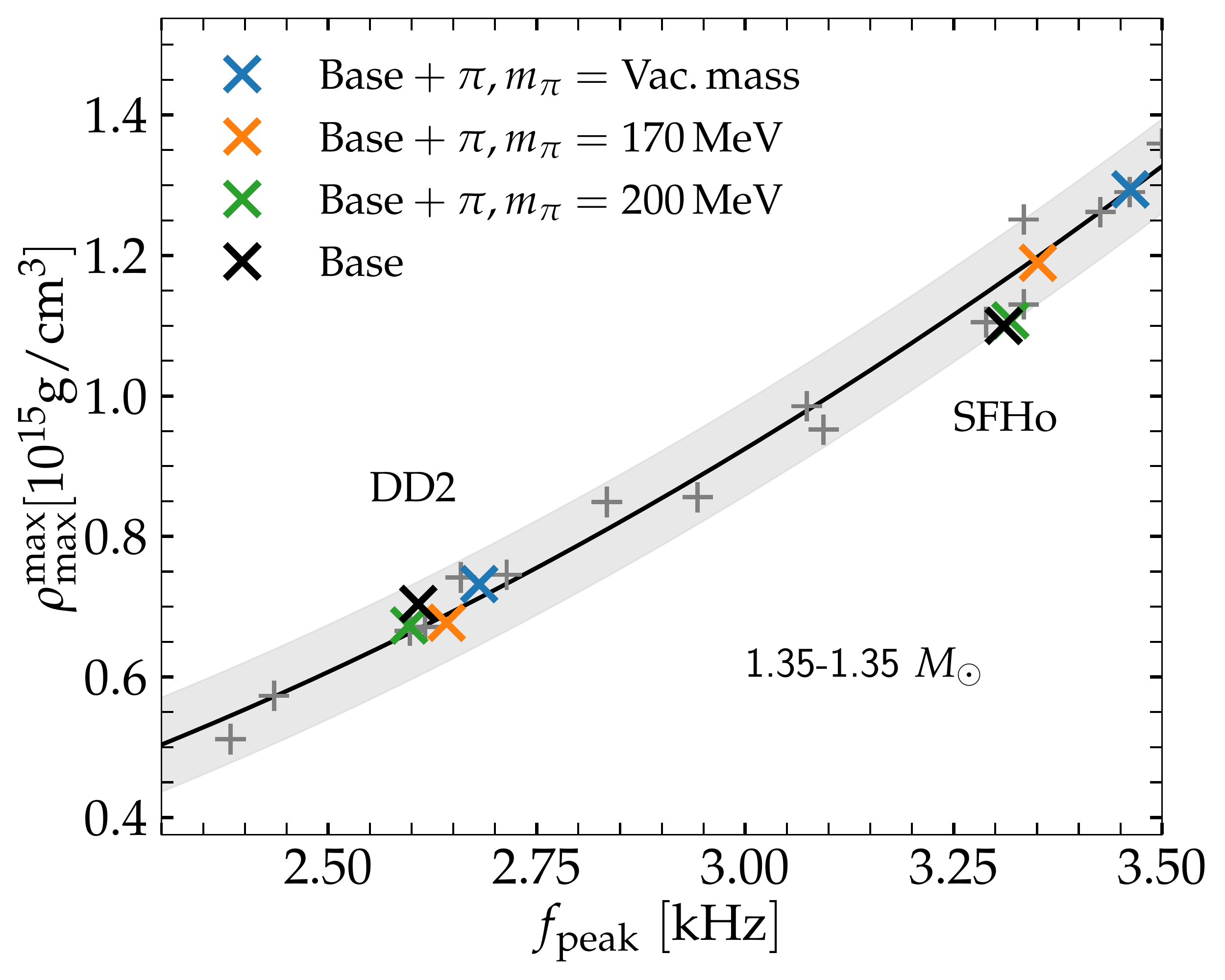}
	\caption{Maximum rest-mass density $\rho^\mathrm{max}_\mathrm{max}$ of the early postmerger evolution as a function of the dominant postmerger GW frequency, $f_\mathrm{peak}$, for a 1.35-1.35$M_\odot$ BNSs using modified SFHo and DD2 EOSs (crosses) compared to models using different microphysical EOSs which do not include pions (plus signs). The solid curve is a least square fit to the data points of EOSs without pions, with the gray shaded area visualizing the largest deviation of the data from the least squares fit.}	
	\label{fig:rhomax_fpeak}
\end{figure}

We furthermore investigate a relation that connects the dominant postmerger frequency and the maximum rest-mass density, $\rho_\mathrm{max}^\mathrm{max}$, which occurs during the early evolution in the postmerger remnant, i.e.~the highest maximum rest-mass density that occurs during the first five milliseconds after merging~\citep{Bauswein2019prl,Blacker_2020}. This correlation is useful because one can estimate the density regime of the central object. Figure~\ref{fig:rhomax_fpeak} reveals that models with pions also follow this relation with good accuracy and that there do not occur any particularly strong deviations. Compared to the base model, the calculation adopting a pion mass equal to the vacuum mass exhibits a slight increase in $\rho_\mathrm{max}^\mathrm{max}$. This is consistent with the behavior in Fig.~\ref{fig:Rhomax_SFHo} and~\ref{fig:Rhomax_DD2}, which also indicate an increased softening of the EOS by the inclusion of pions with this effective mass. This specific model features the largest pion content (mostly condensate pions, see Fig.~\ref{fig:ypiypicypith}) and leads to the most significant shifts in Fig.~\ref{fig:fpeak_R16},~\ref{fig:fpeak_lambda} and~\ref{fig:rhomax_fpeak}, not only parallel to the black solid lines but also perpendicular to it. One may hence not fully exclude that more extreme models with even higher pion content could lead to somewhat stronger deviations from the various empirical relations. However, given that our models provide a decent coverage of the allowed parameter space, substantially stronger deviations seem unlikely.

\subsection{Threshold mass} \label{Threshold}
The high-density EOS determines the outcome of BNS mergers, i.e. the nature, evolution and stability of merger remnants, which can be either a black hole for high total binary masses or a massive, rotating NS if the total mass of the system does not exceed the threshold mass for a prompt gravitational collapse. It is thus important to understand whether pions do have an influence in this context as well.

Specifically, we consider the threshold binary mass $M_\mathrm{thres}$ for prompt black-hole formation, which can also be regarded as a general measure for the stability of the remnant. As in previous work \citep{Bauswein2013,Bauswein_2020,Bauswein_2021} we determine $M_\mathrm{thres}$ by considering simulations with different total binary masses $M_\mathrm{tot}$ for a given EOS model. Within this set of calculations we identify the model with the highest $M_\mathrm{tot}=M_\mathrm{stab}$ that does not undergo a direct collapse, and the lightest system with $M_\mathrm{tot}=M_\mathrm{unstab}$ which directly forms a black hole. The threshold mass is then defined by $M_\mathrm{thres}=0.5 (M_\mathrm{stab}+M_\mathrm{unstab})$. We consider the evolution of the minimum lapse function $\alpha_\mathrm{min}$ to distinguish prompt black-hole formation from a delayed gravitational collapse. If  $\alpha_\mathrm{min}$ continuously decreases after the first contact, a prompt collapse occurs. If $\alpha_\mathrm{min}$ levels off and increases, this indicates a bounce of the merger remnant and we classify this behavior as a delayed collapse. This procedure and the definition of $M_\mathrm{thres}$ imply that $M_\mathrm{thres}$ can only be determined with a finite accuracy apart from other numerical errors in the simulations.  For every EOS model in this study we compute the threshold mass to at least within $\pm 0.02~M_\odot$, i.e. $M_\mathrm{unstab}-M_\mathrm{stab}=0.04~M_\odot$.

\begin{table*}
	\centering
    \caption{Determination of threshold mass. First column gives the
      different EOSs models used in this paper. $M_\mathrm{thres}$ is
      the threshold binary mass for prompt collapse. $M_\mathrm{max}$
      is the maximum mass of the non-rotating NS, $R_\mathrm{1.6}$ is
      the radius of a 1.6~$M_\odot$ nonrotating NS, $R_\mathrm{max}$
      is the radius of the nonrotating maximum mass
      NS. $\Lambda_\mathrm{1.4}$ and $\Tilde{\Lambda}_\mathrm{thres}$
      are the tidal deformability of a 1.4$M_\odot$ NS and the tidal
      deformability of a binary system with a total mass equal to
      $M_\mathrm{thres}$. The last four columns provide an estimated
      $M^\mathrm{fit}_\mathrm{thres}$ using available fit
      formulae~\citep{Bauswein_2021} of the form
      $M^\mathrm{fit}_\mathrm{thres}=M^\mathrm{fit}_\mathrm{thres}(M_\mathrm{max},Y)$
      with $Y$ being either $R_{1.6}$, $R_\mathrm{max}$,
      $\Lambda_{1.4}$ or $\Lambda_\mathrm{thres}$. For the estimate we
      employ the respective values of $M_\mathrm{max}$ and $Y$ for the
      EOS model of the given row. The fit formulae are obtained from
      models without pions and thus the estimates allow to assess the
      performance of these relations for models including pions. The
      difference, $M_\mathrm{thres}$ -
      $M^\mathrm{fit}_\mathrm{thres}(M_\mathrm{max},Y)$, between the
      actual threshold mass (second column) and the estimate by the
      respective fit formula is given in parentheses. We quote the
      maximum residuals of the respective fits in parentheses in the
      third line (see~\citep{Bauswein_2021}). 
	\label{tab:threshold_list}}
    \begin{ruledtabular}
	\begin{tabular*}{\textwidth}{lcccccccccc}
		Model & $M_\mathrm{thres}$ & $M_\mathrm{max}$ & $R_\mathrm{1.6}$ & $R_\mathrm{max}$ & $\Lambda_\mathrm{1.4}$ & $\Tilde{\Lambda}_\mathrm{thres}$ & $M^\mathrm{fit}_\mathrm{thres}$ & $M^\mathrm{fit}_\mathrm{thres}$& $M^\mathrm{fit}_\mathrm{thres}$& $M^\mathrm{fit}_\mathrm{thres}$ \\
        & & & & & & & $(Y=R_{1.6})$ & $(Y=R_\mathrm{max})$ & $(Y=\Lambda_\mathrm{1.4})$ & $(Y=\Tilde{\Lambda}_\mathrm{thres})$    \\
        (Max. dev.$/M_\odot$) & & & & & & & (0.042) & (0.059) & (0.056) & (0.085)    \\
        & $[M_\odot]$ & $[M_\odot]$ & $[\mathrm{km}]$ & $[\mathrm{km}]$ & & & $[M_\odot]$ & $[M_\odot]$ & $[M_\odot]$ & $[M_\odot]$ \\
        \hline
         $\mathrm{SFHo}+\pi, \mathrm{Vac. \, mass}$ & 2.810 & 2.017 & 11.542 & 10.085 & 296.937 & 290.362 & 2.806(0.004) & 2.804(0.006) & 2.784(0.026) & 2.796(0.014) \\ 
         $\mathrm{SFHo}+\pi, \mathrm{170 \, MeV}$ & 2.845 & 2.026 & 11.688 & 10.212 & 324.561 & 292.701 & 2.835(0.010) & 2.832(0.013) & 2.811(0.034) & 2.816(0.029) \\ 
         $\mathrm{SFHo}+\pi, \mathrm{200 \, MeV}$ & 2.855 & 2.038 & 11.741 & 10.277 & 332.950 & 291.953 & 2.851(0.004) & 2.850(0.005) & 2.825(0.030) & 2.832(0.023) \\ 
         $\mathrm{SFHo}$ Base & 2.870 & 2.056 & 11.743 & 10.285 & 332.970 & 282.036 & 2.861(0.009) & 2.859(0.011) & 2.835(0.035) & 2.830(0.040) \\
         $\mathrm{DD2}+\pi, \mathrm{Vac. \, mass}$ & 3.250 & 2.381 & 13.069 & 11.692 & 639.278 & 256.841 & 3.257(-0.007) & 3.271(-0.021) & 3.271(-0.021) & 3.228(0.022) \\ 
         $\mathrm{DD2}+\pi, \mathrm{170 \, MeV}$ & 3.290 & 2.390 & 13.220 & 11.791 & 699.649 & 261.744 &  3.287(0.003) & 3.294(-0.004) & 3.325(-0.035) & 3.256(0.034) \\ 
         $\mathrm{DD2}+\pi, \mathrm{200 \, MeV}$ & 3.310 & 2.403 & 13.246 & 11.865 & 700.166 & 256.079 &  3.298(0.012) & 3.314(-0.004) & 3.333(-0.023) & 3.259(0.051) \\ 
         $\mathrm{DD2}$ Base & 3.322 & 2.417 & 13.246 & 11.899 & 700.146 & 250.548 & 3.306(0.016) & 3.327(-0.005) & 3.341(-0.019) & 3.263(0.059) \\ 
	\end{tabular*}
      \end{ruledtabular}
\end{table*}

\begin{figure}[htb]
    \includegraphics[width=\linewidth]{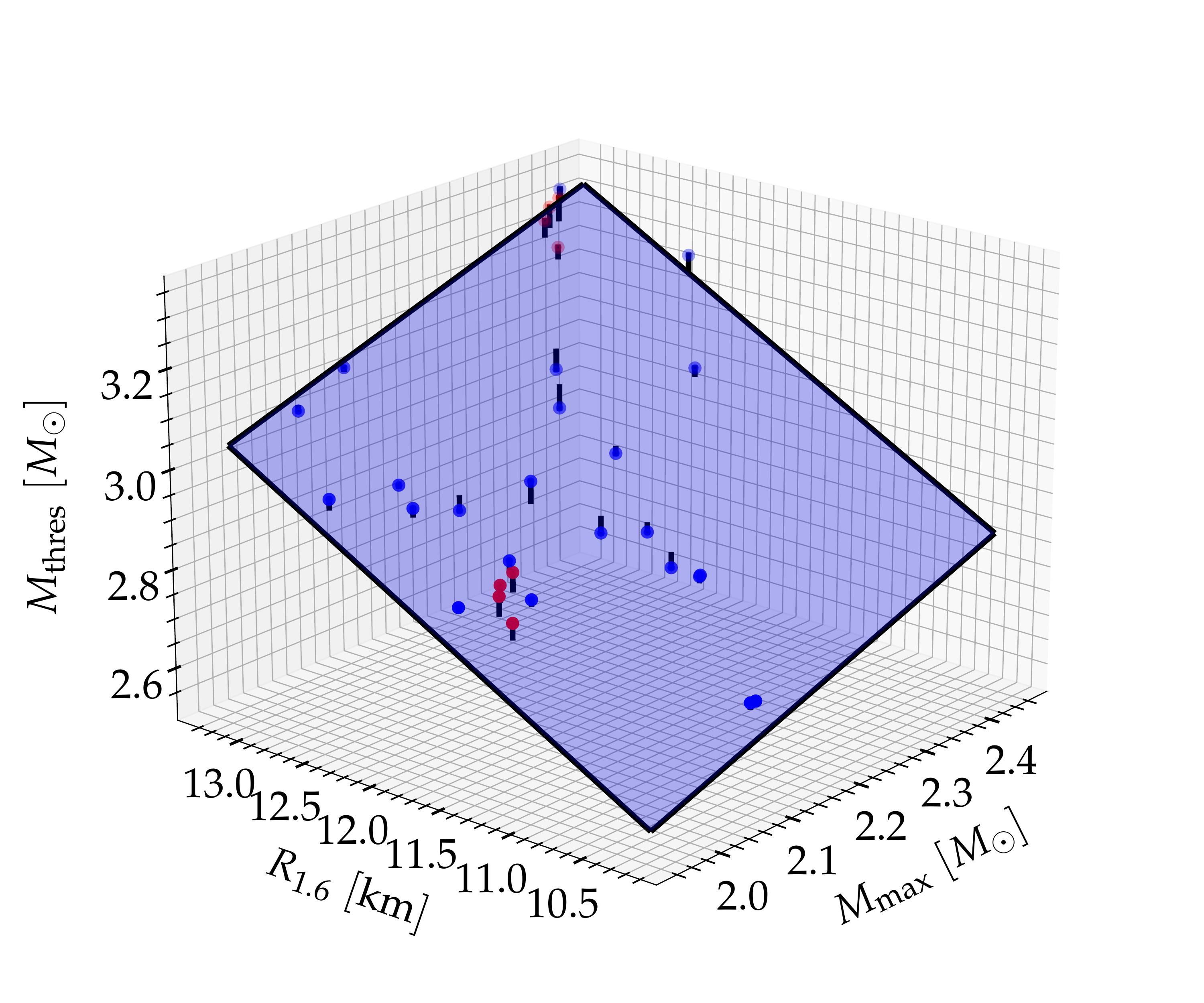}
	\caption{Threshold binary mass for prompt collapse as a function of the radius $R_{1.6}$ of a 1.6~M$_{\odot}$ NS and the maximum mass $M_\mathrm{max}$ of nonrotating NSs. Blue dots display results for EOS models without pions, and the blue plane shows a least square fit to these data (set ``b'' from~\citep{Bauswein_2021}). Vertical lines visualize the deviation between the respective data point and the fit (plane). Red dots show the data for the simulations with the EOSs employed in this study.}	
	\label{fig:Mthres_R16}
\end{figure}

Table~\ref{tab:threshold_list} lists the threshold mass for all EOS models\footnote{The values for $M_\mathrm{thres}$ for our base models slightly deviate from the ones given in~\citep{Bauswein_2021} for the same EOS. The reason is that $M_\mathrm{thres}$ is determined by a bracketing method considering the outcome of binary simulations with different total masses, and in this study we obtain $M_\mathrm{thres}$ with higher accuracy to within $\pm0.02~M_\odot$. We note that the outcome of the respective simulation data are fully consistent with each other.}. For both base EOS models we find that $M_\mathrm{thres}$ is systematically reduced by the inclusion of pions. For a given base model the decrease of $M_\mathrm{thres}$ is stronger for smaller $m_\pi$ and can be as large as $0.08~M_\odot$ (about two per cent of $M_\mathrm{thres}$). This is in line with the behavior already found in the analysis of the GW emission, where smaller effective pion masses lead to a generally more pronounced softening of the EOS. Still, this behavior may not be obvious, since the inclusion of pions actually leads to a stiffening of the EOS in the regime of very high temperatures, which is not strongly depending on the pion mass. Apparently, the overall softening has a stronger impact on the stability, which may be explained by the fact that only a very small volume of the remnant features conditions where a stiffening by pions could occur.

As features of the GW signal, the threshold mass for prompt black-hole formation follows empirical relations, which describe $M_\mathrm{thres}$ as function of stellar parameters of cold, nonrotating NSs. A number of such relations have been put forward based on the consideration of a large set of different EOS models without pions~\citep{Bauswein2013,Bauswein2014,Koeppel2019,Bauswein_2020,Agathos2020,Bauswein_2021,Tootle2021,Kashyap2022,Koelsch2022}. These relations typically depend on two stellar parameters including the maximum mass $M_\mathrm{max}$ of nonrotating NSs. A bi-linear ansatz 
\begin{equation}
M^\mathrm{fit}_\mathrm{thres}(M_\mathrm{max},Y)=a M_\mathrm{max} + b Y + c
\end{equation}
works well with fit parameters $a, b$ and $c$ and $Y$ being either the radius $R_\mathrm{1.6}$ of a 1.6$M_\odot$ NS or radius $R_\mathrm{max}$ of a non-rotating neutron star at its maximum mass or tidal deformability $\Lambda_\mathrm{1.4}$ of a 1.4$M_\odot$ NS or tidal deformability $\Tilde{\Lambda}_\mathrm{thres}$ at the threshold mass~\citep{Bauswein_2020,Bauswein_2021}. All these stellar parameters are listed in Tab.~\ref{tab:threshold_list}.

The inclusion of pions in an EOS model changes simultaneously the stellar parameters of cold, nonrotating NSs and the threshold mass for black-hole formation. It is thus important to check if the models with pions do follow these relations which are obtained by fits to EOS models without pions. We note that these relations are for instance employed to establish constraints on NS properties and are generally used to interpret BNS merger observations. Table~\ref{tab:threshold_list} provides the estimated threshold mass for all models of this study using empirical relations constructed based on EOSs without pions \citep{Bauswein_2021}, where we insert the stellar parameters of our modified and the original EOS tables. Specifically, we list estimates based on the fits numbered 1, 15, 29 and 43 in~\citep{Bauswein_2021}, which employ the EOS set labeled ``b'' therein that contains viable EOS models without phase transition to deconfined quark matter. The EOS models with pions follow these relations with very good accuracy and one cannot identify significant deviations. The differences between the fits and the actual threshold mass are provided in Tab.~\ref{tab:threshold_list}. The deviations are generally small, especially in comparison to the maximum residuals of the fits (see Tab.~\ref{tab:threshold_list}).

The performance of the fits is also visualized in Fig.~\ref{fig:Mthres_R16}, which shows $M^\mathrm{fit}_\mathrm{thres}(M_\mathrm{max},R_{1.6})$. The models with pions (red data points) do not feature any considerable deviations. A similar behavior is found for other empirical relations of \citep{Bauswein_2021}.

Generally, this comparison shows that such type of empirical relations for $M^\mathrm{fit}_\mathrm{thres}$ are accurate and can be used although they neglect pions. This is because pions impact the threshold mass and stellar parameters of NSs such that the resulting effects nearly cancel each other. This also implies that EOS constraints based on relations for the threshold mass as in \citep{Bauswein2017,doi:10.1063/1.5117803,Bauswein_2021}, are not affected by ignoring pions.

\subsection{Mass ejection}

\begin{figure*}
    \centering
    \subfigure[SFHo]{\includegraphics[width=0.48\textwidth]{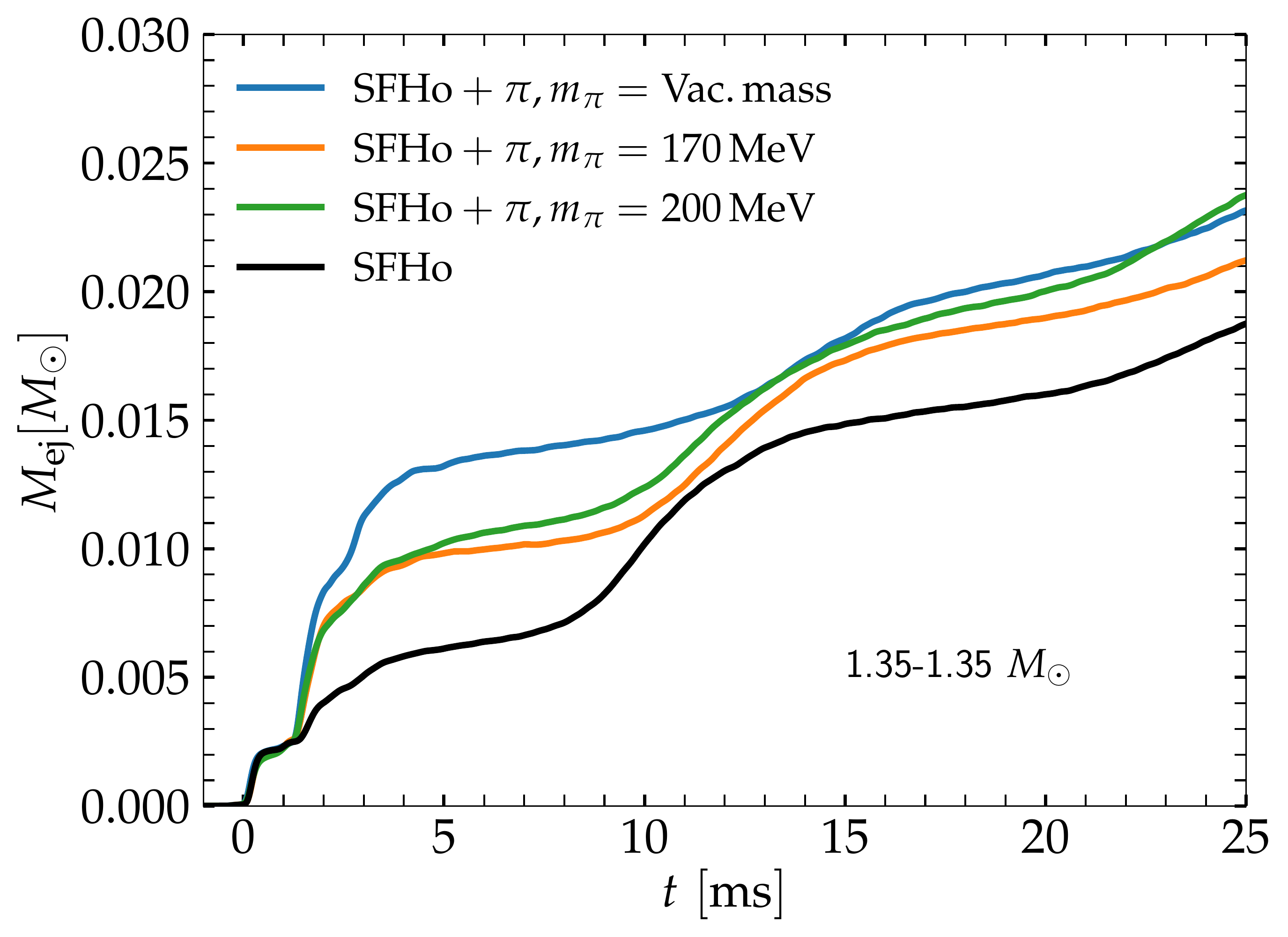}\label{fig:Mej_SFHo}}
    \subfigure[DD2]{\includegraphics[width=0.48\textwidth]{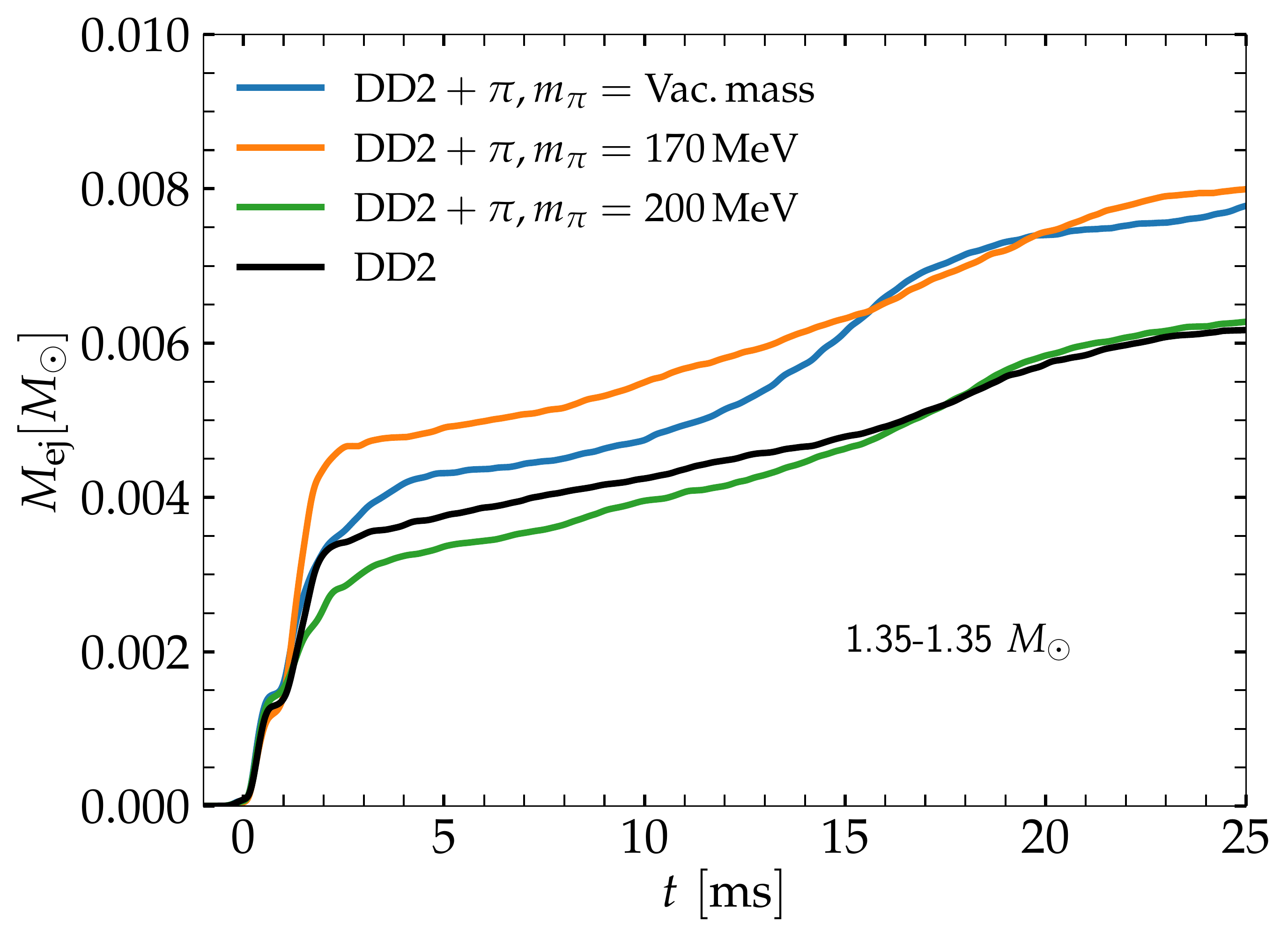}\label{fig:Mej_SDD2}}
    \caption{Time evolution of the ejecta mass for 1.35-1.35$M_\odot$ BNS mergers using modified SFHo EOSs and DD2 EOSs along with their base EOSs. Time zero corresponds to the instant when the minimum lapse function features a first minimum.}
    \label{fig:Mej}
\end{figure*}

We finally address the mass ejection of BNS mergers, which is relevant for the nucleosynthesis of heavy elements through the rapid neutron-capture process and electromagnetic counterparts, so-called kilonovae, e.g.~\citep{Fernandez2016,Baiotti2017,Shibata2019,Metzger2019,Bauswein_2019,Radice2020,Nakar2020,Cowan2021,Rosswog2022,Janka2022}. Clearly, a key quantity is the ejecta mass $M_\mathrm{ej}$, which determines the overall production of heavy elements and the properties of the electromagnetic transient. Mass ejection of BNS mergers includes several phases and with the simulations described here we can only assess the dynamical ejecta, i.e.~the material which becomes gravitationally unbound during the first few 10~milliseconds after merging.

Figure ~\ref{fig:Mej} shows the amount of unbound material as function of time for the EOS models based on SFHo (left panel) and DD2 (right panel) considering 1.35-1.35~$M_\odot$ binaries. We clearly recognize that dynamical mass ejection is enhanced for models with pions and that smaller effective pion masses tend to produce more unbound matter. This behavior may be expected since softer EOSs yielding smaller NS radii lead to larger ejecta masses based on studies which ignore pions~\citep{2013ApJ...773...78B,Hotokezaka2013ejecta}. However, the softening by the inclusion of pions is relatively minor and NS radii do not change by a lot in particular for larger effective pion masses. In this sense the impact of pions visible in Fig.~\ref{fig:Mej} is stronger than expected. Based on previous work one would expect $M_\mathrm{ej}$ to increase by only a few percent if NS radii decrease by one or two per cent, employing fit formulae that link the ejecta mass and NS radii (see e.g.~Fig.22 in~\citep{Janka2022} based on fits from~\citep{Dietrich_2017, Coughlin_2018, Kr_ger_2020, Nedora_2021}). In Fig.~\ref{fig:Mej} we find an increase of more than 20\% in the dynamical ejecta mass for models employ effective pion masses equal to the vacuum values. 

The absolute values of the base models in Fig.~\ref{fig:Mej} at about 10~ms after merging are compatible with the results in~\citep{2013ApJ...773...78B}, where $M_\mathrm{ej}$ is determined at the same time. Also, typical estimates from the existing fit formulae are roughly compatible with $M_\mathrm{ej}$ at about 10~ms after the collision.

We remark that relations between $M_\mathrm{ej}$ and NS parameters generally show a large scatter and that the ejecta mass is very sensitive to numerics and is certainly not fully converged in the simulations in addition to missing physics in the calculations (see e.g. discussion in~\citep{Janka2022}). This may also explain why there is no exact hierarchy with respect to $m_\pi$. SPH simulations with different particle numbers in  \citep{2013ApJ...773...78B} (without neutrinos) indicate that the ejecta mass can vary by several 10\%, where $M_\mathrm{ej}$ does not follow a clear trend with particle number but rather seems to fluctuate statistically. The differences in Fig.~\ref{fig:Mej} are hence only tentative\footnote{We perform simulations with about 600,000 SPH particles for the SFHo EOS without pions and with pions following the evolution into the early postmerger phase. Comparing these simulations, we observe very similar trends, namely that the models with pions yield a significant increase of the ejecta mass compared to the run without pions. Similarly, these higher-resolution runs confirm the behavior of the torus mass discussed below, where again the inclusion of pions with relatively small masses leads to an increase compared to the model without pions. Higher resolution tends to increase the torus mass by $\lesssim10\%$.} but caution that ignoring pions might have a more pronounced impact on the mass ejection than suggested by the change of stellar properties of cold, nonrotating NSs. This might in particular affect attempts to infer EOS constraints from observed kilonovae (e.g.~\citep{Coughlin_2018}). We note that the inclusion of pions does affect the EOS in a complicated non-continuous way in different regimes including softening and stiffening with respect to a base model. This may as well contribute to the behavior seen in Fig.~\ref{fig:Mej}.

We note that at most a very small fraction of the ejecta ($\sim 1 \%$ for SFHo, $0 \%$ for DD2) originates from densities beyond the threshold density where pion condensation sets in at about $3\times 10^{14}~\mathrm{g/cm^3}$ in beta-equilibrium (see Fig.~\ref{fig:Esymm}). Hence, in merger models with pions the initial proton fraction of the ejecta remains basically unaffected compared to the base model. This may suggest that the impact of pions on the composition of the ejecta is only a secondary effect even if weak interactions were taken into account in the simulations.

\begin{figure*}
    \centering
    \subfigure[SFHo]{\includegraphics[width=0.48\textwidth]{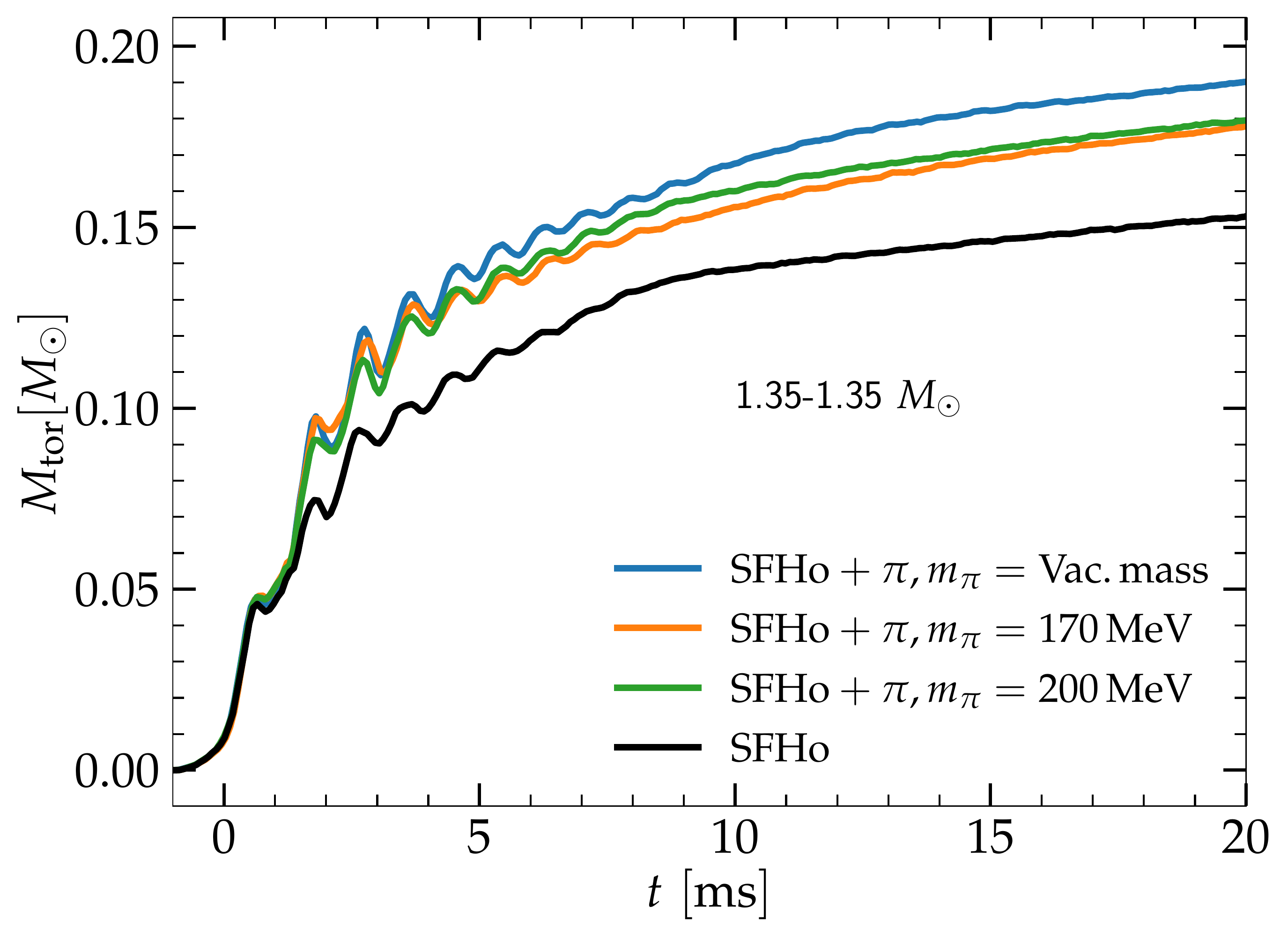}\label{fig:Mtor_SFHo}}
    \subfigure[DD2]{\includegraphics[width=0.48\textwidth]{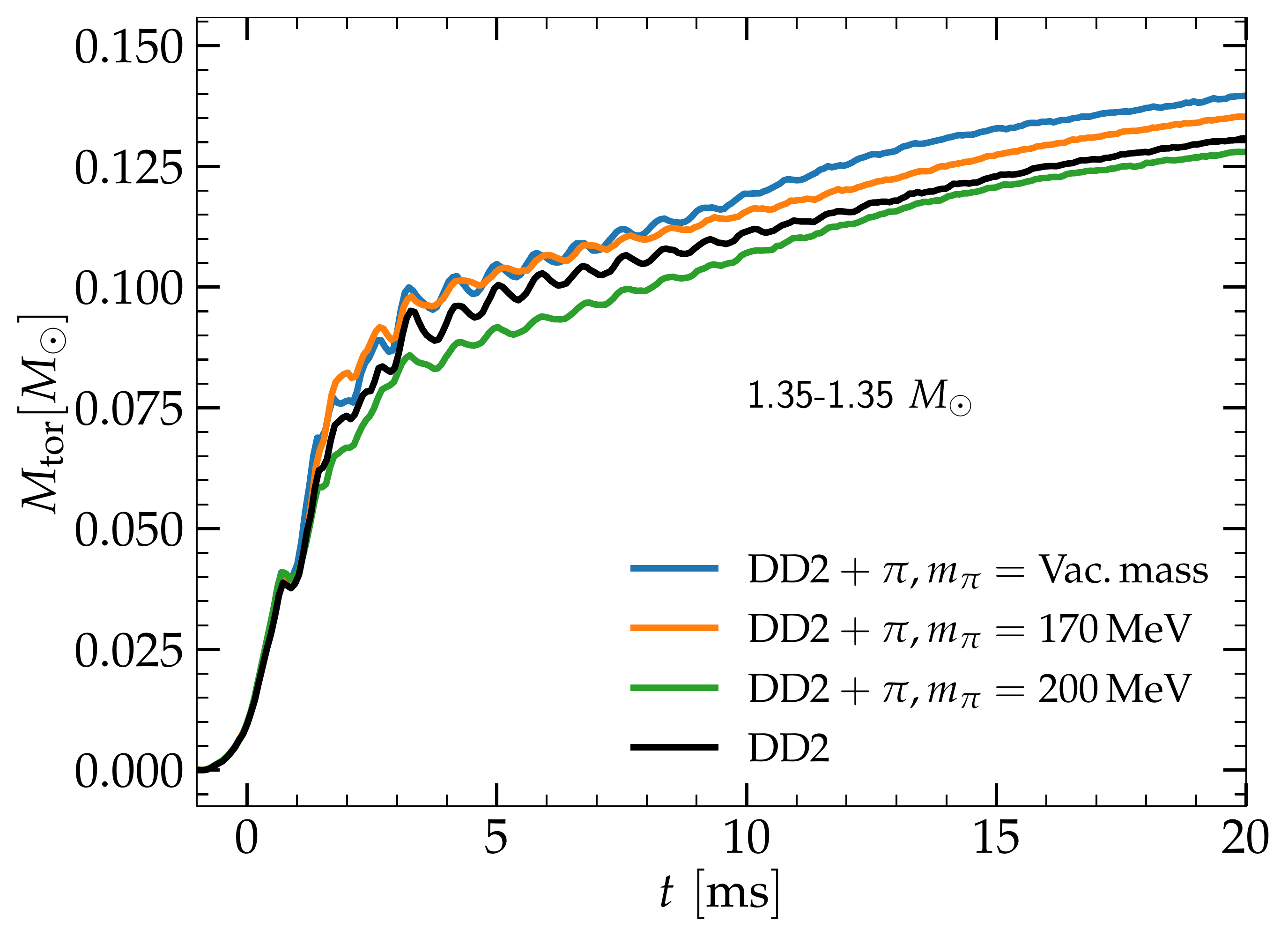}\label{fig:Mtor_DD2}}
    \caption{Time evolution of the torus mass for 1.35-1.35$M_\odot$ BNS mergers using modified SFHo EOSs and DD2 EOSs along with their base EOSs. Time zero corresponds to the instant when the minimum lapse function features a first minimum.}
    \label{fig:Mtor}
\end{figure*}

Figure~\ref{fig:Mtor} shows the evolution of the torus mass as
function of time, where we define the torus as material with a density
below $10^{13} \mathrm{g/cm^3}$ as a coarse estimate. This quantity
can serve as a rough estimate of the amount of secular ejecta, which
consists of a fraction of a few ten per cent of this torus material
(see, e.g.,~\citep{Just2022a} for an extensive analysis of torus
ejecta). As for the dynamical ejecta, we attempt to assess whether the inclusion of pions affects the torus mass. In Fig.~\ref{fig:Mtor}, we observe an increase of the torus mass for models which include pions, i.e. for softer EOSs (except for the DD2 based EOS with $m_\pi=200$~MeV). Alternatively, we can consider fit formulae that relate the torus mass to TOV properties, and are based on simulation results (see Fig.~23 in~\citep{Janka2022} that compiles formulas from~\citep{Dietrich2020,Kr_ger_2020,Nedora_2021}). We notice that these relations are approximate and different fit formulae show sizeable variations. Nevertheless, they indicate that softer EOSs lead to smaller torus masses. Hence, fit formula predict that the change of TOV properties induced by the inclusion of pions would lead to a decrease of the torus mass  contrary to our findings.

\section{Conclusions}
\label{sec:conclusions}
We study the impact of pions on BNS mergers by considering neutral and charged pions as a free non-interacting boson gas, which we add to an existing base EOS. We use the temperature and composition dependent SFHo and DD2 EOSs~\citep{Hempel2010,2012ApJ...748...70H,2013ApJ...774...17S,2005PhRvC..71f4301T, Hempel2010,2010PhRvC..81a5803T}. Although the pion mass may change in a dense medium~\citep{2014PhRvD..90k5027N,2017PhRvD..95c6020J,2021PhRvD.104e4005T,Fore2023}, for this study we assume a constant value of an effective pion mass under all thermodynamical conditions. To assess the impact of changes in the pion mass, we choose different values for a constant effective pion mass. We consider calculations adopting the vacuum values of the pion masses, 170~MeV and 200~MeV.

Pions in dense matter can occur as a condensate, which does not directly add to the pressure, but affects the total pressure at a given $Y_\mathrm{e}$ by changing the equilibrium conditions and thus the baryonic contribution. This leads to a softening of the NS EOS. At finite temperature thermal pions occur, which also affect the chemical equilibrium but may overall stiffen the EOS for very high temperatures. 

For isolated, cold NSs in neutrinoless beta-equilibrium, the inclusion of pions leads to a softening of the EOS compared to the respective model without pions. This affects the TOV solutions of NSs. The inclusion of pions leads to a decrease of the maximum mass of NSs and to smaller NS radii for a given mass. The magnitude of these effects is inversely related to the adopted pion mass. Radii decrease by at most $\sim$200\,m for pion masses close to the vacuum value. For $m_\pi=200$~MeV models, changes of the stellar structure is very small since the condensation is largely suppressed.

The tidal deformability can change by about 10 percent for pion masses equal to their vacuum value. This might affect the EOS inference from a GW analysis study of the inspiral if microphysical parameters of nuclear matter are deduced using microphysical EOS models which neglect pions. 

To understand the influence of pions on the BNS mergers, we conduct relativistic hydrodynamical simulations with the different EOS models with and without pions. We focus on 1.35-1.35~$M_\odot$ mergers but also simulate equal-mass binaries with higher total binary mass.

Concerning the general dynamics of the merger there are no qualitative differences comparing simulations with and without pions. However, we find that depending on the adopted pion mass the electron fraction can be significantly altered, albeit we employ a rather crude treatment of weak interactions by simply advecting the initial electron fraction. For a more sophisticated treatment one should also include muons. We briefly estimate the potential impact of muons in the appendix. In essence, we find that if pions have a sizable impact on the stellar structure for effective pion masses close to the vacuum values, the changes by adding muons are relatively small. For most of our simulated binary configurations we find that pions in the condensate are more abundant than thermal pions. Only for high effective pion masses, thermal pions dominate in the postmerger phase.

We analyze the GW signal in our simulations and observe that the dominant postmerger GW frequency is shifted to higher frequencies by up to 200\,Hz compared to the base model without pions. The shift is more pronounced for smaller effective pion masses. We evaluate how these changes compare to empirical relations between stellar parameters of NSs and the dominant postmerger frequency (e.g.~\citep{Bauswein2012,Bauswein_2012,Hotokezaka2013,Bauswein2014,Bernuzzi2015,Takami2015,Rezzolla2016,Lehner2016,Breschi2019,Tsang2019,Blacker_2020,Vretinaris_2020}). These relations have been obtained in the literature considering large sets of EOS models that neglect pions. We find that these relations remain to good accuracy valid because the inclusion of pions changes both the properties of nonrotating NSs and of postmerger remnants. This justifies to use these relations for the EOS inference in GW observations. Only for small pion masses close to their vacuum values slight deviations may occur.

Similarly, we investigate the impact of pions on the threshold binary mass $M_\mathrm{thres}$ for prompt black-hole formation. Again, we find that the overall softening of the EOS compared to a base model without pions, leads to a reduction of $M_\mathrm{thres}$. The reduction can be as large as about 0.07~$M_\odot$ and is more significant for small pion masses. Again, the inclusion of pions does not lead to significant deviations in empirical relations for $M_\mathrm{thres}$ which have been built using models without pions~\citep{Bauswein_2021}. Similarly, we find relations for the maximum density in the early postmerger evolution to be fulfilled~\citep{Blacker_2020}.

We also briefly assess the consequences for mass ejection in NS mergers. We find that the presence of pions leads to enhanced mass ejection by a few ten per cent compared to the base models. Also ejecta properties such as their mass have been related to stellar parameters of nonrotating NSs~\citep{2013ApJ...773...78B,Hotokezaka2013ejecta}. In this respect we observe that the ejecta mass increase in our models with pions is stronger than one would naively expect based on the changes of stellar parameters of cold, nonrotating stars. This might indicate a systematic bias with consequences for attempts to infer stellar parameters from kilonova observations, but we caution that ejecta properties are generally challenging to resolve and that the uncertainties and the intrinsic scatter in relations describing kilonova features as function of NS properties are rather large.

Future research should improve this first assessment of pion effects in NS mergers in several ways. We describe pions as noninteracting Boson gas and choose a constant pion mass, which allows only a coarse description of the impact on the EOS. A more sophisticated treatment of weak interactions is desirable to understand the impact on the ejecta  composition, where our crude simulations already indicate a potential influence as pions change the electron fraction. Clearly, more EOS models should be considered in future work.

\begin{appendix}

\section{Impact of muons}

\begin{figure*}[htb]
    \centering
    \subfigure[SFHo]{\includegraphics[width=0.48\textwidth]{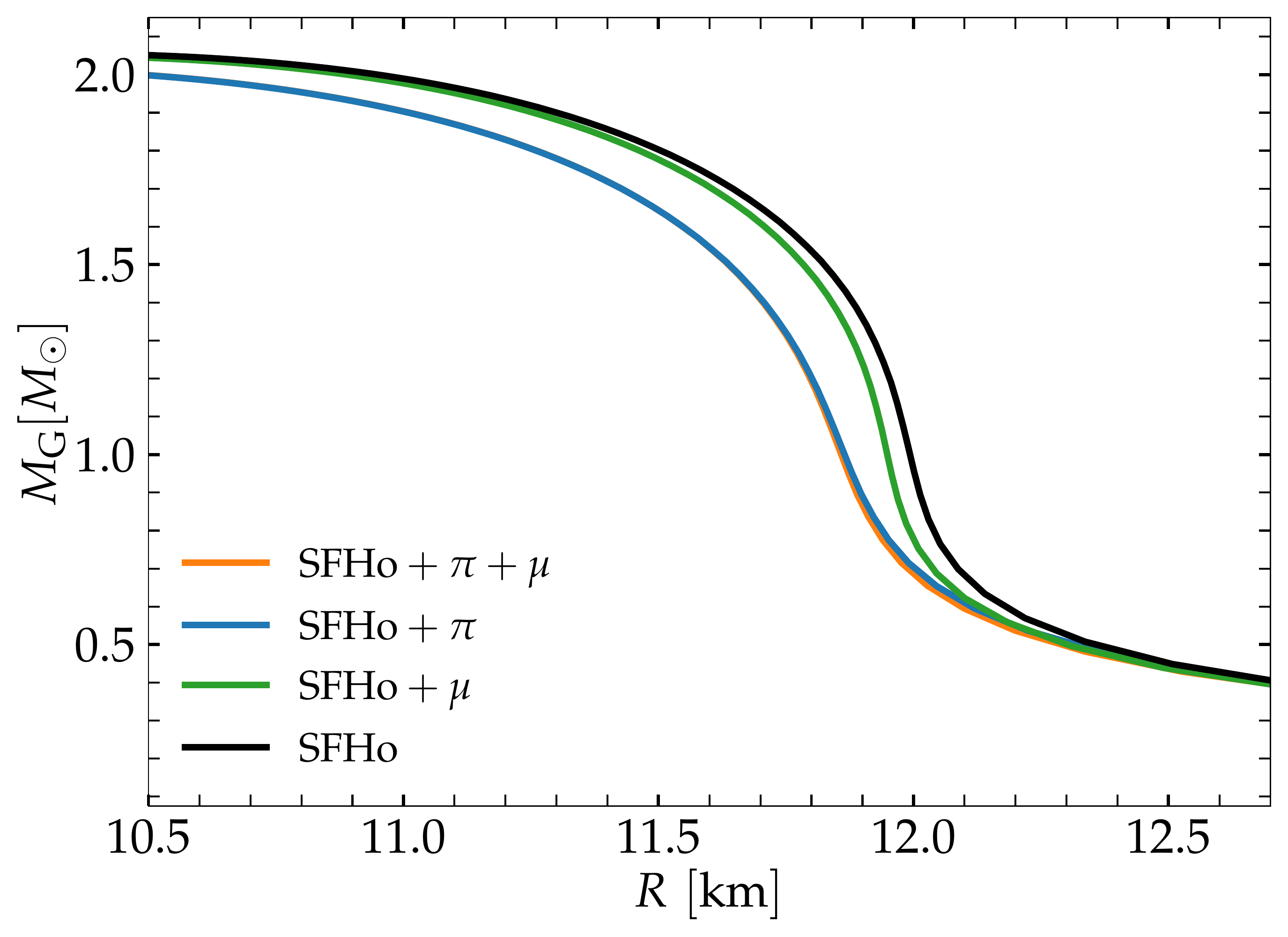}\label{fig:TOV_M_R_Muons_SFHO}}
	  \subfigure[DD2]{\includegraphics[width=0.48\textwidth]{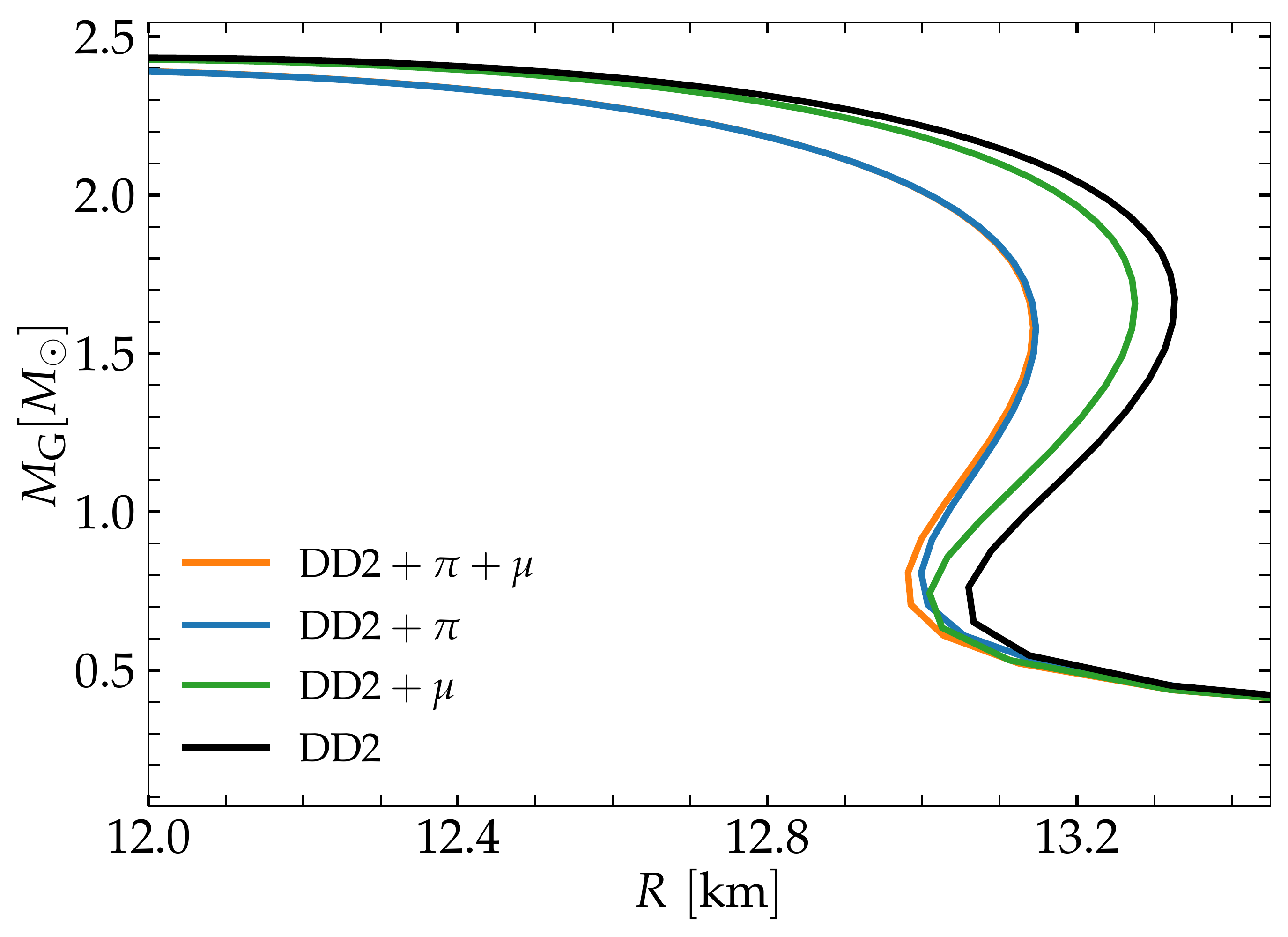}\label{fig:TOV_M_R_Muons_DD2}}
  	\caption{Gravitational mass versus NS radius for the Base, Base+$\pi$, Base+$\mu$ and Base+$\pi$+$\mu$ models with SFHo(left) and DD2(right) as base model.}
  	\label{fig:TOV_M_R_Muons}
\end{figure*}

In this study we focus on the impact that pions have on BNS. We do not consider muons, which we also neglected in many currently available EOS tables, although one should expect that also muons are present in NS matter (see~\citep{Bollig_2017,Loffredo2022,Alford.Harutyunyan.Sedrakian:2022} for the consideration of muons in simulations of core-collapse supernovae and NS mergers). The chemical potential of electrons can reach the rest mass of muons of 105.7~MeV, which can be seen in Fig.~\ref{fig:mun-mup}. Muons are the decay products of pions and should thus in principle be taken into account. We assess the influence of muons by considering stellar equilibrium solutions of isolated stars. We are mostly interested in the impact on the GW signal and the collapse behavior, both of which are determined primarily by the high-density regime of the EOS.

The consideration of muons in EOS tables is in principle straightforward as they can be treated as an ideal Fermi gas similar to electrons. Including muons in the existing tables requires some minor adaptions because the base models determine the contributions of the different constituents by assuming charge neutrality between protons, electrons and positrons. Adding muons and pions thus changes the conditions for charge neutrality. To this end we first remove the contributions from electrons (and positrons for $T>0$) from the base model as we did for adding pions. We then recalculate at every point of the EOS table the contributions of all considered leptons and pions considering the respective new relation for charge neutrality and chemical equilibrium. We consider stellar configurations at zero temperature and neutrinoless $\beta$-equilibrium. Moreover, we investigate stellar models at finite temperature to mimic the behavior of merger remnants. For those we choose a finite temperature of 20~MeV and fix the lepton fraction. At every point of the EOS table with this temperature, we solve the conditions for charge neutrality under the assumption of chemical equilibrium between the nucleons, the leptons (electrons, muons, their anti-particles, and the respective neutrinos and anti-neutrinos), and the pions. We emphasize that this analysis is meant to estimate the impact on GWs and bulk properties of NSs and BNS mergers, which are mostly determined by the high-density EOS, where these conditions are well justified.

\begin{figure*}[htb]
    \centering
    \subfigure[SFHo]{\includegraphics[width=0.48\textwidth]{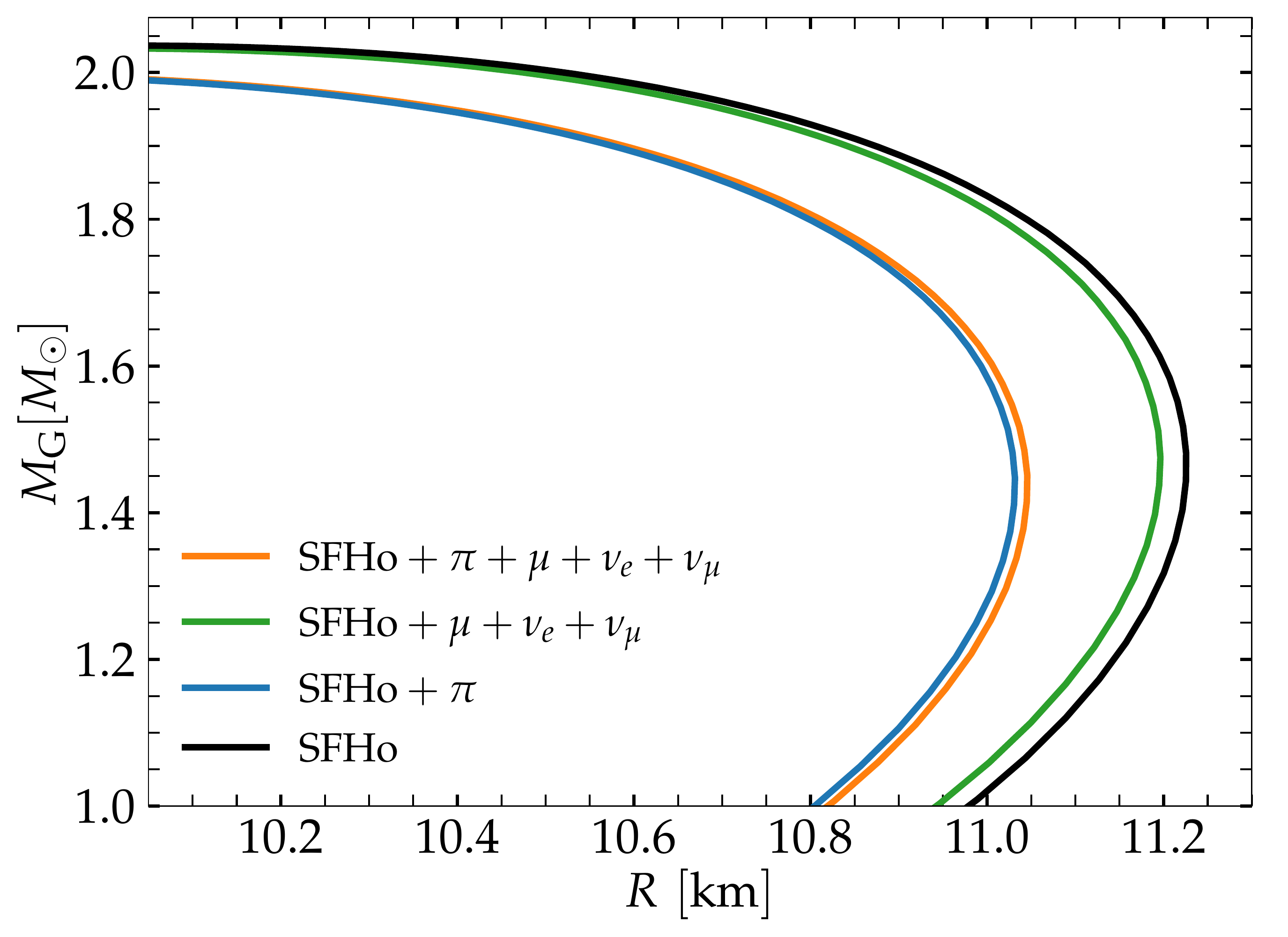}\label{fig:HTOV_M_R_Muons_SFHO}}
	\subfigure[DD2]{\includegraphics[width=0.48\textwidth]{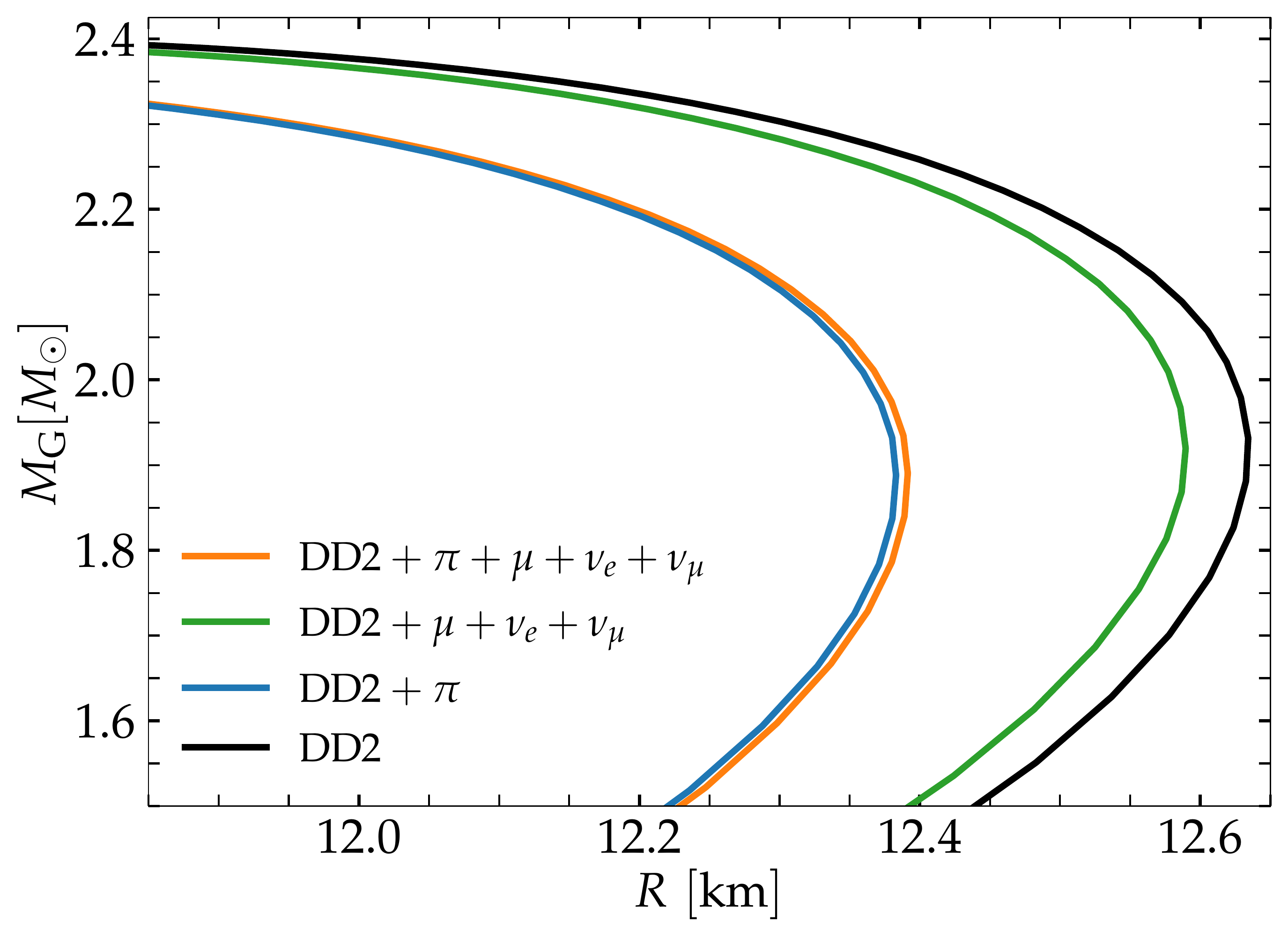}\label{fig:HTOV_M_R_Muons_DD2}}
   	\caption{Gravitational mass versus NS radius for the Base, $\mathrm{Base}+\pi$, Base+$\mu$+$\nu_\mathrm{e}$+$\nu_\mu$, Base+$\pi$+$\mu$+$\nu_\mathrm{e}$+$\nu_\mu$ for a fixed electronic lepton fraction $Y_\mathrm{lep,e}=0.04$, and muonic lepton fraction $Y_\mathrm{lep,\mu}=0.015$ (if muon and muon neutrinos are present) at temperature $T=20\,\mathrm{MeV}$ with SFHo(left) and DD2(right) as base models. For these calculations the surface of the stars is defined at a rest-mass density of $10^{14}\,\mathrm{g/cm^3}$.}	
	\label{fig:HTOV_M_R_Muons}
\end{figure*}

Following the aforementioned procedure we construct different versions of the EOS tables at zero temperature using the SFHo and DD2 base models: the base model with pions only, the base models with muons only, and the base model with muons and pions. For those we determine the beta-equilibrium EOS at zero temperature and compute TOV solutions, which are shown in Fig.~\ref{fig:TOV_M_R_Muons} for SFHo (left panel) and DD2 (right panel). Here we only consider an effective pion mass which equals the vacuum mass.

The inclusion of muons without pions leads to a mild softening of the EOS and yields smaller NS radii. The softening induced by pions is in comparison more significant. For instance, the reduction of the radius and the maximum mass is more pronounced. Interestingly, we find that adding muons to an EOS with pions has only a negligible impact. This is understandable because in the presence of pions the difference $\mu_\mathrm{n}-\mu_\mathrm{p}$ is limited by the pion mass. Hence, under the conditions of neutrinoless beta-equilibrium, $\mu_\mathrm{n}-\mu_\mathrm{p}=\mu_\mathrm{e}=\mu_\pi=\mu_\mu$, the muon production is suppressed as $\mu_\mu$ cannot grow a lot. One may thus argue that neglecting muons is a good approximation in our models with relatively small effective pion masses. However, for the models with higher effective pion mass the suppression of muons will be less restrictive. In this case the inclusion of pions has not a strong impact on the EOS (see Sect. ~\ref{Evolution}), but modifications by muons may become somewhat more significant (see the case where we consider only muons and the base model).

A similar picture arises for finite-temperature EOS models, where we
also consider neutrinos (see Fig.~\ref{fig:HTOV_M_R_Muons}). We fix
the electronic lepton fraction to $Y_\mathrm{lep,e}=0.04$ and the
muonic lepton fraction to $Y_{\mathrm{lep},\mu}=0.015$ (if muon and
muon neutrinos are present), which is coarsely representative of BNS
merger remnants. Here, we assume that the initial electronic and muonic
lepton fractions of the original cold neutron stars are approximately
preserved during the merger at the high densities considered here due
to neutrino-trapping. To better assess the impact on the stellar structure, in the following we consider only the NS core and its radius by computing TOV models for which we disregard densities
below $10^{14}~\mathrm{g/cm^3}$. In this way, we avoid that the stellar models are inflated by the finite temperature at lower densities.

Again, we find that the impact of muons is nearly negligible if pions
are included as can be seen by comparing the blue and orange lines in
Fig.~\ref{fig:HTOV_M_R_Muons}. The comparison between the different
EOS versions also reveals that ignoring neutrinos in our analysis can
be justified as they apparently have only a small influence on the
stellar structure. We thus conclude that considering only pions and
ignoring muons in the BNS merger simulations discussed in the main
text is a reasonable approximation for a first study evaluating the
bulk properties of mergers like GWs and the conditions for black-hole
formation.

\end{appendix}

\begin{acknowledgements}
  We thank Sebastian Blacker, Bengt Friman, Pok Man Lo, Tetyana
  Galatyuk and Jochen Wambach for helpful discussions. VV and AB
  acknowledge support by Deutsche Forschungsgemeinschaft (DFG, German
  Research Foundation) - Project-ID 138713538 - SFB 881 (“The Milky
  Way System”, subproject A10). NR and GMP acknowledge support by the
  European Research Council (ERC) under the European Union’s Horizon
  2020 research and innovation program (ERC Advanced Grant KILONOVA
  No. 885281). AB and VV acknowledges by the European Research Council
  (ERC) under the European Union’s Horizon 2020 research and
  innovation program under grant agreement No. 759253. AB, GMP, ILA
  and NR acknowledge support by the Deutsche Forschungsgemeinschaft
  (DFG, German Research Foundation -- Project-ID 279384907 - SFB 1245
  and MA 4248/3-1. AB and GMP acknowledge support by the State of
  Hesse within the Cluster Project ELEMENTS.
\end{acknowledgements}

%\bibliography{references}

\end{document}